\documentclass[sn-mathphys-num]{sn-jnl}% Math and Physical Sciences Numbered Reference Style 
\usepackage{graphicx}%
\usepackage{multirow}%
\usepackage{amsmath,amssymb,amsfonts}%
\usepackage{amsthm}%
\usepackage{mathrsfs}%
\usepackage[title]{appendix}%
\usepackage{xcolor}%
\usepackage{textcomp}%
\usepackage{manyfoot}%
\usepackage{booktabs}%
\usepackage{algorithm}%
\usepackage{algorithmicx}%
\usepackage{algpseudocode}%
\usepackage{listings}%
\usepackage[export]{adjustbox}
\usepackage{subcaption}
\usepackage{color}
\usepackage{braket}
\usepackage{dsfont}
\usepackage{comment}

\usepackage{tikz,tikz-cd,float}
\usetikzlibrary{positioning, quotes, decorations.markings, decorations.pathmorphing, shapes, calc}

\newcommand{\ol}{\overline}
\newcommand{\wt}{\widetilde}
\newcommand{\Tr}{\mathrm{Tr}}

\newcommand{\Ops}{\mathrm{Ops}}
\newcommand{\End}{\mathrm{End}}
\newcommand{\Hom}{\mathrm{Hom}}

\newcommand*\diff{\mathop{}\!\mathrm{d}}
\newcommand{\pd}{{\partial}}
\newcommand{\id}{{\mathds{1}}}

\newcommand{\BA}{{\mathbf A}}
\newcommand{\BB}{{\mathbf B}}

\newcommand{\BZ}{{\mathbf Z}}

\newcommand{\BPsi}{{\mathbf \Psi}}

\newcommand{\C}{\mathbb C}
\newcommand{\R}{\mathbb R}
\newcommand{\Z}{\mathbb Z}

\newcommand{\fsl}{\mathfrak{sl}}
\newcommand{\fgl}{\mathfrak{gl}}
\newcommand{\fg}{\mathfrak{g}}

\newcommand{\CA}{{\mathcal A}}

\newcommand{\CC}{{\mathcal C}}
\newcommand{\CD}{{\mathcal D}}

\newcommand{\CF}{{\mathcal F}}
\newcommand{\CG}{{\mathcal G}}
\newcommand{\CH}{{\mathcal H}}

\newcommand{\CL}{{\mathcal L}}
\newcommand{\CM}{{\mathcal M}}
\newcommand{\CN}{{\mathcal N}}
\newcommand{\CO}{{\mathcal O}}

\newcommand{\CT}{{\mathcal T}}

\newcommand{\CV}{{\mathcal V}}
\newcommand{\CW}{{\mathcal W}}  
\newcommand{\CX}{{\mathcal X}}
\newcommand{\CY}{{\mathcal Y}}
\newcommand{\CZ}{{\mathcal Z}}

\newcommand{\norm}[1]{{{:\!{#1}\!:}}}
\newcommand{\be}{\begin{equation}}
	\newcommand{\ee}{\end{equation}}

\tikzset{middlearrow/.style={
		decoration={markings,
			mark= at position 0.5 with {\arrow{#1}} ,
		},
		postaction={decorate}
	}
}

\theoremstyle{thmstyleone}%
\theoremstyle{thmstyletwo}%

\theoremstyle{thmstylethree}%

\begin{document}

\title[Line Operators in $U(1|1)$ Chern-Simons Theory]{Line Operators in $U(1|1)$ Chern-Simons Theory}


\author[1,2]{\fnm{Niklas} \sur{Garner}}\email{niklas.garner@maths.ox.ac.uk}

\author*[3]{\fnm{Wenjun} \sur{Niu}}\email{wjniu950925@gmail.com}
\equalcont{These authors contributed equally to this work.}

\affil[1]{\orgdiv{Department of Physics}, \orgname{University of Washington}, \orgaddress{\city{Seattle}, \postcode{98195}, \state{Washington}, \country{USA}}}

\affil[2]{\orgdiv{Mathematical Institute}, \orgname{University of Oxford}, \orgaddress{\street{Woodstock Road}, \city{Oxford}, \postcode{OX2 6GG}, \country{UK}}}

\affil*[3]{\orgname{Perimeter Institute for Theoretical Physics}, \orgaddress{\street{31 Caroline St N}, \city{Waterloo}, \postcode{N2L 2Y5}, \state{Ontario}, \country{Canada}}}


\abstract{We analyze the non-semisimple category of line operators in Chern-Simons gauge theories based off the Lie superalgebra $\mathfrak{gl}(1|1)$. Our proposal is that the category of line operators $\mathcal{C}$ can be identified with the derived category of modules for a boundary vertex operator algebra $\mathcal{V}$ realized as a certain infinite-order simple current extension of the affine current algebra $V(\mathfrak{gl}(1|1))$ by boundary monopole operators. By translating this simple current extension of $V(\mathfrak{gl}(1|1))$ to the unrolled, restricted quantum group $\overline{U}^E(\fgl(1|1))$, we show that our category of line operators admits a second description in terms of a quasi-quantum group $\mathcal{A}$ realized by uprolling. We also compare our results across an expected physical duality with the cyclic orbifold of a free, $B$-twisted hypermultiplet and find a slight discrepancy at the level of braiding and associator. We end with a detailed analysis of coupling to background flat $GL(1, \C)$ connections and the resulting category of non-genuine line operators.}




\maketitle

\section{Introduction}
\label{sec:intro}

Since the pioneering work of Witten \cite{WittenJones}, Chern-Simons quantum field theories (QFTs) have proven to be at a remarkably fertile intersection of quantum topology and quantum algebra. On the one hand, correlation functions of observables in the Chern-Simons theory yield topological invariants of the spacetime 3-manifold; for the case studied by Witten, these generalize the Jones polynomial \cite{Jones} for links in $S^3$ or $\R^3$. On the other hand, these correlation functions are readily computed algebraically by reformulating observables in the 3d bulk QFT categorically by way of a vertex operator algebra (VOA) $\CV$ of local operators on a holomorphic boundary condition. For Chern-Simons theories based on compact, simply connected%
\footnote{Other global forms of the gauge group $G$ can be realized by a simple current extension of the VOA for the simply connected group.} %
Lie group $G$ at positive integer level $k$, one choice of boundary VOA $\CV$ is given by the simple quotient of a $\fg$ current algebra $V_k(\fg)$. This VOA is usually denoted by $L_k(\fg)$. As shown by Reshetikhin and Turaev in \cite{RT}, there is a second categorical realization/extension of the Jones invariants in terms of the representation theory of a quantum group $\CA$, a far less technically complex but still equally rich object; in the case of interest one uses $\CA = U_q(\fg)$ for $q = e^{i\pi/k}$ a $2k$th root of unity.

The first mathematical equivalence between categories of this type was formulated by Kazhdan and Lusztig \cite{KL-I}, where an equivalence between a category of modules for the affine Lie algebra $\widehat{\fg}_k$ and for the quantum group $U_q(\fg)$ was established and is now known as the celebrated Kazhdan-Lusztig correspondence. Unfortunately, this correspondence does not apply to the positive integer levels relevant to Chern-Simons theory. A variant of the Kazhdan-Lusztig correspondence for positive integer levels, i.e. an equivalence between the Kazhdan-Lusztig tensor category for $\widehat{\fg}_k$ at positive integer $k$ and the category used by Reshetikhin-Turaev, was established by Finkelberg in \cite{Finkelberg1, Finkelberg2}, absent a handful of exceptional cases that can be handled separately.%
%
%\footnote{This further raises the question of whether the Kazhdan-Lusztig category for $\widehat{g}_k$ is equivalent to the one arising from the Huang-Lepowsky tensor structure \cite{HL1, HL2, HL3} for $L_k(\fg)$. It is widely believed to be true, but we were unable to find work establishing this result. One result in this direction due to McRae \cite{mcraeaffine} establishes an equivalence between the Huang-Lepowsky tensor category for $L_k(\fg)$ and a Drinfeld category for a suitable ($k$-dependent) truncation of Rep$(\fg)$. Again, it is widely believed that this Drinfeld category is equivalent to the Kazhdan-Lusztig category, but we do not know of any work establishing this result. We thank R. McRae for bringing these points to our attention.}%
%\fixme{Wenjun: I commented out this section since this was actually proven by Zhang.}
%

The boundary VOA $\CV = L_k(\fg)$ is rational for $k > 0$: its category of modules is semisimple%
\footnote{A semisimple (abelian) category is one where any object can be written as a direct sum of subobjects. More generally, there can be objects that have non-trivial subobjects (they are reducible) but nonetheless cannot be written as a direct sum of subobjects (they are indecomposible). See, e.g., Section 1.3 \cite{CDGG} for more details on aspects of non-semisimple categories in the context of QFTs.} %
with a finite number of simple objects. Correspondingly, in the construction of \cite{RT}, only a small portion of the category of representations of $\CA = U_q(\fg)$ is used: one must remove any module with vanishing quantum dimension, resulting in a semisimple category. This process is also called ``semi-simplification." Semisimplicity of the underlying category is imposed by the Chern-Simons theory. A natural question to ask is whether there is yet another, necessarily logarithmic, i.e. non-rational, VOA $\CV$ that realizes the entire category of modules for $\CA$ at positive levels and whether there is a QFT $\CT$ tying these two together.

A logarithmic analog of the Kazhdan-Lusztig correspondence was conjectured by Feigin, Gainutdinov, Semikhatov, and Tipunin \cite{FGST1,FGST2}, whereby the category of modules for the quantum group $U_q(\fg)$ is related to Kausch's triplet algebra $\CW_k$ \cite{Kausch} for $\fg = \fsl(2)$ and the so-called Feigin-Tipunin algebras \cite{FT} for more general $\fg$. Proving this conjectural logarithmic Kazhdan-Lusztig correspondence has been quite difficult, and full proofs only exist for the case of $\fsl(2)$ \cite{CLR, GN}, although we note that S. Lentner has recently provided a proof \cite{Lentner:2025hae} of this conjecture for all simply-laced Lie algebra, conditioned on the existence of a suitable tensor category on the VOA side.%
\footnote{These meager citations do not do justice to the works that lead up to these results. Of particular note is the equivalence of abelian categories of modules for $\CW_k$ and $U_q(\fsl(2))$ established in \cite{Nagatomo:2009xp, mcrae2021structure} as well as the realization that $U_q(\fsl(2))$-mod the naive $R$-matrix does not induce a braiding \cite{KondoSaito} but can be modified to match VOA computations \cite{Gainutdinov:2015lja, Creutzig:2020zvv}. See e.g. Section 1.5 of \cite{CDGG} for a more detailed account.} %
The main barrier in general $\fg$ is that the representation theory of the higher-rank Feigin-Tipunin algebras is quite complicated.

Extensions of the Reshetikhin-Turaev construction to the full, non-semisimple category of modules for the quantum group $U_q(\fg)$ at $q$ an even root of unity have existed in various pieces for nearly as long as the Reshetikhin-Turaev construction itself, including the 3-manifold invariants of Hennings and Lyubashenko \cite{Hennings,Lyubashenko} and the link invariants of Akutsu, Deguchi, and Ohtsuki \cite{ADO}. Work of Costantino, Geer, and Patureau-Mirand (CGP) \cite{CGP} organized these various pieces into an axiomatic TQFT \cite{DGGPR} that now bears their names.%
\footnote{As with the logarithmic Kazhdan-Lusztig correspondence, these citations do not reflect the great body of work that lead to the CGP TQFT. The two main ingredients in the CGP TQFT are the notion of a modified trace (and corresponding renormalized quantum dimension) introduced in \cite{GPT}, which allows one to deal with the vanishing quantum dimensions of modules in $U_q(\fg)$, as well as the notion of a relative modular category due to De Renzi \cite{derenzi2021extended}, which serves as the categorical backbone behind the CGP TQFT as given in \cite{DGGPR}. See Section 1.4 of \cite{CDGG} for more details.} %

The search for the physical QFT that realizes the axiomatic CGP TQFT and that admits the Feigin-Tipunin algebra as a boundary VOA began with work of Gukov, Hsin, Nakajima, Park, Pei, and Sopenko \cite{GHNPPS} which lead to the realization that certain BPS $q$-series reproduced characters of the Feigin-Tipunin algebras \cite{CostantinoGukovPutrov}. A general proposal for the physical QFT, at least when $\fg = \fsl(n)$, recently was given in work of Creutzig, Dimofte, Geer and the first author \cite{CDGG}; see also \cite{FGR} for related work by Feigin, Gukov, and Reshetikhin.

In this paper, we consider a collection of TQFTs that share many of the same features as those described above, in particular possessing rigid braided tensor categories of line operators that can be described from both vertex-algebraic and quantum group perspectives, but are notably simpler and where all of the above connections can be made quite concrete: Chern-Simons theories with the simplest compact supergroup, $U(1|1)$, and variants thereof. In stark contrast to the case of Chern-Simons theories with compact bosonic gauge group, the category of line operators in $U(1|1)$ Chern-Simons theory is non-semisimple: while there are a finite number of simple line operators, a general line operator can not be written as a direct sum of such simple line operators, i.e. there are reducible but indecomposible lines.

\subsection{Line operators in Chern-Simons theories based on $U(1|1)$}

Chern-Simons theories based on $U(1|1)$ have been objects of interest for nearly three decades, starting with the work of Rozansky and Saleur \cite{RSpoly, RSwzw, RStorsion} in trying to find a QFT for the Alexander-Conway polynomial. More recent developments include a detailed treatment of its relation to analytic torsion \cite{Mikhaylov}; a combinatorial approach to understanding the space of states on a torus \cite{AGPS}; and a definition of an axiomatic TQFT using the quantum group $U_q(\fgl(1|1))$ \cite{GYtqft} in the style of CGP. 

In this paper, we extend and compliment these analyses in two ways. First, we develop the vertex-algebraic perspective on these Chern-Simons theories, studying in detail the representation theory of the boundary vertex algebra using recent developments \cite{CR13a, CR13b, CMY20, AW22, BN22} on representation theory of $V(\mathfrak{gl}(1|1))$. Secondly, we connect the boundary VOA approach to the quantum group analyses of \cite{RSpoly, AGPS, GYtqft} via a conjectural equivalence between the representation theory of $V(\mathfrak{gl}(1|1))$ and $\overline{U}_q^E(\fgl (1|1))$. 

Before delving into a detailed synopsis of the paper, we wish to highlight our main results and some of the interesting features of our analysis. 

One of the fundamental ingredients in our approach is a detailed understanding of the affine VOA $V(\fgl(1|1))$. However, in order to obtain the full boundary VOA, we must deal with the topology of the gauge group of interest, not just its Lie algebra. As we are interested in cases where the underlying bosonic group is a compact $U(1)^2$, our ansatz is that the boundary VOA is a simple current extension of $V(\fgl(1|1))$ by a rank two lattice. This is directly analogous to the extension of the Heisenberg VOA to a lattice VOA in the case of abelian Chern-Simons theories. Moreover, as in that case, there are a countable infinity of choices of this rank two lattice to extend by, essentially corresponding to the level of the Chern-Simons theory. The space of all such rank-two extensions is given by triples $(\kappa, \nu, \xi)$, where $\kappa, \nu \in \Z_{> 0}$ are positive integers, and $\xi \in \frac{1}{2}\Z/\kappa \Z$ is a half-integer defined mod $\kappa$; this triple corresponds to a choice of global form of $U(1|1)$ cf. \cite{Mikhaylov}; see also Appendix \ref{sec:u11CS}. We denote the boundary VOA by $\CV_{(\kappa, \nu, \xi)}$ and its category of modules by $\CC_{(\kappa, \nu, \xi)}$. The VOAs $\CV_{(\kappa, \nu, \xi)}$ are not all that different from one another as we can pass from one to any other by a series of finite-order orbifolds and simple current extensions; this corresponds physically to gauging certain finite $0$-form and $1$-form symmetries of the underlying TQFT. For simplicity, we pay most of our attention to the particular example $\CV_k = \CV_{(k,1,\scriptstyle{\frac{k}{2}})}$ and simply state the results for the other cases. 

In Section \ref{sec:lines} we turn to the representation theory of $\CV_k$. We provide a detailed account thereof by exploiting the fact that it is a simple current extension of the well-studied VOA $V(\fgl(1|1))$: the work of Creutzig-Kanade-McRae \cite{CKM17} and Creutzig-McRae-Yang \cite{CMY22} establishes that the category of modules of $\CV_k$ can be identified with the de-equivariantization (see e.g. \cite[Section 8.23]{etingof2016tensor}) of the subcategory of modules that are local with respect to the simple currents we extend by. The category $\CC_k$ that results from this procedure will be our model for line operators in $U(1|1)$ Chern-Simons theory.%
\footnote{More precisely, the category of line operators in the bulk TQFT should instead be the derived category of this abelian category. We will work in the underlying abelian but perform computations relevant for the derived category.} %
Using the technology of simple current extensions, we can completely determine the braided tensor category $\CC_k$, including all indecomposable objects, fusion rules, and braiding. We find that it is a non-semisimple rigid braided tensor category with finitely many simple objects. We compute the derived endomorphism of the identity object, and show that it is isomorphic to functions on the orbifold space $\C^2/\Z_k$. We further explore the relation between $U(1|1)$ Chern-Simons theory and 3d B-model on $\C^2/\Z_k$ in Section \ref{sec:Bmodelorbifold}. In particular, we show that the derived category of $\CC_k$ is equivalent to a category of matrix factorizations. 

On the quantum group side, we perform a completely analogous simple current extension procedure to $\overline{U}_q^E(\fgl (1|1))$. More precisely, we perform de-equivariantization with respect to the corresponding rank-2 lattice of simple $\overline{U}_q^E(\fgl (1|1))$ modules and show that this category can be represented by a quasi-triangular quasi-Hopf algebra $\CA_k$ (which we will call a quasi-quantum group). Thus, we find the quantum group analog of $\CV_k$. This quasi-quantum group is related to the category of matrix factorizations through quadratic Koszul duality. 

We end with an analysis of a category $\CC_k^{\rm ext}$ of nonlocal modules for $\CV_k$ in Section \ref{sec:cext}. The category $\CC_k^{\rm ext}$ models line operators that source a background flat connection for a $U(1)$ flavor symmetry acting by rotations of $\C^2/\Z_k$, cf. \cite{GaiottoTwisted, CDGG}. In VOA terms, these are twisted modules corresponding to the $U(1)$ action on $\CV_k$. 

In contrast to finite-order symmetries, twisted modules for infinite-order symmetries are far from well-understood. In the current set-up, we offer several alternative avenues into the category $\CC_k^{\rm ext}$, all of which provide us with extremely explicit knowledge to twisted sectors in $\CC_k^{\rm ext}$. The first route into $\CC_k^{\rm ext}$ is by considering $\C^\times$-twisted modules for $\CV_k$; the twist parameter is identified with the holonomy of the background flat connection. These modules can be realized by induction of $V(\fgl(1|1))$ modules that are \emph{not} local with respect to the simple currents we extend by. Our second approach to $\CC_k^{\rm ext}$ is by viewing it as a category of modules for a vertex algebra $\CV_k^{\rm ext}$ that differs from $\CV_k$ by a large center; the generator of the center captures the background flat connection \cite{Feigin:2025gfj}. We access this category by realizing $\CV_k^{\rm ext}$ as a large-level limit of a VOA $\CV_k^\delta$. The third and final route we take is by realizing that we can arrive at $\CV_k^{\rm ext}$ by another route, by appending an auxiliary commutative boson to $V(\fgl(1|1))$ and then extending in the same way we got $\CV_k$ from $V(\fgl(1|1))$. Thus, $\CV_k$ provides a nice playground for studying the gauging of infinite-order symmetries on vertex operator algebras. 

In the remainder of this Introduction, we expand on the above points and briefly summarize the central aspects of our analysis.

\subsection{Boundary VOAs}
\label{sec:introVOA}

One way to reformulate Chern-Simons theory categorically is to associate to it a VOA $\CV$ whose (derived) category of modules is identified with the category of line operators in the bulk TQFT. %In Section \ref{sec:bdyVOA} we introduce a holomorphic Dirichlet boundary condition for each of the above global forms and analyze the associated boundary VOA $\CV_{(\kappa, \nu, \xi)}$.
The VOA $\CV$ is realized as the algebra of local operators on a suitable holomorphic boundary condition. Given such a holomorphic boundary condition, the space of local operators at the endpoint of a given bulk line operator $\CL$ intersecting the boundary has the structure of a module $\CM_\CL$ for $\CV$ via collision of boundary local operators; colliding a local operator $\CO$ joining two line operators $\CL, \CL'$ with the boundary commutes with collision with boundary local operators and hence defines a morphism of $\CV$ modules; colliding and braiding of line operators is identified with fusion and half-monodromy of the corresponding $\CV$ modules. In this way, the boundary VOA encodes the 3d bulk theory, at least as far as line operators are concerned.%
\footnote{Strictly speaking, it is possible that some aspects of bulk line operators are not encoded faithfully by the boundary VOA $\CV$. For example, there could line operators that cannot end on the boundary condition; any such line operator would be represented by the trivial, $0$-dimensional $\CV$ module. The process just outlined, whereby bulk line operators are identified with modules for $\CV$, defines a functor $\CC \to \CV\textrm{-mod}$. We will assume that this functor is fully faithful, i.e. the map from the space of local operators $\Hom_\CC(\CL, \CL')$ joining two lines $\CL, \CL'$ and the corresponding space of (derived) $\CV$-module morphisms $\Hom_{\CV\textrm{-mod}}(\CM_\CL, \CM_{\CL'})$ is an isomorphism, and conservative, i.e. no non-zero line operator is sent to the 0-dimensional module.} %
We note it is the derived category of modules that is of physical relevance. For example, local operators are realized by derived morphisms between the corresponding modules, rather than simply morphisms. Throughout this paper we focus on the underlying \emph{abelian} category of modules, except in Section \ref{sec:locops} where we compute the algebra of bulk local operators realized by derived endomorphisms of the vacuum module.

The boundary VOA can be derived in a fashion analogous to Chern-Simons theories with compact, abelian gauge group. We start with the perturbative subalgebra of affine currents $V(\fgl(1|1))$ and extend by certain modules corresponding to boundary monopole operators. The modules one extends by encode the non-perturbative aspects of the theory of interest: there are many Chern-Simons theories based on the same Lie algebra. For example, the Heisenberg algebra shows up in all $U(1)$ Chern-Simons theory and the non-perturbative information is encoded in the choice of a lattice of simple currents to extend by. This lattice is identified with the image of the cocharacter lattice with respect to the (map induced by) the metric underlying the Chern-Simons theory.%
\footnote{In general, one expects the boundary VOA to be realized as the cohomology of a certain VOA bundle $\CV_{\textrm{Gr}_G}$ over the affine Grassmannian $\textrm{Gr}_G$, cf. \cite[Section 7]{CDGG}. Our boundary VOA admits the relatively simple description as a simple current extension due entirely to the fact that the (closed points of the) affine Grassmannian for abelian $G = (\C^\times)^r$ is simply a copy of $\Z^r \simeq \Hom(\C^\times, G).$ For non-abelian $G$, one is forced to contend with the non-trivial geometry and topology of $\textrm{Gr}_G$.} %
Similarly, for $U(1|1)$ Chern-Simons theories we must choose a rank 2 lattice of simple currents to extend $V(\fgl(1|1))$ by. As we will see, the choice of such a lattice is parameterized by a triple $(\kappa, \nu, \xi)$, where $\kappa, \nu$ are positive integers and $\xi$ is a half-integer defined modulo $\kappa$. This choice is in one-to-one correspondence with a choice of the additional non-perturbative data need to define is meant by $U(1|1)$ Chern-Simons theory, i.e. there is a distinct TQFT $\CT_{(\kappa, \nu, \xi)}$ that gives rise to each of the VOAs $\CV_{(\kappa, \nu, \xi)}$.

% As detailed in Section \ref{sec:bdyVOA}, $\CV_{(\kappa, \nu, \xi)}$ is realized as an extension of $V(\fgl(1|1))$ by a rank-2 lattice of simple currents, each lattice vector realized by a spectral flow of the vacuum module. The choice of spectral flows we extend by is dictated by the global form of the gauge group -- there is a spectral flow automorphism for each cocharacter of the bosonic subgroup $U(1)^2$ or, better, each (closed) point on the affine Grassmannian $\textrm{Gr}_{(\C^\times)^2}$. Each cocharacter lattice $\Lambda_{(\kappa, \nu, \xi)}$ will thus lead to an interesting rank-2 simple current extension of the \emph{same} perturbative current algebra.

Explicitly, 
\be
\label{eq:introVOA}
	\CV_{(\kappa, \nu, \xi)} = \bigoplus\limits_{\mathfrak{m} \in \Z} \Bigg( \widehat{A}_{\frac{\mathfrak{m} \kappa}{\nu},0} \oplus \bigoplus_{\mathfrak{n} \in \Z_{>0}} \Pi^{2\mathfrak{n} \xi}\bigg(\Pi \widehat{A}_{\frac{\mathfrak{m} \kappa}{\nu} + \mathfrak{n}\big(\frac{\xi}{\nu} - \frac{\nu}{2}\big)+\frac{1}{2}, \mathfrak{n} \nu} \oplus \widehat{A}_{\frac{\mathfrak{m} \kappa}{\nu} - \mathfrak{n}\big(\frac{\xi}{\nu} - \frac{\nu}{2}\big)-\frac{1}{2}, -\mathfrak{n} \nu}\bigg) \Bigg)~,
\ee
where $\widehat{A}_{n, e}$ are atypical modules of $V(\fgl(1|1))$ obtained by spectral flow of the vacuum and $\Pi$ denotes a shift in parity. Note that each VOA $\CV_{(\kappa, \nu, \xi)}$ can be related to the reference VOA $\CV_{(1, 1, 0)}$ by a sequence of finite-order simple current extensions and orbifolds. The second author considered an analogous simple current extension of $V(\fgl(1|1))$, albeit only extending by a rank-1 lattice, as a boundary VOA for the topological $B$-twist of 3d $\CN=4$ QED \cite{BN22} and compared the resulting category of modules to that of the boundary VOA for its 3d mirror theory, a free hypermultiplet; general abelian gauge theories are discussed by the authors of loc. cit. together with T. Dimofte and T. Creutzig in the more recent \cite{BCDN}. The simple current extension studied in \cite{BN22} belongs to a 2-parameter family of VOAs $\mathfrak{W}_{n+\frac{1}{2},l}$ introduced by Creutzig and Ridout in \cite{CR13b}. The VOA $\mathfrak{W}_{n+\frac{1}{2},l}$ is realized by extending the perturbative current algebra $V(\fgl(1|1))$ by the rank-1 lattice generated by $\widehat{A}_{n+\frac{1}{2},l}$, from which we immediately see that $\CV_{(\kappa, \nu, \xi)}$ can also be interpreted as a rank-1 simple current extension of $\mathfrak{W}_{\frac{\xi}{\nu}+\frac{1-\nu}{2}, \nu}$.%
\footnote{In fact, it can actually be realized as a rank-1 simple current extension of infinitely many different $\mathfrak{W}_{n+\frac{1}{2},l}$. For example, if we identify $l = \nu$, we can identify $n$ with $\frac{\mu \kappa}{\nu} + \frac{\xi}{\nu} - \frac{\nu}{2}$ for any integer $\mu$; this corresponds to the subalgebra with $\mathfrak{m} = \mu\mathfrak{n}$.} %
The work \cite{BN22} considers $\mathfrak{W}_{0,1} \simeq V_{\beta\gamma} \otimes V_{bc}$, which is identified with a pair of symplectic bosons $V_{\beta\gamma}$ and a pair of complex fermions $V_{bc}$.

We describe the category of line operators in $\CT_{(\kappa, \nu, \xi)}$, i.e. the category of modules for $\CV_{(\kappa, \nu, \xi)}$ in Section \ref{sec:lines}. Using the realization of $\CV_{(\kappa, \nu, \xi)}$ as an infinite-order simple current extension, we apply techniques developed by Creutzig-Kanade-McRae \cite{CKM17} and Creutzig-McRae-Yang \cite{CMY22} to understand the category in terms of the category of modules for the underlying affine current algebra $V(\fgl(1|1))$. Explicitly, we consider the Kazhdan-Lusztig category $KL$ of finite-length, grading-restricted generalized $V(\fgl(1|1))$ modules; importantly, this category contains modules where the zeromode of the stress tensor acts non-semisimply but where generalized conformal weights are bounded from below. The simple current extension $\CV_{(\kappa, \nu, \xi)}$ is a VOA object in the category $KL$%
\footnote{Strictly speaking, the VOAs we consider don't belong to $KL$ because they are infinite-order simple current extensions. We must instead pass to a direct-limit completion of $\textrm{Ind}(KL)$ that allows for such simple current extensions. We will mostly suppress the distinction between $KL$ and its completion $\textrm{Ind}(KL)$; see \cite{CMY22} for more details.}, %
i.e. it is a $V(\fgl(1|1))$-module that itself possesses the structure of a VOA. Our proposal for the category of line operators is simply the (derived category of the) category $\CC_{(\kappa, \nu, \xi)}$ of $\CV_{(\kappa, \nu, \xi)}$-modules in $KL$. This can be expressed in terms of the Kazhdan-Lusztig category $KL$ as
\be
	\CC_{(\kappa, \nu, \xi)} = KL^0/\Lambda_{(\kappa, \nu, \xi)}
\ee
where $KL^0$ denotes the subcategory of modules that are local with respect to $\CV_{(\kappa, \nu, \xi)}$, i.e. genuine $\CV_{(\kappa, \nu, \xi)}$ modules and not twisted modules, and $KL^0/\Lambda_{(\kappa, \nu, \xi)}$ is the de-equivariantization thereof by the rank-2 lattice $\Lambda_{(\kappa, \nu, \xi)}$ of simple currents we extend by. Roughly speaking, the de-equivariantization identifies modules that differ by fusion with the simple currents and can be thought of as the mathematical avatar of screening of Wilson lines. Although modules in $\CC_{(\kappa, \nu, \xi)}$ generally have unbounded (generalized) conformal weights, the fact that we restrict to modules that also belong to $KL$ means Theorem 1.4 of \cite{CMY22} can be applied to show that $\CC_{(\kappa, \nu, \xi)}$ naturally has the structure of a braided tensor category defined by the $P(z)$-intertwiners of Huang-Lepowsky-Zhang \cite{HLZ1, HLZ2, HLZ3, HLZ4, HLZ5, HLZ6, HLZ7, HLZ8}.

We comment finally that although the definition of $KL^0$ is highly non-trivial, we will show that it is in fact something rather concrete. It will turn out that $KL^0$ is precisely the subcategory where the action of the zero-modes of $V(\fgl (1|1))$ integrates to an action of the group $U(1|1)$. This ultimately leads to a complete understanding of the category $\CC_{(\kappa, \nu, \xi)}$.

\subsection{Quantum groups}
\label{sec:introqgroup}

A second categorical realization of line operators in Chern-Simons theories is via the (derived) category of modules for a suitable quasi-quantum group. For each of the triple $(\kappa, \nu, \xi)$ there is a corresponding quasi-quantum group $\CA_{(\kappa, \nu, \xi)}$ that we describe in Section \ref{sec:qgroup}. We determine this quasi-quantum group in two ways. First, we can follow the analysis in Section 5 of \cite{BN22} to relate the category $\CC_{(\kappa, \nu, \xi)}$ to a category of modules for some finite-dimensional algebra, namely $\CA_{(\kappa, \nu, \xi)}$. Alternatively, the same algebra can be derived by realizing that simple current extensions have a direct analog in the world of quantum groups, namely, the process of uprolling, cf. \cite{creutzig2022uprolling}. As described in \cite{creutzig2024KL}, it is possible to use this procedure to explicitly describe quantum groups realizing the category of lines for an arbitrary ($A$- or $B$-twisted) 3d $\CN=4$ abelian gauge theory of ($\CN=4$) vector multiplets and hypermultiplets.%
\footnote{The Chern-Simons gauge theories considered in this paper are not quite in the class of theories studied in \cite{BCDN}, although they are not too far from them either. As we saw above, the theory for $(\kappa, \xi, \nu) = (k,0,1)$ is very nearly the $\Z_k$ orbifold of a $B$-twisted hypermultiplet. This orbifold is 3d mirror to the $A$-twist of an $\CN=4$ $U(1)$ gauge theory with a hypermultiplet of charge $k$. The abelian category of lines in this $A$-twisted theory should agree with the category $\CC_{(k,1,0)}$ described above, but they will have different braiding morphisms.} %

Explicitly, we start with the unrolled, restricted quantum group $\ol{U}^{E}(\fgl(1|1))$ (the quantum group analog of $V(\fgl(1|1))$). This algebra has three bosonic generators $N, E, K$, the latter being invertible, and two fermionic generators $\Psi_\pm$ with the following non-vanishing commutation relations:
\be
[N, \Psi_\pm] = \pm \Psi_\pm \qquad \{\Psi_+, \Psi_-\} = K - 1
\ee
cf. Section 2.2 of \cite{GYtqft}, and we restrict to modules where $K = e^{2\pi i E}$.%
\footnote{In comparing to loc. cit., one should identify the fermionic generators $\Psi_\pm$ with $X,Y$ and the bosonic generator $N$ by $G$. We have also rescaled one of the fermionic generators (e.g. their $X$) by the unit $(q - q^{-1})K$ and redefined $K \to K^{1/2}$. Because $q \neq 0, \pm 1$ and we retain the logarithm $E$, these redefinitions are inconsequential.} %
It has been conjectured, and recently proven in \cite{creutzig2023algebraic, creutzig2024KL}, that the braided tensor category associated to $\ol{U}^{E}(\fgl(1|1))$ and $V(\fgl(1|1))$ are equivalent. 

We then consider the following module
\be
\label{eq:introqgroup}
\CM_{(\kappa, \nu, \xi)} = \bigoplus_{\mathfrak{m},\mathfrak{n}} A_{\mathfrak{m}\frac{\kappa}{\nu} + \mathfrak{n}(\frac{\xi}{\nu}-\frac{\nu}{2})+\epsilon(\mathfrak{n}), \mathfrak{n}\nu}
\ee
where $A_{n,e}$ is the 1-dimensional representations of the quantum group corresponding to $\widehat{A}_{n,e}$; see Section \ref{sec:uprolling} for more details. We think of this as an extension of the unrolled quantum group by currents realizing the cocharacter lattice $\Lambda_{(\kappa, \nu, \xi)}$; this rank-2 lattice is generated by the simple currents $A_{\frac{\kappa}{\nu},0}$ and $A_{\frac{\xi}{\nu}+\frac{1-\nu}{2},\nu}$, cf. Eq. \eqref{eq:introVOA}. The module $\CM_{(\kappa, \nu, \xi)}$ and corresponding quasi-quantum group $\CA_{(\kappa, \nu, \xi)}$ don't explicitly show up in the CGP-style TQFT described in \cite{GYtqft}. Instead, the data of this module goes into a free realization of the cocharacter lattice $\Lambda_{(\kappa, \nu, \xi)}$ in the category of modules for the unrolled quantum group $\ol{U}^{E}(\fgl(1|1))$, an essential ingredient in the CGP construction. The authors of loc. cit. consider several cases of interest, e.g. where the quantum parameter $q \in \C^\times$ is generic and when it is a root of unity; the ones most directly related to the present analysis corresponds to those where the quantum parameter $q$ is a even root of unity, cf. Section 2.7.2 of \cite{GYtqft}, corresponding to the theories with $(\kappa, \nu, \xi) = (k,1,0)$. 

$\CM_{(\kappa, \nu, \xi)}$ itself has the structure of a commutative (super)algebra object in the category of modules for the unrolled, restricted quantum group, in the same way simple current extensions define commutative algebra objects. By analyzing local modules of $\CM_{(\kappa, \nu, \xi)}$, we identify  $\CA_{(\kappa, \nu, \xi)}$ as a subquotient of $\ol{U}^{E}(\fgl(1|1))$. Explicitly, because the generating currents $A_{\frac{\kappa}{\nu},0}$ and $A_{\frac{\xi}{\nu}-\frac{\nu}{2}+\frac{1}{2},\nu}$ change the eigenvalues of the bosonic generators $N,E$, they no longer have a well-defined action after extending. Nonetheless, we get a well-defined action of the subalgebra generated by the fermionic generators $\psi_\pm$ as well as the exponentials (and their inverses)
\be
K_1 = \exp\bigg(\frac{2\pi i}{\kappa}\big(\nu N - (\tfrac{\xi}{\nu}+\tfrac{\nu}{2}) E\big)\bigg) = q^{2\nu \big(N - \frac{1}{2}E\big)-\frac{2\xi}{\nu}E} \qquad K_2 = \exp\bigg(\frac{2\pi i}{\nu}E\bigg) = q^{\tfrac{2\kappa}{\nu} E}
\ee
where $q = e^{i \pi/\kappa}$. These generators satisfy the relations
\be
\label{eq:introqgroup1}
K_1 \psi_\pm = q^{\pm 2\nu}	\psi_\pm K_1 \qquad \{\psi_+, \psi_-\} = (K_2)^{\nu} - 1
\ee
inside the unrolled, restricted quantum group. 

Modules for $\CA_{(\kappa, \nu, \xi)}$ can be induced from modules of the unrolled quantum group by tensoring with the module $\CM_{(\kappa, \nu, \xi)}$ in direct analogy with how we obtain modules for $\CV_{(\kappa, \nu, \xi)}$ from $KL$. Not all modules for the unrolled quantum group will be honest modules for $\CA_{(\kappa, \nu, \xi)}$: we must restrict to modules that have trivial monodromy with the simple currents that generate the extension. Restricting to local modules further imposes the relations
\be
\label{eq:introqgroup2}
(K_1)^\kappa = (K_2)^{2\xi} \qquad (K_2)^\kappa = 1
\ee
whence $\CA_{(\kappa,\nu, \xi)}$ is a quotient of a subalgebra of the unrolled, restricted quantum group. As with VOAs $\CV_{(\kappa, \nu, \xi)}$, the quasi-quantum groups $\CA_{(\kappa, \nu, \xi)}$ can all be related to one another via a sequence of finite-order orbifolds and simple current extensions. We note that $\CA_{(1,1,0)}$ is simply an exterior algebra with generators $\psi_\pm$ -- each of the bosonic generators $K_1,K_2$ is required to act as $1$ on local modules. Indeed, the boundary VOA $\CV_{(1,1,0)}$ is simply a pair of symplectic fermions $V_{\chi_+, \chi_-}$ times a decoupled lattice VOA $V_\rho \otimes V_\theta$. In fact, for the choice $(k,1,0)$, one can always trivialize the associator and obtain a genuine quantum group $\CA_{(k,1,0)}$; see Section \ref{subsubsec:Ak10}.

\subsection{Deforming by background flavor connections}
\label{sec:introbackground}

Each of the QFTs $\CT_{(\kappa, \nu, \xi)}$ has a $U(1)^{(0)}$ 0-form flavor symmetry that can be coupled to background flat, complexified connections.%
\footnote{At the level of the field theory, this flavor symmetry simply rescales the hypermultiplets. This is not a gauge transformation due to the non-zero Chern-Simons terms -- both the scalars \emph{and} monopoles transform under gauge transformations. Indeed, this flavor symmetry is gauge-equivalent to the topological flavor symmetry.} %
As detailed in \cite{Mikhaylov}, the partition function of these QFTs on a 3-dimensional manifold equipped with such a flat connection yields a version of analytic torsion of the background, cf. Eq. (4.12) of loc. cit. Theories that admit deformations by background flat, complexified connections for a (0-form, complexified) flavor symmetry $G$ are particularly special; for example, they naturally give boundary conditions for, or interfaces between, the $B$-twist (a.k.a. the geometric Langlands twist of \cite{KWemduality} with canonical parameter $\psi = \infty$) of 4d $\CN=4$ super Yang-Mills with gauge group $G$. As emphasized in \cite{CDGG}, but dating back as far as the work of Kasheav and Reshetikhin \cite{KashaevReshetikhin}, these background flat connections have a natural place in the categorical reformulation of the Chern-Simons theory in terms of a boundary VOA or a quantum group: line operators $\CL$ that source a flat connection with holonomy $g \in G$ correspond to non-local modules where the monodromy of $\CL$ and the trivial line $\id$ is exactly this holonomy $g$. This relaxed notion of a line operator, sometimes called a non-genuine line operator, can be use to define an extended category $\CC^{\textrm{ext}}$ of line operators sourcing flat connections: the objects are line operators sourcing a (possibly nontrivial) flat $G$ background on its complement and the morphisms are local operators that join two such line operators. Note that two lines with different holonomies cannot be connected by any local operator, hence this extended category decomposes into blocks
\be
	\CC^{\textrm{ext}} = \bigoplus_{g \in G} \CC^{\textrm{ext}}_g
\ee
where $\CC^{\textrm{ext}}_{g}$ is the subcategory of lines sourcing a flat connection with holonomy $g$. The subcategory $\CC^{\textrm{ext}}_1$ coincides with the usual category of line operators $\CC$.

At the level of the quasi-quantum group $\CA_{(\kappa, \nu, \xi)}$, line operators with non-trivial background flat, complexified connections with holonomy $g \in G = \C^\times$ can be identified with (non-local) modules on which $(K_2)^\kappa$ acts by multiplication by the scalar $g$. The unrolled versions of these additional modules are at the heart of the construction of axiomatic TQFTs appearing in \cite{CGP, GYtqft}. Accessing the extended category $\CC^{\textrm{ext}}_{(\kappa, \nu, \xi)}$ from the perspective of the boundary VOA $\CV_{(\kappa, \nu, \xi)}$ is a bit more subtle and we address this problem in Section \ref{sec:flatconns}. In Section \ref{sec:flatdeform}, we describe how to couple $\CV_{(\kappa, \nu, \xi)}$ to a background flat connection realized as a (commutative) bosonic field $A(z)$ identified with the background connection; this coupled VOA can also be realized from a certain large-level limit $\delta \to \infty$ of an auxiliary VOA $\CV_{(\kappa, \nu, \xi)}^{\delta}$, cf. Section 6.2 of \cite{CDGG}. The recent paper \cite{FLcenter} describes some general aspects of these large-level limits, focusing on the Feigin-Tipunin algebras $\CW_p(\fg)$ that are expected to correspond to the quantum groups $\mathfrak{u}_q\fg$ for $q = e^{i \pi/p}$ a $2p$th root of unity.

With the deformation of the boundary VOA $\CV_{(\kappa, \nu, \xi)}$ by a background flat, complexified connection, we turn to the extended category $\CC_{(\kappa, \nu, \xi)}^{\textrm{ext}}$ in Section \ref{sec:cext}. We present three complementary perspectives on this extended categories, with varying degrees of generality, and highlight the aspects that each perspective makes manifest. The first perspective (Section \ref{sec:nonlocal}) is via simply including certain non-local modules for $\CV_{(\kappa, \nu, \xi)}$: instead of requiring local modules, we simply allow modules where the monodromy with $\CV_{(\kappa, \nu, \xi)}$ is given by the action of $g \in \C^\times$. This perspective has the benefit of being entirely categorical, but does not offer much utility for performing explicit computations. The second perspective (Section \ref{sec:largelevellimit}) uses the above large-level limit: it realizes the extended category of modules as the large-level limit of the category of (local!) modules for the auxiliary VOA $\CV_{(\kappa, \nu, \xi)}^{\delta}$. The third perspective (Section \ref{sec:shifting}) is conceptually the most simple, but also the least general: we use the fact that deforming by a flat connection $A$ can be realized by shifting the central current $E \to E - A$. Passing this shift though our free-field computations yields a third realization of the extended category $\CC_{(\kappa, \nu, \xi)}^{\textrm{ext}}$. In Section \ref{sec:largelevelvsnonlocal}, we show the relation between the first and second approaches to the category $\CC^{\textrm{ext}}_{(\kappa, \nu, \xi)}$.

\subsection{Outlook}
We end the Introduction with a collection of future directions that to build on the present work.

1) Cohomological extension of \cite{AGPS} by way of Hochschield (co)homology of $\CC_{(\kappa, \nu, \xi)}$
2) Study vertex algebras appearing in other abelian Gaiotto-Witten theories, connect to the work \cite{Garner:2024yin}
3) Studying coupling to flat connections for general groups other than $U(1)$, anomaly, etc. 

\subsection*{Acknowledgments}

We would like to thank Thomas Creutzig for his patience in teaching us about many aspects of quantum groups and VOAs. We would also like to thank T. Creutzig and Tudor Dimofte for their collaboration on related projects as well as their invaluable insights into problems relating QFT, quantum groups, and VOAs. We would also like thank Mat Bullimore, Nathan Geer, Justin Hilburn, Robert McRae, Cris Negron, Natalie Paquette, Ingmar Saberi, and Matthew Young for useful conversations during the development of this project. N.G. acknowledges support from the University of Washington.

\section{Boundary VOA for $U(1|1)$ Chern-Simons Theory}
\label{sec:bdyVOA}

In this section we determine a boundary VOA for the Chern-Simons theories based on the compact supergroup $U(1|1)$. As described in Section \ref{sec:global}, and emphasized in Section 4.2 of \cite{Mikhaylov}, there are a countably infinite number of Chern-Simons theories $\CT_{(\kappa, \nu, \xi)}$ based on the Lie superalgebra $\fgl(1|1)$ even when we require the bosonic gauge group is the compact torus $U(1)^2$.
All of the corresponding VOAs are built upon affine VOA $V(\fgl(1|1))$, generated by bosonic fields $N(z)$, $E(z)$ and fermionic fields $\psi_\pm(z)$ with the following OPEs.
\begin{gather}
	\label{eq:opegl11standard}
	N(z) N(w) \sim  \frac{1}{(z-w)^2} \qquad  N(z) E(w) \sim \frac{1}{(z-w)^2}\nonumber \\
	N(z) \psi_\pm(w) \sim \frac{\pm \psi_\pm(w)}{z-w}\\
	\psi_+(z) \psi_-(w) \sim \frac{1}{(z-w)^2} + \frac{E(w)}{z-w} \nonumber
\end{gather}
%As usual, the coefficients of the $(z-w)^{-2}$ terms in the OPEs of the bosonic fields $N,E$ exactly reproduces the `t Hooft anomaly for the boundary $U(1) \times U(1)$ flavor symmetry, i.e. the effective level in Eq. (3.32) of \cite{DimofteGaiottoPaquette}.
We note that this is not the standard bilinear form on $V(\fgl(1|1))$, but is related to the usual one by a field redefinition from $N$ to $\wt{N} = N - \tfrac{1}{2}E$, cf. \cite[Remark 2.1.1]{CMY20}. This change of basis doesn't alter the commutation relations of $\fgl(1|1)$ because $E$ is central. It is worth noting that the Sugawara stress tensor takes a particularly simple form in our original basis:
\be
	T_{\textrm{Sug}} = \tfrac{1}{2}\big(\norm{N E + E N - \psi_+ \psi_- + \psi_- \psi_+}\big)
\ee

To avoid undue clutter, we will focus on the specific theories $\CT_k := \CT_{(k,1,\frac{k}{2})}$. %with the exception of Section \ref{sec:VOAorbifold} where we compare $\CT_{(k,1,0)}$ to the $\Z_k$ orbifold of a $B$-twisted hypermultiplet.
The analysis of the general case is essentially equivalent and we remark on the differences along the way. In the following, we take the ansatz that the full non-perturbative algebra $\CV_k:= \CV_{(k,1,\frac{k}{2})}$ takes the form of a simple current extension of the perturbative current algebra $V(\fgl(1|1))$. %Again, this ansatz will be sufficient for our purposes because the gauge group is abelian; this simple ansatz will not work for any Chern-Simons theory with non-abelian bosonic gauge group.%
%
%\footnote{The reason for this is simple: if we assume the gauge group is simply connected, the boundary VOA is a quotient of the corresponding affine current algebra, rather than an extension thereof; passing from the simply-connected form to other compact global form involves a current extension. In the abelian setting, the current algebra is itself simple and we only need to consider the extension.} %
%
This is entirely analogous to the case of $U(1)$ Chern-Simons theories where the perturbative algebra is a Heisenberg VOA, a.k.a. $V(\fgl(1))$, and non-perturbative corrections extend it to a suitable lattice VOA. Detailed analyses of some simple current extensions of $V(\fgl(1|1))$ appear in work of Creutzig and Ridout \cite{CR13a, CR13b}, with a more recent flurry of works by Allen and Wood \cite{AW22}; Creutzig, McRae, and Yang \cite{CMY20}; and Ballin and the second author \cite{BN22}. These references study simple current extensions $\mathfrak{W}_{n+\frac{1}{2},l}$ that correspond to introducing boundary monopole operators, i.e. vertex operators/Fock modules, for a single cocharacter $X_{n+\frac{1}{2},l}$, corresponding to a rank 1 bulk gauge group.%
\footnote{Explicitly, the generator takes the form $$X_{n+\frac{1}{2},l} = \big(n+\tfrac{l}{2}\big) E + l N$$ for fixed $\ell \in \Z$ and $n \in \tfrac{1}{2l}\Z$. The VOA for $n + \tfrac{l}{2} = 0$, i.e. $\mathfrak{W}_{\frac{1-l}{2},l}$, should arise at the boundary of $B$-twisted $\CN=4$ theory $U(1)$ gauge theory with a hypermultiplet of charge $l$. For example, the current extension $\mathfrak{W}_{0,1}$ is identified with the $\beta\gamma$ VOA (times a pair of complex fermions) \cite[Section 4.5]{CR13b} and can be interpreted as the boundary VOA for the $B$ twist of 3d $\CN=4$ SQED with a single flavor, cf. \cite{BN22}. The theory for general $n$ should arise from adding a Chern-Simons term with level $l(2n + l)$.} %
The main difference we will encounter arises from the fact that we will extend by a full-rank lattice of simple currents.

In Section \ref{sec:spectralflow}, we review the spectral flow automorphisms of $V(\fgl(1|1))$ and identify the modules associated to each cocharacter of the bosonic gauge group. In Section \ref{sec:halfindex} we compare the graded character of our proposal for $\CV_k$ to the half-index, cf. \cite{DimofteGaiottoPaquette}, enumerating the local operators present on the holomorphic Dirichlet boundary condition described in Appendix \ref{sec:u11CS}, finding agreement modulo some parity subtleties we describe below.

In Section \ref{sec:freefield} we provide a free-field realization of the boundary VOA $\CV_k$ and in Section \ref{sec:VOAorbifold} use an analogous free-field realization for the theory $\CT_{(k,1,0)}$ to show that the corresponding boundary VOA $\CV_{(k,1,0)}$ is an orbifold. From the perspective of the free-field realization, the aforementioned parity issues are identical to those appearing in Section 4.4 of \cite{CR13b}. Namely, when $k$ is odd it is not possible to consistently make parity assignments for the vertex operators in the free-field realization of $\CV_k$.%
\footnote{The boundary VOA $\CV_{(k,1,0)}$ does not suffer from these parity subtleties. More generally, we find that it is impossible to make consistent parity assignments in the free-field realization of $\CV_{(\kappa, \nu, \xi)}$ when $2\xi$ is an odd integer, i.e. if $\xi \in \frac{1}{2} + \Z$.} %
The resolution described in Section 4.5 of loc. cit. is to dress the vertex operators with additional operator-valued functions that rectify the parity issues.

\subsection{$\CV_k$ from spectral flow}
\label{sec:spectralflow}

As first noted in \cite{SS06}, the affine VOA $V(\fgl(1|1))$ admits a family of spectral flow automorphisms. From the perspective of the bulk Chern-Simons theory, these spectral flow automorphisms arise from performing large gauge transformations in the neighborhood of a point on the boundary, i.e. inserting a vortex line ending at that point on the boundary. Explicitly, the spectral flow automorphism $\sigma_{l,\lambda}$ is given by
\be
	\sigma_{l,\lambda}(N) = N - \frac{\lambda}{z} \qquad \sigma_{l,\lambda}(E) = E - \frac{l}{z} \qquad \sigma_{l,\lambda}(\psi_\pm) = z^{\mp l} \psi_\pm
\ee
where $l \in \Z$ and $\lambda \in \C$. In terms of the mode algebra, this automorphism takes the form
\be
	\sigma_{l,\lambda}(N_r) = N_r - \lambda \delta_{r,0} \qquad \sigma_{l,\lambda}(E_r) = E_r - l \delta_{r,0} \qquad \sigma_{l,\lambda}(\psi_{\pm, r}) = \psi_{\pm, r \mp l}
\ee
Note that $\sigma_{l,\lambda} \circ \sigma_{l',\lambda'} = \sigma_{l+l', \lambda + \lambda'}$. %one should think of this as performing a singular gauge transformation $A^{N}_z \to A^{N}_z - \tfrac{l}{z}$ and $A^{E}_z \to A^{E}_z - \tfrac{\lambda-l}{z}$.
This automorphism transforms the Sugawara stress tensor $T_{\textrm{Sug}}$ as
\be
	\sigma(T_{\textrm{Sug}}) = T_{\textrm{Sug}} + z^{-1}\big((l - \lambda) E - l N\big) + z^{-2}\big(\tfrac{1}{2}l(2\lambda - l)\big)
\ee
The original spectral flow automorphism of \cite{SS06} arises as the special case when $\lambda = \frac{l}{2}$.

The boundary VOA $\CV_k$ is given by extending the perturbative current algebra by spectral flows of the vacuum, although only certain spectral flow modules should be included. %Under the identification of spectral flow and large gauge transformations, its clear that the included spectral flow modules must be compatible with the global structure of the gauge group. For the case at hand, we are thus interested in the spectral flow modules for corresponding to
%\be
%	A_N \to A_N - \frac{\mathfrak{n}}{z} \qquad A_E \to A_E - \frac{\mathfrak{m} k + \mathfrak{n}\tfrac{k}{2}}{z}
%\ee
%for $\mathfrak{m},\mathfrak{n} \in \Z$. Translating this to the currents%
%
%\footnote{For Chern-Simons theory with gauge Lie algebra $\fg$, the quantum-corrected identification between the affine currents $B_a$ and the gauge fields $A^a_z$ is $B_a = K^{\textrm{eff}}_{ab} A^b_z$. Here we have $$K^{\textrm{eff}}_{ab} = \begin{pmatrix}
%		1 & 1\\ 1 & 0
%	\end{pmatrix} \qquad \Rightarrow \qquad N := B_N = A_N + A_E,\, E := B_E = A_N$$} %
%
%
%$N,E$, we find
The analysis of Appendix \ref{sec:u11CS} leads us to the lattice of spectral flows corresponding to
\be
	N \to N - \frac{(\mathfrak{m}k + \mathfrak{n}(\tfrac{k}{2}+1))}{z} \qquad E \to E - \frac{\mathfrak{n}}{z}~.
\ee
The basic spectral flows $(\mathfrak{m},\mathfrak{n}) = (1,0)$ and $(0,1)$ correspond to the spectral flow automorphisms with $(l,\lambda) = (0,k)$ and $(1, \tfrac{k}{2}+1)$, respectively. We denote the corresponding spectral flows $\sigma_1, \sigma_2$. In the traditional generators $\wt{N}, E, \psi_\pm$ these basic spectral flows act as
\be
\label{eq:spectral}
\begin{aligned}
	\sigma_1(\wt{N}) & = \wt{N} - \frac{k}{z} \qquad & \sigma_1(E) & = E \qquad & \sigma_1(\psi_\pm) & = \psi_\pm\\
	\sigma_2(\wt{N}) & = \wt{N} - \frac{(k+1)/2}{z} \qquad & \sigma_2(E) & = E - \frac{1}{z} \qquad & \sigma_2(\psi_\pm) & = z^{\mp 1} \psi_\pm
\end{aligned}
\ee

It is a straightforward task to identify the modules obtained via spectral flow of the vacuum. See Appendix \ref{sec:reptheory} for a review of relevant aspects of the representation theory of $\fgl(1|1)$ and $V(\fgl(1|1))$, or, e.g., \cite[Section 3]{BN22} for a more thorough discussion. Consider first the spectral flow automorphism $\sigma_1$. Let $\ket{0}$ denote the vacuum vector; the state $\sigma_1(\ket{0})$ is a highest weight vector of $V(\fgl(1|1))$, with weights $n = k$ and $e = 0$. Since spectral flow respects submodules, we conclude that $\sigma_1(V(\fgl(1|1))) = \widehat{A}_{k, 0}$.

Now consider the spectral flow $\sigma_2(V(\fgl(1|1)))$. Once again $\sigma_2(\ket{0})$ is a highest weight vector but now with $n = \frac{k+1}{2}$ and $e = 1$. We conclude that $\sigma_2(V(\fgl(1|1))) = \widehat{A}_{\frac{k}{2}, 1}$. More generally, forgetting about parity for the moment, we find:
\be
	\sigma_1^{\mathfrak{m}} \sigma_2^{\mathfrak{n}}(V(\fgl(1|1))) = \begin{cases}
		\widehat{A}_{\mathfrak{m}k + \mathfrak{n}\big(\frac{k-1}{2}\big) + \frac{1}{2}, \mathfrak{n}} & \mathfrak{n} > 0\\
		\widehat{A}_{\mathfrak{m}k,0} & \mathfrak{n} = 0\\
		\widehat{A}_{\mathfrak{m}k + \mathfrak{n}\big(\frac{k-1}{2}\big) - \frac{1}{2}, \mathfrak{n}} & \mathfrak{n} < 0
\end{cases}
\ee
The more precise treatment of parity we describe below gives us the extension by (the lattice generated by) the modules $\widehat{A}_{k,0}$ and $\Pi^{k+1} \widehat{A}_{\frac{k}{2},1}$%
\footnote{More generally, we find that the VOA $\CV_{(\kappa, \nu, \xi)}$ is realized as an extension by the lattice of simple currents generated by $\widehat{A}_{\frac{\kappa}{\nu}, 0}$ and $\Pi^{2\xi+1}\widehat{A}_{\frac{\xi}{\nu} + \frac{1-\nu}{2}, \nu}$ and their contragredients.} %
\be
\label{eq:VOA}
	\CV_k = \bigoplus\limits_{\mathfrak{m} \in \Z} \Bigg( \widehat{A}_{\mathfrak{m}k,0} \oplus \bigoplus_{\mathfrak{n} \in \Z_{>0}}\Pi^{\mathfrak{n} k}\bigg(\Pi\widehat{A}_{\mathfrak{m} k + \mathfrak{n}\big(\frac{k-1}{2}\big)+\frac{1}{2}, \mathfrak{n}} \oplus \widehat{A}_{\mathfrak{m} k - \mathfrak{n}\big(\frac{k-1}{2}\big)-\frac{1}{2}, -\mathfrak{n}}\bigg) \Bigg)
\ee
where $\Pi$ denotes a parity shift. Proposition 2.2.4 of \cite{CMY20} implies that the dual, i.e. contragredient, module of $\widehat{A}_{n,e}$ is $\Pi \widehat{A}_{-n,-e}$ when $e$ is nonzero and $\widehat{A}_{-n,0}$ otherwise. Thus the VOA $\CV_k$ is a self-dual extension of $V(\fgl(1|1))$.

This strange parity assignment cannot be seen solely from the perspective of these monopoles as modules for the perturbative algebra $V(\fgl(1|1))$. This should not be particularly surprising: if one considers pure $U(1)$ Chern-Simons theory at level $k$, corresponding to a lattice VOA $V_{\varphi}$ with pairing $(\varphi, \varphi) = k$, the boundary monopole with magnetic charge 1 $\norm{e^\varphi}$ has parity $(-1)^F = (-1)^k$. The parity assignments in Eq. \eqref{eq:VOA} will become apparent when we describe their free-field realization in terms of vertex operators below.

\subsubsection{The half-index}
\label{sec:halfindex}

We can check that the above extension to the perturbative algebra agrees with physical expectations by comparing its supercharacter with a half-index computation. Somewhat more precisely, if we view our Chern-Simons theory as a twist of a certain%
\footnote{Explicitly, our theory can be identified with twisted 3d $\CN=2$ theory of two $\CN=2$ abelian vectormultiplets with gauge fields $A_N, A_E$ having Chern-Simons kinetic terms coupled to two chiral multiplets $Z^+, Z^-$.} %
$\CN=2$ theory, the work \cite{DimofteGaiottoPaquette} provides a concrete expression for the character of the vacuum module for the boundary VOA, viewed as a half-index of the corresponding $\CN=(0,2)$ boundary condition. We find:
\be
\label{eq:vacchar}
\begin{aligned}
	I\!\!I(q;s,t) & = \Tr (-1)^R q^J s^{N_0} t^{E_0}\\
	&= \sum_{\mathfrak{m},\mathfrak{n} \in \Z} q^{\mathfrak{n}\mathfrak{m} k + \mathfrak{n}^2\big(\frac{k+1}{2}\big)} s^{\big(\mathfrak{m}k + \mathfrak{n}(\frac{k}{2}+1)\big)} t^{\mathfrak{n}} \frac{(s q^{1+\mathfrak{n}};q)_\infty (s^{-1}q^{1-\mathfrak{n}};q)_\infty}{(q;q)_\infty^2}
\end{aligned}
\ee
In this expression, $R$ measures the $R$-charge/cohomological grading; $q$ is a fugacity for twisted spin $J$, which is the same conformal dimension with respect to $T_{\textrm{Sug}}$; $s$ (resp. $t$) is a fugacity for the boundary $U(1)$ flavor symmetry generated by $N_0$ (resp. $E_0$); the integers $\mathfrak{m},\mathfrak{n}$ denote the magnetic charge of the corresponding boundary local operator; and $(x;q)_\infty$ is the $q$-Pochhammer symbol, defined as
\be
(x;q)_\infty = \prod\limits_{n \geq 0}(1-x q^n)\,.
\ee

In order to connect to the character formulae of \cite{CR13a, CR13b}, we perform the change of variables
\be
\sigma = s\,, \quad \tau = t s^{1/2}\,,
\ee
so that $\sigma$ is a fugacity for the generator $\wt{N}_0$, to express the half-index as
\be
\label{eq:vacchar3}
	I\!\!I(q;\sigma,\tau) = \sum_{\mathfrak{m},\mathfrak{n} \in \Z} q^{\mathfrak{n}\mathfrak{m} k + \mathfrak{n}^2\big(\frac{k+1}{2}\big)} \sigma^{\mathfrak{m}k + \mathfrak{n}\big(\frac{k+1}{2}\big)} \tau^{\mathfrak{n}} \frac{(\sigma q^{1+\mathfrak{n}};q)_\infty (\sigma^{-1}q^{1-\mathfrak{n}};q)_\infty}{(q;q)_\infty^2}\,.
\ee
In order to compare this half-index to the graded supercharacter, we need to replace the $(-1)^R$ in the definition of the half-index with the fermionc parity operator $(-1)^F$; as noted in Footnote 12 of \cite{DimofteGaiottoPaquette}, this can be accomplished by sending $q^{\frac{1}{2}}$ with $-q^{\frac{1}{2}}$ yielding
\be
\label{eq:vaccharfin}
	\sum_{\mathfrak{m},\mathfrak{n} \in \Z} (-1)^{\mathfrak{n}^2(k+1)} q^{\mathfrak{n}\mathfrak{m} k + \mathfrak{n}^2\big(\frac{k+1}{2}\big)} \sigma^{\mathfrak{m}k + \mathfrak{n}\big(\frac{k+1}{2}\big)} \tau^{\mathfrak{n}} \frac{(\sigma q^{1+\mathfrak{n}};q)_\infty (\sigma^{-1}q^{1-\mathfrak{n}};q)_\infty}{(q;q)_\infty^2}\,.
\ee
Alternatively, we could choose $R$-charge assignments so that fermionic parity agrees with $R$-charge mod $2$, cf. Section 5.2.3 of \cite{descent}; this would involve shifting the standard $R$-charge assignments in the $B$-twist by the topological flavor symmetry to correctly account for the parity of boundary monopole operators.%
\footnote{Explicitly, the natural cohomological grading in the $B$-twist is given by the weight with respect to the diagonal torus in the Higgs branch $R$-symmetry $SU(2)_H$. This grading isn't compatible with the fermionic parity of local operators, e.g. the hypermultiplet scalars $Z^\pm$ are given degree 1, so one typically uses the $U(1)$ symmetry that acts on $Z^\pm$ with weight $\pm 1$ to make $Z^+$ degree 0 and $Z^-$ degree 2, cf. Eq. (2.26) of \cite{twisted}. The $R$ appearing in Eq. \eqref{eq:vacchar} corresponds to this choice. Unfortunately, this degree assignment doesn't match the parity of the boundary local operators and we need to consider $R' = R + (k+1) E_0$ to remedy this. This shift precisely accounts for the $(-1)^{\mathfrak{n}^2(k+1)} = (-1)^{\mathfrak{n}(k+1)}$ in Eq. \eqref{eq:vaccharfin}.} %

We want to compare Eq. \eqref{eq:vaccharfin} to the supercharacter of the VOA $\CV_k$ identified in the previous subsection. The supercharacter of the atypical modules for the current algebra $V(\fgl(1|1))$ can be obtained from the supercharacter of the Verma modules $\widehat{V}_{n,e}$ and splicing the exact sequences presented in Section \ref{sec:introVOA}, cf. Section 3.2 of \cite{CR13a}; the resulting supercharacters are given by
\be
\begin{aligned}
	\chi\big[\widehat{A}_{n, e}\big](q;\sigma, \tau) & = \Tr (-1)^F q^{L_0} \sigma^{\wt{N}_0} \tau^{E_0}\\
	& = \begin{cases}
		(-1)^{e-1} q^{e(n + e - \epsilon(e))} \sigma^{n + e - \epsilon(e)} \tau^{e}  \frac{(\sigma q^{1+e};q)_\infty (\sigma^{-1} q^{1-e};q)_\infty}{(q;q)_\infty^2} & e > 0\\
		(-1)^{e} q^{e(n + e - \epsilon(e))} \sigma^{n + e - \epsilon(e)} \tau^{e}  \frac{(\sigma q^{1+e};q)_\infty (\sigma^{-1} q^{1-e};q)_\infty}{(q;q)_\infty^2} & e \leq 0
	\end{cases}
\end{aligned}
\ee
where $\epsilon(e) = \frac{1}{2}\textrm{sign}(e)$. Inserting the formula for the supercharacters into Eq. \eqref{eq:vaccharfin}, we see that the half-index $I\!\!I$ counting local operators at the boundary is in full agreement with the vacuum character $\chi[\CV_k]$ of the boundary VOA identified in the previous section.

\subsection{Free field realization of $\CV_k$}
\label{sec:freefield}

Having determined the boundary VOA $\CV_k$ as a module for the perturbative algebra $V(\fgl(1|1))$, we can understand it more explicitly with a free field realization. This free field realization, once we make the suitable change in extension, will allow us to describe any of the $\CV_{(\kappa, \nu, \xi)}$ and will give us direct access to the category of line operators, to be studied in Section \ref{sec:lines}.

Let us first recall the free field realization of $V(\fgl(1|1))$ described in \cite[Section 2]{creutzig2021duality}. Consider a VOA of three free bosons $X,Y,Z$ with pairings $(X,Y) = 1 = (Z,Z)$. The currents $\pd X$, $\pd Y$, and $\pd Z$, with OPEs
\begin{equation}
	\pd X\pd Y\sim \frac{1}{(z-w)^2}\,, \qquad \pd Z\pd Z\sim \frac{1}{(z-w)^2}\,,
\end{equation}
generate a Heisenberg VOA that we denote $\CH$. For each linear combination $\mu=aX+bY+cZ$, there is a simple Fock module $\mathbb{F}_\mu$ of $\CH$ generated by a single vector $\ket{\mu}$. Explicitly, $\ket{\mu}$ is annihilated by the positive modes of the abelian currents $\pd X, \pd Y, \pd Z$; the negative modes act freely; and the zero modes act via their paring with $\mu$
\be
	X_0 \ket{\mu} = b \ket{\mu}\,, \qquad Y_0 \ket{\mu} = a \ket{\mu}\,, \qquad  Z_0 \ket{\mu} = c \ket{\mu}\,.
\ee
Here, and in the following, $X_0$ denotes the zero mode of a Heisenberg generator $\pd X$, and similarly the other Heisenbergs. The vacuum module $\CH$ is identified with $\mathbb{F}_0$. We then consider the following module:
\begin{equation}
	\CV_Z:=\bigoplus\limits_{n\in \Z} \mathbb{F}_{nZ}
\end{equation}
$\CV_Z$ has the structure of a vertex operator superalgebra and is realized by extending $\CH$ by the fermionic vertex operators $\norm{e^{\pm Z}}$ with OPE
\begin{equation}
	\norm{e^Z}\norm{e^{-Z}}\sim \frac{1}{z-w}.
\end{equation}
The assignments
\begin{equation}
	\begin{aligned}
		N(z) & \mapsto \pd X(z)+\pd Z(z) \qquad & E(z) & \mapsto \pd Y(z)\\
		\psi_+(z) & \mapsto \norm{e^{Z(z)}}\qquad & \psi_{-}(z) & \mapsto \pd \norm{e^{-Z(z)}}+\pd Y \norm{e^{-Z(z)}}
	\end{aligned}
\end{equation}
define an embedding $V(\fgl(1|1))\hookrightarrow \CV_Z$. Define the screening operator:
\begin{equation}
	S:=\oint \norm{e^{Z-Y}}: \CV_{Z,\nu}\longrightarrow \CV_{Z,\nu-Y}.
\end{equation}
One can show that:
\begin{equation}
	V(\fgl(1|1))=\text{Ker}\left(S\big\vert_{\CV_{Z,0}}\right).
\end{equation}

We now use this free-field realization to understand the extended VOA $\CV_k$, which is the extension of $V(\fgl(1|1))$ by the lattice of simple currents generated the spectral flows $\sigma_1(V(\fgl(1|1))), \sigma_2(V(\fgl(1|1)))$, where the action of $\sigma_1, \sigma_2$ was given in Eq. \eqref{eq:spectral}. For each linear combination $\nu=aX+bY$, the Fock module $\CH_\nu$ can be lifted to a module of $\CV_Z$, which we call $\CV_{Z,\nu}$. Explicitly, this is given by:
\begin{equation}
	\CV_{Z,\nu}=\bigoplus\limits_{n\in \Z} \mathbb{F}_{aX+bY+nZ}\,.
\end{equation} Consider the linear combinations $\lambda_1=kY$ and $\lambda_2=X + \frac{k}{2}Y$.%
\footnote{For general $(\kappa, \nu, \xi)$ we take the linear combinations $\lambda_1 = \frac{\kappa}{\nu}Y$ and $\lambda_2 = \nu X + \frac{\xi}{\nu}Y$.} %
As described in \cite[Section 4.3]{CR13b}, we have the following identifications:
\begin{equation}
	\sigma_1 (V(\fgl(1|1)))=\text{Ker}\big(S\big\vert_{\CV_{Z,\lambda_1}}\big)\,, \qquad \sigma_2(V(\fgl(1|1)))=\text{Ker}\big(S\big\vert_{\CV_{Z,\lambda_2}}\big)\,.
\end{equation}
Under this equivalence, the images of the vacuum vector are given by:
\be
	\sigma_1 (\ket{0})=\ket{kY},~\sigma_2(\ket{0})=\ket{X + \tfrac{k}{2}Y + Z}.
\ee
One can see that these vectors are in the kernel of $S$ and have the correct weights under $N_0$ and $E_0$. Let $\CV_{Z,k}$ be the extension of $\CV_Z$ by the lattice of simple modules spanned by $\CV_{Z,\lambda_1}$ and $\CV_{Z,\lambda_2}$. In other words:
\be
	\CV_{Z,k}=\bigoplus\limits_{\mathfrak{m}, \mathfrak{n} \in \Z} \CV_{Z,\mathfrak{m}\lambda_1 + \mathfrak{n}\lambda_2}.
\ee
Then the above considerations imply that there is an embedding $\CV_k\hookrightarrow \CV_{Z,k}$ such that
\begin{equation}
	\CV_k=\mathrm{Ker}\left(S\big\vert_{\CV_{Z,k}}\right).
\end{equation}

As mentioned above, we can identify the proper parity assignments mentioned in Eq. \eqref{eq:VOA} using this free field realization. Namely, the fermionic parities of the states $\ket{kY}$ and $\ket{X+\tfrac{k}{2}Y+Z}$ are determined by the lengths of the corresponding lattice vectors: the length of the former is $0$ and the latter is length $k+1$, whence $\ket{kY}$ is always a boson and $\ket{X+\tfrac{k}{2}Y+Z}$ is a boson when $k$ is odd and a fermion when $k$ is even. Since both of these states are highest weight vectors for $V(\fgl(1|1))$, their parity determines the whether we have $\widehat{A}$ or $\Pi \widehat{A}$: the state $\ket{kY}$ generates $\widehat{A}_{k,0}$ and the state $\ket{X + \tfrac{k}{2}Y + Z}$ generates $\Pi^{k+1} \widehat{A}_{\frac{k}{2},1}$. A similar analysis can be applied to the remaining modules arising in $\CV_k$, yielding the result quoted in Eq. \eqref{eq:VOA}.

Note that, just as in \cite[Section 4.4]{CR13b}, when $k$ is odd one needs to be careful when defining the VOA structure. The problem can be seen from the above free field realization: the vertex operators $e^{X-\frac{k}{2} Y + Z}$ and $e^{kY}$ are bosonic with themselves but are mutually fermionic. The solution, as in loc. cit., is to include operator-valued functions to the vertex operators. In our case, we would define the vertex operator $Y(\ket{nX - \tfrac{nk}{2} Y + n Z}, z)$ corresponding to a state $\ket{nX - \tfrac{nk}{2} Y + n Z}$ by the following:
\be
	Y(\ket{n X - \tfrac{nk}{2} Y + n Z}, z)=e^{\pi i n\big(X_0 - \frac{k}{2}Y_0 + Z_0\big)}\norm{e^{nX - \frac{nk}{2} Y + n Z}}\,. 
\ee
This does not change the commutation relations of $Y(\ket{X - \frac{k}{2}Y + Z},z)$ with itself, but the commutation relations of $Y(\ket{\pm (X - \frac{k}{2}Y + Z)}, z)$ with $e^{\pm Y}$ become anti-commutation relations. This modification makes $V_{Z,k}$, and consequently $\CV_k$, a vertex operator superalgebra.

\subsubsection{A closer look at $\CV_{(k,1,0)}$}
\label{sec:VOAorbifold}
We now take a brief detour into different example that will be relevant to us later: the theory $\CT_{(k,1,0)}$ and its boundary VOA $\CV_{(k,1,0)}$. Using (a slight modification of) the above free-field realization, we can provide a more concrete description of the $\CV_{(k,1,0)}$. The category of modules for this VOA realizes the category of lines in the theory $\CT_{(k,1,0)}$, which was predicted by \cite{Mikhaylov} to be equivalent to the $\Z_k$ orbifold of a $B$-twisted hypermultiplet. Line operators in a $B$-twisted hypermultiplet, prior to the $\Z_k$ orbifold, are described by the symplectic fermion VOA $V_{\chi_+,\chi_-}$. Correspondingly, line operators in the orbifold theory are reproduced by the orbifold VOA $V^{\Z_k}_{\chi_+, \chi_-}$. In the following, we will see that $\CV_{(k,1,0)}$ is quite close to this orbifold, but they are actually distinct. As we will see in Section \ref{sec:Bmodelorbifold}, module categories for these VOAs are only equivalent up to a change in the braiding morphisms.

We start with the following field redefinition, which was used in \cite[Section 4.2.2]{BN22}:
\be
	\phi_1 = Z-Y\,, \qquad \phi_2 = X+Z\,, \qquad \phi_3 = X-Y+Z.
\ee
These have OPEs
\begin{equation}
	\pd\phi_1\pd\phi_1\sim \pd\phi_2\pd\phi_2\sim -\pd\phi_3\pd\phi_3\sim \frac{1}{(z-w)^2}.
\end{equation}
The VOA $\CV_{(k,1,0)}$ arises by extending the Heisenberg VOA $\CH$ by the lattice spanned by $X$, $kY$, and $Z$ and we can express these in the new variables as:
\be
	\phi_3 - \phi_1\,, \qquad k(\phi_2 - \phi_3)\,, \qquad \phi_1 + \phi_2 - \phi_3\,.
\ee 
For convenience, we will denote by $\mathbb{F}_{n\phi_1}$ the Fock module corresponding of $n\phi_1$, and similarly for $\phi_2, \phi_3$. The screening operator $S=\oint \norm{e^{Z-Y}}=\oint \norm{e^{\phi_1}}$ only acts on the Fock modules of $\phi_1$ and we will denote by $M_{n\phi_1}$ the kernel of the screening operator restricted to $\mathbb{F}_{n\phi_1}$. It follows that $\CV_{(k,1,0)}$ takes the following form:
\be
	\CV_{(k,1,0)}=\bigoplus_{\mathfrak{l},\mathfrak{m}, \mathfrak{n}\in \Z} M_{\mathfrak{l}\phi_1}\otimes \mathbb{F}_{(\mathfrak{l} + \mathfrak{m}k + \mathfrak{n})\phi_2}\otimes \mathbb{F}_{(-\mathfrak{l}-\mathfrak{m}k)\phi_3}. 
\ee
Let us denote by $a,b,c$ the following integers:
\be
	a=\mathfrak{l} \qquad b = \mathfrak{l} + \mathfrak{m}k + \mathfrak{n} \qquad c = -\mathfrak{l}-\mathfrak{m}k.
\ee
We do not sum over all values of the integers $(a,b,c)$ since the change of summation variables from $(\mathfrak{l}, \mathfrak{m}, \mathfrak{n})$  is not onto: because $a+c = \mathfrak{m}k$, the sum $a+c$ is divisible by $k$. In other words,
\be\label{eq:ffCVpd}
	\CV_{(k,1,0)}=\bigoplus\limits_{\substack{a,b,c\in \Z\\ a+c \in k \Z}} M_{a\phi_1}\otimes \mathbb{F}_{b\phi_2}\otimes \mathbb{F}_{c\phi_3}.
\ee
Note that there is free field realization of a pair of symplectic fermions $\chi_\pm$ given by
\be
	V_{\chi_+,\chi_-}=\bigoplus\limits_{a \in \Z} M_{a\phi_1}.
\ee
Let us denote by $V_{\phi_2}$ (resp. $V_{\phi_3}$) the lattice VOA associated to $\phi_2$ (resp. $\phi_3$). The VOA $V_{\chi_+\chi_-}\otimes V_{\phi_2} \otimes V_{\phi_3}$ has an action of $\mathbb{Z}_k$ given by
\begin{equation}
	\exp \left(\frac{2\pi i}{k}\left(\phi_{1,0}-\phi_{3,0}\right)\right)\,,
\end{equation}
and Eq. \eqref{eq:ffCVpd} implies that $\CV_{(k,1,0)}$ is then identified with the $\mathbb{Z}_k$ orbifold of this product:
\be
	\CV_{(k,1,0)}=\left(V_{\chi_+,\chi_-}\otimes V_{\phi_2} \otimes V_{\phi_3}\right)^{\mathbb{Z}_k} = (V_{\chi_+,\chi_-}\otimes V_{\phi_3})^{\Z_k} \otimes V_{\phi_2}.
\ee
In terms of the original Heisenberg VOA $\CH$, the action of $\mathbb{Z}_k$ is generated by:
\begin{equation}
	\exp \left(-\tfrac{2\pi i}{k} X_0\right).
\end{equation}

\section{The Category of Line Operators}
\label{sec:lines}

If we view the boundary VOA $\CV_k$ as an object in (a suitable completion of) the Kazhdan-Lusztig category $KL$ of $V(\mathfrak{gl}(1|1))$, the category $\CC_k := \textrm{Rep}^0_{KL} \CV_k$ of generalized modules for our boundary VOA $\CV_k$ that belong to $KL$, i.e. our candidate%
\footnote{As noted in the Introduction, the physical category of line operators should be identified with the derived category $D^b\CC_k$. We will mostly focus on the abelian category $\CC_k$, except when we consider bulk local operators in Section \ref{sec:locops}.} % 
for the category of line operators in the bulk Chern-Simons theory, can be identified as a braided tensor category with a de-equivariantization, cf. \cite[Section 8.23]{etingof2016tensor}, of the subcategory $KL^0$ of modules in $KL$ that are local for $\CV_k$ \cite{CKM17, CMY22}. Namely, there is a functor $\CF: KL^0 \to \CC_k$ taking a $\CV_k$-local module $M$ to a generalized $\CV_k$ module via fusion with $\CV_k$:
\be
	\CF(M) = \CV_k \times M
\ee
This functor is not faithful as we can fuse $M$ with the simple currents used to extend the perturbative current algebra and obtain the same result. Instead, the image of $\CF$ can be described by identifying $V(\fgl(1|1))$ modules that differ by fusion with these simple currents, i.e. by de-equivariantizing the action of $\Z^2$ generated by fusion with the simple currents $\widehat{A}_{k,0}$ and $\Pi^{k+1}\widehat{A}_{\frac{k}{2},1}$ or their contragredients. Thus, our candidate for the category $\CC_k$ is this de-equivariantization:
\be
	\CC_k \simeq KL^0/\Lambda_k
\ee
where we denote the lattice generated by these simple currents by $\Lambda_k$, which can in turn be identified with the cocharacter lattice as described in Appendix \ref{sec:u11CS}.

Any module $M$ in the Kazhdan-Lusztig category $KL$ admits a decomposition into generalized eigenspaces for the central generator $E_0$, leading to a decomposition of the category itself into blocks
\be
	KL = \bigoplus_{e \in \C} KL_{e}
\ee
where $KL_{e}$ denotes the subcategory of modules on which $E_0$ has generalized eigenvalue $e$. The category of $\CV_k$-local modules $KL^0$ inherits an identical decomposition. When $e \in \Z$, the subcategory $KL_{e}$ is generated by the atypicals $\widehat{A}_{n,e}$ and it is generated by the typicals $\widehat{V}_{n,e}$ otherwise; these modules exhaust the simple objects of $KL$ \cite{CR13a}. In other words, any module in $KL_e$ can be realized as an iterated extension of these simple modules or, equivalently, any such module has a composition series consisting of these modules. Since fusion with the simple current $\widehat{A}_{n,e}$ is an equivalence $\widehat{A}_{n,e} \times - : KL_{e'} \to KL_{e'+e}$, with inverse given by fusion with the contragredient module $\Pi \widehat{A}_{-n,-e}$ (or, if $e = 0$, $\widehat{A}_{-n,0}$), we automatically see that $\CC_k$ inherits a decomposition into blocks
\be
	\CC_k = \bigoplus\limits_{[e] \in \C/\Z} \CC_{k,[e]}
\ee
where $\CC_{k,[e]}$ denotes the subcategory of modules of $\CV_k$ where the generalized eigenvalues of the central generator $E_0$ are contained in $[e] = e + \Z$. In fact, we will find that $\CC_{k,[e]}$ is nonempty only when $[e] \in \frac{1}{k}\Z/\Z \subset \C/\Z$.

In Section \ref{sec:Wilson}, we use the realization of $\CC_k$ as a de-equivariantization of $KL^0$ to provide an concrete description of the categories $\CC_{k, [e]}$. To do so, we use the free-field realization of Section \ref{sec:freefield} to analyze the locality constraint going into the definition of $KL^0$; the result is quite simple, $KL^0$ only contains modules where the zeromodes $kE_0$ and $N_0 + \frac{k}{2}E_0$ act semisimply with integer eigenvalues, but has a somewhat technical derivation that we relegate to Appendix \ref{sec:reptheory}. We note in Section \ref{sec:1form} the presence of a $\Z_k^{(1)}$ 1-form global symmetry and describe its action on the category $\CC_k$. For example, the category $\CC_{k,[e]}$ corresponds to line operators with 1-form charge $[e]$.

Section \ref{sec:fibercat} applies results of \cite{BN22} to provide a simplified description of $\CC_{k,[e]}$ as an abelian category and uses this to derive the algebra of local operators in $\CT_k$, finding the algebra of holomorphic functions on the ``Higgs branch'' $\C^2/\Z_k$.

In Section \ref{sec:qgroup} we relate $\CC_k$ to the category of modules for a (quasi-)quantum group $\CA_k$. We arrive at $\CA_k$ in two ways: first, we consider a mild modification of the argument in \cite[Section 5]{BN22} to establish an equivalence of categories between $\CC_k$ and a category of modules for an, a priori unknown, algebra $\CA_k$. Our second route to $\CA_k$ is by mimicking the simple current extension described in the previous section: we realize $\CA_k$ by uprolling the unrolled, restricted quantum group $\ol{U}^{E}(\fgl(1|1))$, cf. \cite{creutzig2022uprolling}.

Finally, in Section \ref{sec:Bmodelorbifold} we return to the theory $\CT_{(k,1,0)}$ and more closely compare the category $\CC_{(k,1,0)}$ of line operators therein to that of the 3d $B$-model (a.k.a Rozansky-Witten theory) with orbifold target $\C^2/\Z_k$. We verify that these theories indeed have the same abelian categories of line operators from various perspectives, but we find that this equivalence does not preserve the braiding morphisms, so that line operators in $\CT_{(k,1,0)}$ and this 3d $B$-model are \emph{not} equivalent. It is presently unclear how to reconcile this with the results of \cite{Mikhaylov}.

Although we won't use it here, we note that $\CV_k$ can also be viewed as a simple current extension of the VOAs appearing in \cite{CR13b}. For example,  the subVOA with $\mathfrak{n} = -2 \mathfrak{m}$ is the simple current extension $\mathfrak{W}_{-\frac{1}{2},2}$ introduced in loc. cit., where it was identified with (the simple quotient of) an $\fsl(2|1)$ current algebra at level $-\tfrac{1}{2}$ 
\be
	\mathfrak{W}_{-\frac{1}{2},2} \simeq V_{-\frac{1}{2}}(\mathfrak{sl}(2|1))
\ee
Our boundary VOA $\CV_k$ is an extension of $V_{-\frac{1}{2}}(\mathfrak{sl}(2|1))$ in (a completion of) the category $\textrm{Rep}_{KL}^0V_{-\frac{1}{2}}(\mathfrak{sl}(2|1))$ of generalized $V_{-\frac{1}{2}}(\mathfrak{sl}(2|1))$ modules that belong to in $KL$. If $\CG$ denotes the induction functor from $V_{-\frac{1}{2}}(\mathfrak{sl}(2|1))$-local modules in $KL$ to $\textrm{Rep}^0_{KL} V_{-\frac{1}{2}}(\mathfrak{sl}(2|1))$, we see that $\CV_k$ is given by extending by the infinite-order simple currents $\CG(\Pi^{k+1}\widehat{A}_{-\frac{k}{2},1})$ and $\CG(\Pi^{k+1}\widehat{A}_{\frac{k}{2},1})$:
\be
	\CV_k = V_{-\frac{1}{2}}(\mathfrak{sl}(2|1)) \bigoplus_{m >0} \CG(\Pi^{k+1}\widehat{A}_{\frac{k}{2},1})^{\boxtimes m} \oplus \CG(\Pi^{k+1}\widehat{A}_{-\frac{k}{2},1})^{\boxtimes m}
\ee
Of course, as mentioned in Section \ref{sec:introVOA}, the subalgebra of $\CV_k$ with $r \mathfrak{m} = s \mathfrak{n}$ ($r > 0$) can also be identified with one of the simple current extensions of \cite{CR13b}, namely $\mathfrak{W}_{(s+\frac{r}{2})k+\frac{r-1}{2}, r}$, and thus $\CV_k$ a simple current extension thereof.

\subsection{Wilson lines and local modules}
\label{sec:Wilson}

A dramatic simplification to the determination of what modules in $KL$ are local with respect to the simple current extension $\CV_k$ follows from the fact that any module in $KL$ is a sub-quotient, i.e. a quotient of a submodule, of a module realizable via the free-field realization described in Section \ref{sec:freefield}; see Appendix \ref{sec:reptheory} for a proof of this fact.%
\footnote{Strictly speaking, the free-field realization is able to realize specific parities of $V(\fgl(1|1))$ modules. For example, the module $\widehat{A}_{\lambda,0}$ can be realized via the vertex operator $\norm{e^{\lambda Y}}$; $\Pi \widehat{A}_{\lambda, 0}$ is a perfectly good module but cannot be realized in terms of free fields. It is unclear whether restricting attention to those modules of $V(\fgl(1|1))$ that can be realized in terms of free fields is physically meaningful, i.e. whether the free-field realization is merely a calculational tool or has physical meaning in and of itself. We note that this restriction could be related to the additional $\Z_2$ appearing in the translation groups used in \cite{GYtqft} and to the fact that the underlying supersymmetric Chern-Simons-matter theory has $k^2$ Bethe vacua, cf. Section 5 of \cite{CDGG}, in contrast to the $2k^2$ simples we find by including both parities.} %
In particular, we can perform a computation with free fields to determine the desired monodromies. We note that it suffices to determine the monodromy of the simple modules in $KL$ with the simple currents $\widehat{A}_{k,0}$ and $\Pi^{k+1}\widehat{A}_{\frac{k}{2},1}$; as shown in Section \ref{sec:compmonodromy}, the monodromy of these simple currents and a given module $M$ is given by
\be
\begin{aligned}
	\CM(\widehat{A}_{k,0}, M) & = \id_{\widehat{A}_{k,0}} \times \exp\big(2\pi i(\overset{v_1}{\overbrace{k E_0}})\big)\\
	\CM(\Pi^{k+1}\widehat{A}_{\frac{k}{2},1}, M) & = \id_{\Pi^{k+1}\widehat{A}_{\frac{k}{2},1}} \times \exp\big(2\pi i(\overset{v_2}{\overbrace{N_0+\tfrac{k}{2} E_0}})\big)
\end{aligned}
\ee
viewed as maps on the tensor product. From this expression for the monodromy, we immediately see that for $M$ to be $\CV_k$-local the generators $v_1, v_2$ must act semisimply with integer eigenvalues.
%This implies if $M = \widehat{U}$ is induced from a $\fgl(1|1)$ module $U$, then $U$ must integrate to a representation of the global form of the gauge group for $\CT_k$.
Explicitly, we find that the category $KL^0$ can be identified as
\be
	KL^0 = \bigoplus_{e \in \frac{1}{k}\Z} KL^{N+\frac{k}{2}E, E}_{e}
\ee
where $KL^{N+\frac{k}{2}E, E}_{e}$ denotes the subcategory of $KL$ where $N_0 + \frac{k}{2}E_0$ acts semisimply with integer eigenvalues and $E_0$ acts semisimply with eigenvalue $e$. Thus, the category of line operators $\CC_k$ can be realized as the following de-equivariantization:
\be
	\CC_k = \bigg(\bigoplus_{e \in \frac{1}{k}\Z} KL^{N+\frac{k}{2}E, E}_{e}\bigg)/\Lambda_k\,.
\ee

It will be useful to make the category $KL^0$ of $\CV_k$-local modules a bit more explicit. Due to the fact that the trivial-monodromy constraint implies both $N$ and $E$ act semisimply, it suffices to consider the atypical simples $\widehat{A}_{n,e}$ and typical simples $\widehat{V}_{n,e}$. All of the remaining modules are given by extensions of these simples on which the modes $N, E$ act semisimply, such as the indecomposible modules $\widehat{P}_{n,e}$, which for $e = 0$ are induced from the four-dimensional projective indecomposible $\fgl(1|1)$ module $P_n$. First consider the atypicals $\widehat{A}_{n,e}$. Because fusion with $\Pi^{k+1}\widehat{A}_{\frac{k}{2},1}$ and its contragredient can always be used to shift $e \to 0$, it suffices to consider this case.%
\footnote{In the general case it suffices to consider atypicals with $e = 0, ..., \nu -1$.} %
The monodromy of the generators and the simple current $\widehat{A}_{n,0}$ are given by
\be
\begin{aligned}
	\CM(\widehat{A}_{k,0}, \widehat{A}_{n,0}) & =  \id_{\widehat{A}_{k,0}} \times \id_{\widehat{A}_{n,0}}\\
	\CM(\Pi^{k+1}\widehat{A}_{\frac{k}{2},1}, \widehat{A}_{n,0}) & = \exp\big(2\pi i n\big) \id_{\Pi^{k+1}\widehat{A}_{\frac{k}{2},1}} \times \id_{\widehat{A}_{n,0}}
\end{aligned}
\ee
from which it follows that $\widehat{A}_{n, 0}$ and $\CV_k$ are mutually local if and only if $n$ is an integer. The modules $\CF\big(\widehat{A}_{\ell,0}\big)$ for $\ell \in \Z$ arise naturally from bulk Wilson lines associated to the 1-dimensional $\mathfrak{gl}(1|1)$ module $A_{\ell}$. In particular, we find that 
\be
	\CF\big(\widehat{A}_{\ell,0}\big) = \bigoplus\limits_{\mathfrak{m} \in \Z} \Bigg( \widehat{A}_{\ell + \mathfrak{m}k,0} \oplus \bigoplus_{\mathfrak{n} \in \Z_{>0}}\Pi^{\mathfrak{n}k}\bigg(\Pi\widehat{A}_{\ell+\mathfrak{m} k + \mathfrak{n}\big(\frac{k-1}{2}\big)+\frac{1}{2}, \mathfrak{n}} \oplus \widehat{A}_{\ell+\mathfrak{m} k - \mathfrak{n}\big(\frac{k-1}{2}\big)-\frac{1}{2}, -\mathfrak{n}}\bigg) \Bigg)
\ee
whose graded supercharacter is given by
\be
	\chi[\CF\big(\widehat{A}_{\ell,0}\big)] = \sum_{\mathfrak{m},\mathfrak{n} \in \Z} (-1)^{\mathfrak{n}^2(k+1)}(q^{\mathfrak{n}}s)^{\ell} q^{\mathfrak{n}\mathfrak{m} k + \mathfrak{n}^2\big(\frac{k+1}{2}\big)} s^{\mathfrak{m}k + \mathfrak{n}(\frac{k}{2}+1)} t^{\mathfrak{n}} \frac{(s q^{1+\mathfrak{n}};q)_\infty (s^{-1}q^{1-\mathfrak{n}};q)_\infty}{(q;q)_\infty^2}
\ee
exactly matching the half-index in the presence of the Wilson line $\CW_{A_{\ell}}$. Note that the module $\CF(\widehat{A}_{\ell,0})$ only depends on the value of $\ell$ up to a shift by integer multiples of $k$; this is the source of the need to de-equivariantize and reflects the fact that Wilson lines can be screened by gauge vortex lines in the presence of non-zero Chern-Simons terms.

The analysis for the typical modules $\widehat{V}_{n, e}$ is nearly identical and we find that the monodromy is trivial if and only if the eigenvalues of $k E_0$ and $N_0 + \frac{k}{2}E_0 = \wt{N}_0 + \frac{k+1}{2}E_0$ are integral. Correspondingly, we must require $e \in \frac{1}{k}\Z \backslash \Z$ and $n+\frac{1}{2} + \frac{(k+1)e}{2} \in \Z$. Just as above, we can identify the induced modules $\CF\big(\widehat{V}_{n, e}\big)$ with Wilson lines associated to $V_{n, e}$, and screening by gauge vortex lines implies many of these Wilson lines should be identified; for example, it suffices to consider $\widehat{V}_{n,e}$ with $0 < e < 1$.

Putting these results together, we see that the category $\CC_k$ only has support in the subcategories $KL_e$ for $e \in \frac{1}{k}\Z \subset \C$. At the level of the block decomposition mentioned above, we find
\be
	\CC_k = \bigoplus_{[e] \in \frac{1}{k}\Z/\Z} \CC_{k,[e]} \qquad \CC_{k,[e]} = \bigg(\bigoplus_{\ell\in \Z} KL^{N+\frac{k}{2}E, E}_{\ell+e}\bigg)/\Lambda_k
\ee
The category $\CC_{k,[0]}$ contains the vacuum module $\CV_k$ and is generated by the (modules induced by) atypicals $V(\fgl(1|1))$ modules $\CF(\widehat{A}_{0,0}), ..., \CF(\widehat{A}_{k-1,0})$, identified with Wilson lines for the atypical $\fgl(1|1)$ modules $A_0, ..., A_{k-1}$ up to screening. When $e = \frac{l}{k} \neq 0$ for $l \in \Z$, the category is generated by the (modules induced by the) typicals $\CF(\widehat{V}_{\frac{1}{2}+\frac{l(k+1)}{2k}, \frac{l}{k}}), ..., \CF(\widehat{V}_{k-\frac{1}{2}+\frac{l(k+1)}{2k}, \frac{l}{k}})$, identified with Wilson lines for the corresponding typical $\fgl(1|1)$ modules.

\subsubsection{1-form global symmetry}
\label{sec:1form}

Before moving on, we note that the Wilson lines $\CW_{A_{\ell}}$ generate a $\Z_k^{(1)}$ 1-form global symmetry. In particular, using the fusion rules (see e.g. Theorems 3.2.2 and 3.2.3 of \cite{CMY20})
\be
\label{eq:gl11fusion}
	\widehat{A}_{n,e} \times \widehat{A}_{n',e'} = \widehat{A}_{n + n' - \epsilon(e,e'), e+e'} \qquad \widehat{A}_{n,e} \times \widehat{V}_{n, e'} = \widehat{V}_{n + n'- \epsilon{e}, e+e'}
\ee
where $\epsilon(e,e') = \epsilon(e) + \epsilon(e') - \epsilon(e+e')$ for $\epsilon(e) = \frac{1}{2}\textrm{sign}(e)$, we see that fusing the simple currents $\widehat{A}_{n, 0}$ and $\widehat{A}_{n', 0}$ results in the simple current $\widehat{A}_{n + n', 0}$. Moreover, since the functor $\CF$ identifies the simple currents $A_{n,e}$ with $n$ values differing by integer multiples of $k$, we see that the fusion rules for the modules $\CF(\widehat{A}_{n, 0})$ ($n = 0, ..., k-1$) exactly match the group law of $\Z_k$. This 1-form symmetry is generated by, e.g., the single Wilson line $\CW := \CW_{A_{1}}$.%
\footnote{More generally, we find 1-form symmetry group with two generators $x,y$ satisfying $$x^\kappa = 1 \qquad y^d x^{2\xi} = 1$$ where, as above, $d = \gcd(\kappa,\nu)$. The generator $x$ is identified with the module $\CF(\widehat{A}_{\frac{1}{\nu},0})$ and $y$ is identified with $\CF(\widehat{A}_{\frac{1}{2} - (\frac{\xi}{\nu}+\frac{\nu}{2})\frac{1}{d}, \frac{\nu}{d}})$. As in $\CT_k$, the generator $x$ corresponds to a Wilson line associated to the atypical representations $A_{\frac{1}{\nu}}$. The line operator $y$ is not quite a Wilson line, but can be viewed as a subobject of the Wilson line associated to the typical representation $V_{\frac{1}{2} - \frac{\xi}{d \nu}, \frac{\nu}{d}}$.} %

\begin{figure}[H]
	\centering
	\begin{tikzpicture}
		\draw (-1,1) node {$\CW$};
		\draw (-0.5,1) -- (-0.5,-1);
		
		\draw (1,1) node {$\CL$};
		\draw (0.5,1) -- (0.5,-1);
		
		\draw[-latex] (-0.4,0.5) -- (-0.1,0.5);
		\draw[-latex] (-0.4,-0.5) -- (-0.1,-0.5);
	\end{tikzpicture}
	\raisebox{1cm}{$\rightsquigarrow$}
	\begin{tikzpicture}
		\draw (-0.75,1) node {$\CW \times \CL$};
		\draw (0,1) -- (0,-1);
	\end{tikzpicture}
	\hspace{3cm}
	\begin{tikzpicture}
		\draw (0.75,0) arc (0:180:0.75 and 0.225);
		
		\draw (0.5,1) node {$\CL$};
		\draw[white,fill = white] (-0.05,1) -- (0.05, 1) -- (0.05,-1) -- (-0.05,-1) -- cycle;
		\draw (0,1) -- (0,-1);
		
		\draw (-0.75,0) arc (180:360:0.75 and 0.225);
		
		\draw[-latex] (-0.65,0) -- (-0.25,0);
		\draw[-latex] (0.65,0) -- (0.25,0);
		
		\draw (-1,0.25) node {$\CW$};
	\end{tikzpicture}
	\raisebox{1cm}{\quad $\rightsquigarrow$ \quad}
	\begin{tikzpicture}
		\draw (-0.5,1) node {$\CL$};
		\draw (0,1) -- (0,-1);
		\draw (0,0) node {$\bullet$};
		\draw (0.75,0) node {$S_{\CW, \CL}$};
	\end{tikzpicture}
	\caption{The two actions of a 1-form symmetry generator $\CW$ on the line operator $\CL$. Left: Action of the 1-form symmetry by fusion with the generator $\CW$. This sends the line operator $\CL$ to the line operator $\CW \times \CL$. Right: Action of the 1-form symmetry generator by wrapping the generator $\CW$. This results in a specific local operator $S_{\CW,\CL}$ on the line operator $\CL$.}
	\label{fig:1formactions}
\end{figure}

Such a 1-form global symmetry has two natural actions on the category of line operators; see Figure \ref{fig:1formactions} for an illustration. The first action comes directly from fusion: given another line operator $\CL$, this action produces the line operator $\CW \times \CL$. (Left side of Figure \ref{fig:1formactions}.) Using the above fusion rules, we find that this action on the simple Wilson lines is given by
\be
	\CW \times \CW_{A_{n}} = \CW_{A_{n + 1}}\,, \qquad \CW \times \CW_{V_{n, e}} = \CW_{V_{n + 1, e}}\,.
\ee
Note that this not quite the tensor product of the corresponding $\fgl(1|1)$ modules due to the $\mod k$ identification of $n$. More generally, we find that fusion with $\CW$ is an auto-equivalence of the fiber categories $\CC_{k,[e]}$ with inverse given by fusing with $\CW^{k-1}$.

The second action of the 1-form global symmetry comes from wrapping the generating line operator $\CW$ around a given line. (Right side of Figure \ref{fig:1formactions}.) Given a second line operator $\CL$, this action produces a local operator $S_{\CW,\CL}$ on $\CL$, i.e. an element of the endomorphism algebra $\End(\CL)$ of $\CL$. This local operator is particularly special because it is central in $\End(\CL)$ due to the fact that we can slide the wrapped $\CW$ along $\CL$. More generally, the ``open Hopf link'' map $S_{\CW, -}$ commutes with every morphism in the following sense: if $\CL_1, \CL_2$ are two line operators and $\CO \in \Hom(\CL_1, \CL_2)$ a local operator joining them, then
\be
	S_{\CW, \CL_2} \CO = \CO S_{\CW,\CL_1}
\ee
is an identity in $\Hom(\CL_1, \CL_2)$. See Figure \ref{fig:central} for an illustration. In addition, the maps $S_{\CW,\CL}$ furnish a representation of the group algebra $\C[\Z_k]$ in $\End(\CL)$.%
\footnote{In general, the maps $S_{-,\CL_2}: \CL_1 \to S_{\CL_1,\CL_2}$ yield a representation of the full fusion ring on the endomorphism algebra $\End(\CL_2)$.}

\begin{figure}[H]
	\centering
	\begin{tikzpicture}
		\draw (-0.5,1) node {$\CL_2$};
		\draw (0,1) -- (0,-1);
		\draw (0,0) node {$\bullet$};
		\draw (0.75,0) node {$\CO S_{\CW, \CL_1}$};
		\draw (-0.5,-1) node {$\CL_1$};
	\end{tikzpicture}
	\raisebox{1cm}{\quad \reflectbox{$\rightsquigarrow$} \quad}
	\begin{tikzpicture}
		\draw (0.75,-0.5) arc (0:180:0.75 and 0.225);
		
		\draw (-0.5,1) node {$\CL_2$};
		\draw[white,fill = white] (-0.05,1) -- (0.05, 1) -- (0.05,-1) -- (-0.05,-1) -- cycle;
		\draw (0,1) -- (0,-1);
		
		\draw (-0.75,-0.5) arc (180:360:0.75 and 0.225);
		
		\draw (-1,-0.25) node {$\CW$};
		
		\draw (0,0) node {$\bullet$};
		\draw (0.25,0) node {$\CO$};
		
		\draw (-0.5,-1) node {$\CL_1$};
	\end{tikzpicture}
	\raisebox{1cm}{\quad $\leftrightsquigarrow$ \quad}
	\begin{tikzpicture}
		\draw (0.75,0.5) arc (0:180:0.75 and 0.225);
		
		\draw (-0.5,1) node {$\CL_2$};
		\draw[white,fill = white] (-0.05,1) -- (0.05, 1) -- (0.05,-1) -- (-0.05,-1) -- cycle;
		\draw (0,1) -- (0,-1);
		
		\draw (-0.75,0.5) arc (180:360:0.75 and 0.225);
		
		\draw (-1,0.25) node {$\CW$};
		
		\draw (0,0) node {$\bullet$};
		\draw (0.25,0) node {$\CO$};
		
		\draw (-0.5,-1) node {$\CL_1$};
	\end{tikzpicture}
	\raisebox{1cm}{\quad $\rightsquigarrow$ \quad}
	\begin{tikzpicture}
		\draw (-0.5,1) node {$\CL_2$};
		\draw (0,1) -- (0,-1);
		\draw (0,0) node {$\bullet$};
		\draw (0.75,0) node {$S_{\CW, \CL_2} \CO$};
		
		\draw (-0.5,-1) node {$\CL_1$};
	\end{tikzpicture}
	\caption{An illustration showing that $S_{\CW, -}$ is central.}
	\label{fig:central}
\end{figure}

We can determine this action by using the monodromy map one again; monodromy with $\widehat{A}_{1,0}$ simply measures the eigenvalue of $E_0$ mod $\Z$:
\be
	\CM(\widehat{A}_{1,0}, \widehat{A}_{n, 0}) = \id_{\widehat{A}_{1,0}} \times \id_{\widehat{A}_{n,0}} \,, \qquad \CM(\widehat{A}_{1,0}, \widehat{V}_{n, e}) = \exp(2 \pi i e) ~ \id_{\widehat{A}_{1,0}} \times \id_{\widehat{V}_{n,e}}\,.
\ee
This implies that the open Hopf link maps are simply scalar multiples of the identity operator
\be
	S_{\CW, \CW_{A_{n}}} = \id_{\CW_{A_{n}}} \qquad S_{\CW, \CW_{V_{n, e}}} = \exp(2 \pi i e) \id_{\CW_{V_{n, e}}}\,.
\ee
In particular, $\CW$ has a trivial action on itself: the $\Z_k^{(1)}$ 1-form symmetry is non-anomalous.%
\footnote{The 1-form symmetry of $\CT_{(\kappa, \nu, \xi)}$ is generically anomalous. We find that the generator $\CF(\widehat{A}_{\frac{1}{\nu},0})$ has trivial monodromy with itself but monodromy $e^{\frac{2\pi i}{d}}$ with $\CF(\widehat{A}_{\frac{1}{2}-(\frac{\xi}{\nu}+\frac{\nu}{2})\frac{1}{d},\frac{\nu}{d}})$. The monodromy of $\CF(\widehat{A}_{\frac{1}{2}-(\frac{\xi}{\nu}+\frac{\nu}{2})\frac{1}{d},\frac{\nu}{d}})$ and itself is $e^{-\frac{4\pi i \xi}{d^2}}$. It follows that the 1-form flavor symmetry is non-anomalous if and only if $d = 1$, i.e. when $\kappa$ and $\nu$ are coprime.} %
More generally, if $\CL \in \CC_{k, [e]}$ we find that $S_{\CW,\CL}$ is simply multiplication by the $k$th root of unity $\exp(2 \pi i e)$. Note that the centrality of $S_{\CW,-}$ immediately implies there can be no morphisms between modules with different values of this phase factor. In particular, there can be no morphisms between $\CC_{k, [e]}$ and $\CC_{k, [e']}$ unless $[e] = [e']$.

\subsection{The category $\CC_{k, [e]}$}
\label{sec:fibercat}

In this subsection we provide a concise description for the subcategory of line operators $\CC_{k, [e]}$, at least at the level of abelian categories, in terms of the Lie superalgebra $\fgl(1|1)$. 

We start with the identification of $\CC_{k,[e]}$ with the de-equivariantization of $\bigoplus_{n\in\Z} KL^0_{n+e}$ by fusion with the simple currents $\Pi^{k+1}\widehat{A}_{k/2,1}$, $\widehat{A}_{k,0}$, and their duals. We can simplify this de-equivariance by performing it in two steps. First, we de-equivariantize with respect to fusion with the currents $\Pi^{k+1}\widehat{A}_{k/2,1}$ and $\Pi^{k}\widehat{A}_{-k/2,-1}$; we can use this de-equivariantization to restrict our attention to $KL^0_{e}$ with $0 \leq e < 1$. The second step will then be de-equivariantization with respect to fusion with the currents $\widehat{A}_{\pm k,0}$.

Note that when $e\notin \mathbb{Z}$ or $e=0$, there is an equivalence between $KL_{e}$ and the category of modules for $\fgl(1|1)$ on which $E = \{\psi_+, \psi_-\}$ acts with generalized eigenvalue $e$, cf. \cite[Proposition 3.2]{BN22}. Restricting to local modules $KL^0_{e}$, we are forced to require $N_0$ and $E_0$ act semisimply, whence $E_0$ acts as scalar multiplication by $e$. Moreover, the generator $N_0 + \frac{k}{2}E_0$ is required to have integer eigenvalues.%
\footnote{For more details, see Appendix \ref{sec:reptheory}.} %
As described in \cite[Section 5.1]{BN22}, we can alternatively think of such a representation of $\mathfrak{gl}(1|1)$ as a $\Z$-graded module for the $\Z$-graded superalgebra generated by the two odd generators $\psi_\pm$ (degrees $\pm 1$) with anti-commutator $\{\psi_+, \psi_-\} = e$. We denote this $\Z$-graded superalgebra $\mathrm{Cl}_e$; the algebra $\mathrm{Cl}_0$ is an exterior algebra on two generators for whereas $\mathrm{Cl}_e$ for $e \neq 0$ is a Clifford algebra.

If we were to stop here, morphisms would be required to have degree $0$ to mirror the fact that morphisms of $\mathfrak{gl}(1|1)$ modules respect the grading. Instead, we must contend with the de-equivariantization arising from fusion with $\widehat{A}_{\pm k,0}$. Since fusion with these modules simply shifts the grading by $\pm k$, leaving $e$ fixed, we see that this de-equivariantization amounts to reducing the $\Z$ grading on $\mathrm{Cl}_e$ and its modules to a grading by $\Z_k$. If we think of the grading on modules and morphisms as arising from an action of $\C^\times$ and restricting to degree $0$ morphisms as taking $\C^\times$ equivariance, we conclude that the effect of the de-equivariantization is to instead take $\Z_k$-equivariant morphisms, i.e. restricting morphisms to have degree $0 \mod k$. Putting this together, we find that the subcategory of line operators $\CC_{k,[e]}$ is identified as an abelian category with the $\Z_k$-equivariant category of $\mathrm{Cl}_e$ modules:
\be
	\CC_{k, [e]} \simeq \mathrm{Cl}_e\textrm{-mod}^{\Z_k}
\ee

\subsubsection{Bulk local operators}
\label{sec:locops}
From the above identification of the category of line operators, we can determine the algebra of bulk local operators as the derived endomorphisms of the trivial line operator $\id$, which belongs to the subcategory $\CC_{k, [0]} \simeq \mathrm{Cl}_0\textrm{-mod}^{\Z_k}$. Viewing our Chern-Simons theory $\CT_k$ as the $B$-twist of an $\CN=4$ Gaiotto-Witten theory, these local operators should be interpreted as the algebra of functions on the Higgs branch $\CM_H$ of this $\CN=4$ theory.

From the perspective of the boundary VOA $\CV_k$, the trivial line operator $\id$ is identified with the vacuum module. Passing to the category $\mathrm{Cl}_0\textrm{-mod}^{\Z_k}$, we identify the trivial line $\id$ with the trivial 1-dimensional module $A_0$ where both $\psi_+, \psi_-$ act as zero. Prior to taking $\Z_k$-equivariance, it is a classical result that the derived endomorphisms of $A_0$ is a polynomial algebra in two even variables:
\be
	\textrm{Ext}_{\mathrm{Cl}_0\textrm{-mod}}(A_0, A_0) = \C[x_\pm]
\ee
where $x_\pm$ has degree $\pm 1$. This is the standard Koszul duality between $\bigwedge^\bullet \C^2$ and $\textrm{Sym}^\bullet (\C^2)^*$. Once we take into account the $\Z_k$-equivariance, we must restrict ourselves to morphisms that have degree $0 \mod k$. We conclude that the algebra of local operators is merely the $\Z_k$-invariants of $\C[x_\pm]$, exactly matching the algebra of functions on the Kleinian singularity $\C^2/\Z_k$
\be
	\textrm{Ext}_{\CC_{k, [0]}}(\id, \id) = \textrm{Ext}_{\mathrm{Cl}_{0}\textrm{-mod}^{\Z_k}}(A_0, A_0) = \C[\CM_H] \qquad \CM_H = \C^2/\Z_k
\ee
in agreement with the index computations and geometric quantization analysis in Section \ref{sec:locopsGQ}.

It is worth noting that our result slightly differs from that of Kapustin and Saulina in \cite[Section 3.2]{KapustinSaulina-CSRW}. The main difference in the two analyses comes from how one imposes gauge invariance in a derived setting. The authors of loc. cit. regard gauge invariant local operators as classes in $\C \oplus \C = \textrm{Lie}(U(1) \times U(1))$ Lie algebra cohomology with values in the algebra of monopoles $M_{m_+, m_-}$ and the bosonic fields $Z^\pm$ (denoted $A,B$ in \cite{KapustinSaulina-CSRW}); this corresponds to taking derived invariants for \emph{infinitesimal} gauge transformations, i.e. derived Lie algebra invariants. The issue is that they should instead have taken invariants with respect to \emph{finite} gauge transformations. Since the complexified gauge group $\C^\times \times \C^\times$ is reductive, taking group invariants is actually an exact functor and, in particular, should not be taken using ghosts, cf. \cite[Section 6.2.1]{CostelloDimofteGaiotto-boundary}. Taking this modification into account, the analysis of Kapustin-Saulina exactly reproduces our algebra of local operators.

%\subsubsection{The category $\CC_{(k,1,0),[e]}$}
% Before moving to quantum groups, we want to mention that the above analysis and results are nearly identical for the theory $\CT_{(k,1,0)}$: mutual-locality with the simple currents $\widehat{A}_{k,0}$ and $\widehat{A}_{0,1}$ requires the generators $k E_0$ and $N_0$ act with integer eigenvalues; there is a non-anomalous $\Z_k^{(1)}$ 1-form realized by the (modules for $\CV_{(k,1,0)}$ induced by) the simple currents $\widehat{A}_{\ell,0}$; and for each 1-form charge $[e] \in \frac{1}{k}\Z/\Z$ we can described the subcategory $\CC_{(k,1,0),[e]}$, at least as an abelian category, in terms of $\Z_k$-equivariant modules for the exterior/Clifford algebra $\textrm{Cl}_e$
% \be
% 	\CC_{(k,1,0),[e]} \simeq \textrm{Cl}_e\textrm{-mod}^{\Z_k} \simeq \CC_{k,[e]}
% \ee
% We see that the categories $\CC_k$ and $\CC_{(k,1,0)}$ of line operators in the theories $\CT_k$ and $\CT_{(k,1,0)}$ are equivalent as abelian categories! It is important to note that this equivalence is only true at the level of abelian categories and, in particular, does not preserve the braided tensor structure induced from the Kazhdan-Lusztig category $KL$.

\subsection{The quasi-quantum group $\CA_k$}
\label{sec:qgroup}

In this section we relate the category $\CC_k$ to the category of modules for a quasi-quantum group $\CA_k$ in two ways. The first way, which is conceptually quite simple, is to follow the argument in Section 5.2 of \cite{BN22} to use the above description of $\CC_{k,[e]}$ in terms of $\Z_k$-equivariant modules for the algebra $\textrm{Cl}_e$. The drawback to this approach is that it only accesses aspects of line operators at the level of an abelian category, i.e. it only utilizes the fact that $\CA_k$ is an algebra, and so we will be fairly brief. Our second route to the quasi-quantum group $\CA_k$ is via uprolling the unrolled quantum group $\ol{U}^{E}(\fgl(1|1))$, cf. \cite{creutzig2022uprolling}, and is contained in Section \ref{sec:uprolling}. This approach is technically more difficult than the above but is, by design, a direct analog of the realization of $\CV_k$ as a simple current extension of $V(\fgl(1|1))$ and gives us access to the entirety of $\CA_k$. This second approach also more readily generalizes to the quasi-quantum group $\CA_{(\kappa, \nu, \xi)}$ for the theory $\CT_{(\kappa, \nu, \xi)}$.

\subsubsection{$\CC_{k,[e]}$ and $\CA_k$}
Recall that $\CC_{k,[e]}$ can be described, at least as an abelian category, as the $\Z_k$-equivariant category of modules for the exterior/Clifford algebra $\textrm{Cl}_e$ generated by the fermions $\psi_\pm$. Choose a module $M$ in $\textrm{Cl}_e\textrm{-mod}^{\Z_k}$; this is simply a $\Z_k$-graded vector space $M = \bigoplus_{[n] \in \Z_k} M_{[n]}$ with an action of the algebra $\textrm{Cl}_e$ compatible with the grading. We define an action of (invertible) bosonic generators $K_1^{\pm 1}, K_2^{\pm 1}$ and fermionic generators $\Psi_\pm$ via the formulae
\be
\begin{aligned}
	K_1^{\pm1} \big|_{M_{[n]}} & = q^{\pm 2 n} \id_{M_{[n]}} \qquad & K_2 & = q^{2k e} \id_{M}\\
	\Psi_+ & = \psi_+ \qquad & \Psi_- & = \tfrac{1}{e}(q^{2ke} - 1) \psi_-
\end{aligned}
\ee
where $q = \exp(i\pi/k)$ is a $2k$th root of unity. In these expressions, the $\frac{1}{e}(q^{2ke}-1)$ should be interpreted as its limit $2\pi i$ for $e = 0$. The relations satisfied by these generators are that $K_2$ is central and
\be\label{eq:A_kcommutation}
	K_1^k = 1 \qquad K_2^k = 1 \qquad K_1 \Psi_\pm = q^{\pm 2} \Psi_\pm K_1 \qquad \{\Psi_+, \Psi_-\} = K_2-1
\ee
We call this algebra $\CA_k$.

With the above action of $\CA_k$, any $\Z_k$-equivariant morphism of $\textrm{Cl}_{e}$ modules naturally defines a morphism of modules for $\CA_k$ and hence we get a functor from $\textrm{Cl}_{e}\textrm{-mod}^{\Z_k}$ to the category of modules for $\CA_k$ where $K_2$ acts as $q^{2ke}$. As with \cite{BN22}, this is an equivalence of categories with inverse being the obvious one that identifies the generators as $\psi_+ = \Psi_+$ and $\psi_- = \frac{e}{1-q^{2ke}} \Psi_-$ and the $\Z_k$-gradings with the eigenvalues of $K_1$. Together with the previous section, this implies that the category $\CC_k$ of line operators in our Chern-Simons theory $\CT_k$ is equivalent to the category of $\CA_k$ modules 
\be
	\CC_k \simeq \bigoplus_{e} \textrm{Cl}_e\textrm{-mod}^{\Z_k} \simeq \CA_k\textrm{-mod}
\ee
as abelian categories.

\subsubsection{$\CA_k$ from uprolling}
\label{sec:uprolling}

We now present a second route to $\CA_k$ via uprolling. This is a similar extension procedure as we saw with the boundary VOA $\CV_k$, but now for quantum groups; see \cite{creutzig2022uprolling} for more general properties of uprolling. In particular, we will find that this procedure selects a subquotient of the unrolled quantum group $\ol{U}^{E}(\fgl(1|1))$. The idea was first described in \cite[Section 4]{creutzig2020quasi} and was applied in \cite[Section 8.3]{BCDN} for a similar situation.

To set the stage, let us recall yet another well-known conjectural equivalence between a quantum group and a VOA. Let $\mathfrak{M}(2)$ denote the singlet VOA, which can be quickly defined as the $\C^\times$ orbifold of a pair of symplectic fermions $\mathfrak{M}(2) = V_{\chi_+\chi_-}^{\mathbb{C}^\times}$, where $\mathbb{C}^\times$ acts on $\chi_\pm$ with weight $\pm 1$. The corresponding quantum group is the unrolled restricted quantum group $U=\overline{U}_i^H(\mathfrak{sl}(2))$ and is generated by five bosonic generators $E, F, H$ and $K^\pm$ satisfying
\be
	K K^{-1} = K^{-1} K = 1, \qquad [H,E]=2E,\qquad [H,F]=-2F,\qquad [E,F]=\frac{K-K^{-1}}{2i}
\ee
together with the relations $E^2=F^2=0$ and $i^H=K$. Here $i=\exp(\pi i /2)$ is a fourth root of unity. This is a quasi-triangular Hopf algebra with universal R matrix
\be
	R=i^{ H\otimes H/2}(1+ 2i E\otimes F).
\ee
A convenient way to understand this quasi-triangular Hopf algebra is via the relative double construction. If we consider the Hopf subalgebra generated by $\C[H, K, E]$, then $U$ is its relative double over $\C[H, K]$. One can also treat $\C[E]$ as a Hopf algebra object in modules of $\C[H, K]$, and reconstruct $U$ from the category of Yetter-Drinfeld modules:
\be
U\text{-Mod}\simeq {}^{\C[E]}_{\C[E]}\CY\CD (\C[H, K]\text{-Mod}). 
\ee
See \cite{creutzig2023algebraic} for some details. It is conjectured \cite{creutzig2014false, costantino2015some, creutzig2018logarithmic}, and proven for the weighted subcategory in \cite{creutzig2023algebraic}, that there is equivalence of braided tensor categories:
\be
	\mathfrak{M}(2)\textrm{-mod}\simeq U\textrm{-mod}. 
\ee

The symplectic fermion VOA $V_{\chi_+,\chi_-}$ is a simple current extension of the singlet VOA $\mathfrak{M}(2)$: we extend by the subspaces of $V_{\chi_+, \chi_-}$ with non-zero $\C^\times$ weight, which are naturally modules for the invariant subalgebra. The corresponding extension in $U\textrm{-mod}$ is as follows. For each $\mathfrak{l}\in \mathbb{Z}$, there is a one-dimensional representation $S_\mathfrak{l}$ of $U$ where $E$ and $F$ act trivially and $H$ acts with weight $2\mathfrak{l}$. The direct sum of modules
\be
	X=\bigoplus\limits_{\mathfrak{l}\in \mathbb Z} S_\mathfrak{l}
\ee
is the quantum group analogue of $V_{\chi_+\chi_-}$, viewed as a $\mathfrak{M}(2)$ module, and is a commutative superalgebra object in $U\textrm{-mod}$. Just as the case of simple current extensions of VOAs, local modules of the simple current extension $X$ can be realized as lifts of $U$ modules that satisfy a certain locality condition. Namely, a module $M$ of $U$ can be lifted a module for $X$ if and only if the monodromy
\be
\begin{tikzcd}
	S_\mathfrak{l}\otimes M\rar{\tau \circ R} & M\otimes S_\mathfrak{l} \rar{\tau \circ R}& S_\mathfrak{l}\otimes M
\end{tikzcd}
\ee
is trivial for all $\mathfrak{l}$. Here $R$ is as above and $\tau$ is the natural flipping map $\tau(a\otimes b)= b\otimes a$. From the definition of $S_\mathfrak{l}$ and the above expression for $R$, this monodromy can be computed easily to be $1\otimes i^{2\mathfrak{l}H}$. Locality means that $i^{2\mathfrak{l}H}=1$ for all $\mathfrak{l}$ and, in particular, $i^{2H}=1$. This is equivalent to requiring that $H$ acts semisimply with eigenvalues in $2\mathbb Z$.

The lifting functor $M \to X \otimes M$ identifies $M$ with $S_\mathfrak{l}\otimes M$. In other words, the action of $H$ is no longer well-defined if we restrict our attention to $M$. The actions of $E$ and $F$ survive the extension; the operator $i^H$ isn't preserved by tensoring, but we view it as the operator measuring fermionic parity (so that $S_{\mathfrak{l}}$ is fermionic when $\mathfrak{l}$ is odd). If we define $\psi_+=i^H E$ and $\psi_-=F$, we find the following anticommutator:
\be
	\{\psi_+,\psi_-\}=i^{2H}-1=0. 
\ee 
Namely, the superalgebra generated by $\psi_\pm$ and the parity operator $i^H$ is simply an exterior algebra on two variables. This confirms that the quantum group associated to the symplectic fermion VOA $V_{\chi_+, \chi_-}$ is simply this exterior algebra. 

Simple current extension could potentially produce a quasi-Hopf algebra structure on this exterior algebra. This can be checked at the level of $\C[H]$ modules since $U\text{-Mod}$ is a category of Yetter-Drinfeld modules over $\C[H]$ and Yetter-Drinfeld modules behave well under de-equivariantization \cite[Theorem 10.2]{creutzig2023algebraic}. However, since we extend by a self-dual lattice of $\C[H]$ modules, no associator appears. 

Let us go a bit further and see how to obtain the quantum group associated to $V(\fgl(1|1))$. We first recall the free field realization of $V(\fgl(1|1))$ using the symplectic fermion VOA $V_{\chi_+,\chi_-}$ together with a Heisenberg VOA generated by bosons $A$ and $B$ with pairing $(A,B)=1$ by the assignment
\be
	N = \partial A+\tfrac{1}{2}\partial B, \qquad E = \partial B, \qquad \psi_+ = e^{B}\chi_+,\qquad \psi_- = e^{-B}\chi_-.
\ee
This is equivalent to the free field realization we used in Section \ref{sec:freefield}.%
\footnote{Explicitly, we introduce a third bosonic field $C$ with pairing $(C,C) = 1$ and identify the symplectic fermions as $\chi_+ = e^{C}, \chi_- = \pd e^{-C}$. We can then identify the bosons $X,Y,Z$ as $$X = A - \tfrac{1}{2}B - C \qquad Y = B \qquad Z = C+B.$$} %
Importantly, we can view $V(\fgl(1|1))$ as an extension of the singlet algebra $\mathfrak{M}(2)$ times the Heisenberg VOA generated by $A,B$ \cite{CLR}. 

To find the quantum group analogue of $V(\fgl(1|1))$, we need the quantum group analogue of the Heisenberg VOA. This quantum group is the algebra $U'$ generated by two commuting variables that we denote, by an abuse of notation, $A$ and $B$ with universal $R$ matrix
\be
	R'=\exp\bigg(i \pi \big(A\otimes B+B\otimes A\big)\bigg)
\ee
encoding the pairing $(A,B) = 1$ on the Heisenberg algebra. We will denote by $\mathbb{C}_{a,b}$ the module of $U'$ where the weight of $A$ and $B$ are $a$ and $b$, respectively. The affine VOA $V(\fgl(1|1))$ corresponds to the following superalgebra extension of $U\otimes U'$:
\be
	\CX=\bigoplus S_{\mathfrak{l}}\otimes \mathbb{C}_{\mathfrak{l},0}.
\ee
The quantum group corresponding to $V(\fgl(1|1))$ follows from by repeating the above analysis. First, for a module $M$ of $U\otimes U'$ to be mutually local with $\CX$ imposes
\be
	(i^He^{i \pi B})^2=e^{i \pi (H+2B)}=1.
\ee
Namely, $H/2+B$ must act semisimply with integer eigenvalues on $M$. Moreover, since we are free to tensor with $(S_\mathfrak{l} \otimes \mathbb{C}_{\mathfrak{l},0})$,  the actions of $H$ and $A$ are not well-defined on $M$ after the extension. However, the actions of $\wt{N}=H/2-A$ and $E=-B$ survive the extension. Define $\Psi_\pm=\psi_\pm$ as above, we find the following anticommutator:
\be
	\{\Psi_+,\Psi_-\}=i^{2H}-1=e^{-2 \pi i B}-1=e^{2\pi i E} - 1. 
\ee
We identify $i^He^{-i \pi E}$ as the parity operator, and so we arrive at the superalgebra generated by $\wt{N}, E, \Psi_\pm$. This is readily identified with the unrolled restricted quantum group $\overline{U}^{E}(\fgl(1|1))$. We note that the braiding of $U \otimes U'$ modules induces one for modules of $\overline{U}^{E}(\fgl(1|1))$. Checking the corresponding extension at the level of $\C[H, A, B]$ modules, one easily finds that no associator appears. 

In principle, one can compute an $R$ matrix for $\overline{U}^{E}(\fgl(1|1))$ that matches the braiding coming $U \otimes U'$, but it is very complicated and will not be particularly useful for our discussion here. Instead, we simply compute monodromy between modules of $\overline{U}^{E}(\fgl(1|1))$ by choosing pre-images from modules of $U \otimes U'$.

We now extend once more to obtain the quantum group analog of $\CV_k$. To do so, we would like to identify the module of the quantum group $\overline{U}^{E}(\fgl(1|1))$ corresponding to $\widehat{A}_{n, e}$, for each $n\in \C$ and $e \in \C$; the free-field realization in Section \ref{sec:freefield} will be our guide. In \cite{BN22}, the module $\widehat{A}_{n-e/2+\epsilon(e), e}$ is shown to be the unique submodule of the free-field module generated by $\vert eX+n Y\rangle$. The vector $\vert eX + nY +eZ\rangle$ belongs to the kernel of the screening operator and is a generator of $\widehat{A}_{n-e/2+\epsilon(e), e}$; it is identified with the spectral flow of the vacuum vector. The corresponding module of $U\otimes U'$ is simply $\C_{n+e/2, e}$. Upon lifting, this module corresponds to the module of $\overline{U}^{E}(\fgl(1|1))$ where $N$ acts with weight $-n-e/2$ and $E$ with weight $-e$, and $\Psi_\pm$ act trivially. We denote this module by $A_{n-e/2+\epsilon (e), e}$ for $e\ne 0$. In other words, $A_{n, e}$ is the one-dimensional module of $\overline{U}^{E}(\fgl(1|1))$  where the action of $E$ is $-e$ and the action of $N$ is $-n-e+\epsilon (e)$. The action of $\Psi_\pm$ is trivial. Note that trivially acting $\Psi_\pm$ is consistent with the above anticommutator only when $e\in \mathbb{Z}$. One can show that under this definition, the fusion rule of $A_{n, e}$ is given by:
\be
	A_{n, e}\otimes A_{n', e'}\cong A_{n+n'-\epsilon(e, e'), e+e'}
\ee
just as for the simple currents of the VOA module.

The lattice of simple currents corresponding to the extension $\CV_k$ is generated by $A_{k, 0}$ and $A_{\frac{k}{2}, 1}$. Therefore, we consider an extension of the following form:
\be
	\CM_k:=\bigoplus\limits_{\mathfrak{m}, \mathfrak{n}} A_{\mathfrak{m}k+\mathfrak{n}\big(\frac{k-1}{2}\big)+\epsilon (\mathfrak{n}), \mathfrak{n}}
\ee
Note that although we did not write out the parity shifts here, the parity of these modules should match the one used for $\CV_k$, namely $\Pi^{k+1} A_{\frac{k}{2},1}$ and $A_{k,0}$. From the above discussions, $\CM_k$ can be written as a module of $U\otimes U'$ as
\be
	\CM_k=\bigoplus\limits_{\mathfrak{l}, \mathfrak{m}, \mathfrak{n}}S_{\mathfrak{l}-\mathfrak{n}}\otimes \mathbb{C}_{\mathfrak{l}+\mathfrak{m}k+\mathfrak{n}\big(\frac{k-1}{2}\big),\mathfrak{n}},
\ee
cf. the free-field realization in Section \ref{sec:freefield}.%
\footnote{Explicitly, we identify the weights of $X$,$Y$, and $Z$ with those of $A-\frac{B}{2}-\frac{H}{2}$, $B$, and $B+\frac{H}{2}$, respectively. The module $S_{\mathfrak{l}-\mathfrak{n}}\otimes \mathbb{C}_{\mathfrak{l}+\mathfrak{m}k+\mathfrak{n}\big(\frac{k-1}{2}\big),\mathfrak{n}}$ can then be seen to have weight $(\mathfrak{m}k + \mathfrak{n}\frac{k}{2}, \mathfrak{n}, \mathfrak{l})$, in agreement with weight the vector $\ket{\mathfrak{n}(X+\frac{k}{2}Y) + \mathfrak{m} (k Y) + \mathfrak{l}Z}$.} %
Lifting modules from $\overline{U}^{E}(\fgl(1|1))$ to $\CM_k$ imposes the following trivial-monodromy conditions:
\be
	\exp\bigg(2\pi i \big(k E\big)\bigg)=1 \qquad \exp\bigg(2\pi i \big((\wt{N}-\tfrac{E}{2}) + \tfrac{k}{2}E\big)\bigg)=1
\ee
This can be computed from the $R$ matrix of $U\otimes U'$ as above. Moreover, the freedom to tensor with $A_{k,0}$ and $A_{\frac{k}{2},1}$ means that neither the action of $\wt{N}$ nor of $E$ is well-defined. However, the exponentials
\be
	K_1 = \exp\bigg(\frac{2\pi i}{k} \big((\wt{N}-\tfrac{E}{2}) + \tfrac{k}{2}E\big)\bigg), \qquad K_2 = \exp\bigg(2\pi i E\bigg)
\ee 
survive the extension. The above locality constraint implies $K_1^{k}=K_2^{k}=1$. The algebra generated by $K_1, K_2$ and $\Psi_\pm$ is exactly the superalgebra $\CA_k$ we are after, whose algebra relations are given in Eq. \eqref{eq:A_kcommutation}, if we identify the quantum parameter as $q = e^{i\pi/k}$. This confirms that $\CA_k$ is the quantum group analog of $\CV_k$. 

A different feature from the $V(\fgl (1|1))$ case is that $\CA_k$ acquires an associator. This can be seen from the extension procedure at the level of $\C[H, A, B]$ modules. The lattice $\Lambda_k$ we extend by is spanned by vectors
\be
\Lambda_k=\mathrm{Span}_\Z\left\{(1,1,0), (0, k, 0), (0, \frac{k+1}{2}, 1)\right\},
\ee
and it is not difficult to check that its dual lattice $\Lambda_k'$ is spanned by
\be
\Lambda_k'=\mathrm{Span}_\Z\left\{(1,1,0), (0, 1, 0), (-\frac{1}{k}, -\frac{k+1}{2k}, \frac{1}{k})\right\}.
\ee
The quotient $\Lambda_k'/\Lambda_k=\Z_k\times \Z_k$ and corresponds to representations of $K_1, K_2$. The quadratic form induced on the quotient is given by\footnote{The expression $\exp (\pi i/k a^2)$ is only defined up to $\pm 1$ when $k$ is odd, which is related to the fact that the extension is a super-algebra extension.}
\be
Q(a, b)=\exp (\pi i/k (a^2-2ab)),\qquad (a, b)\in \Z_k\times \Z_k. 
\ee
Due to the term $\exp (\pi i/k a^2)$, there is no choice of braiding $\sigma$ that is multiplicative. Therefore, $\CA_k$ acquires an associator. This associator can be chosen to be non-trivial only on the first copy of $\Z_k$.

 % Again, we will not compute the induced $R$ matrix here. It can be computed from the extension procedure and the Yetter-Drinfeld presentation. 

\subsection{Comparison between $\CT_{(k,1,0)}$ and 3d $B$-model to $\C^2/\Z_k$}
\label{sec:Bmodelorbifold}

In this section we briefly turn our attention to a slightly different theory, namely $\CT_{(k,1,0)}$. Work of Mikhaylov identifies $\CT_{(k,1,0)}$ with a $\Z_k$ orbifold of the $B$ twist of a free hypermultiplet \cite{Mikhaylov}, i.e. Rozansky-Witten theory (alias 3d $B$-model) with target $\C^2/\Z_k$. It is relatively straightforward to derive the category of line operators from this 3d $B$-model perspective, and we can compare the resulting category to the category $\CC_{(k,1,0)}$ of modules for $\CV_{(k,1,0)}$.

Line operators in the $B$ twist of a free hypermultiplet can be described as follows, see e.g. \cite[Section 2.3]{CDGG} for more details. Consider the 3d $B$-model on a spacetime $\R \times \C^\times$, the complement of a line in $\R^3 \simeq \R \times \C$. If we compactify the theory around the circle in $\C^\times \simeq \R_{>0} \times S^1$, we can identify line operators of the 3d $B$-model with boundary conditions of a 2d $B$-model whose target is the loop space $\CL \C^2$ of the 3d target. The kinetic terms around $S^1$ are viewed as a superpotential for the 2d $B$-model, whence the category of line operators in 3d can be identified with a (derived) category of matrix factorizations, cf. \cite[Equation (2.38)]{CDGG}:
\be
	\textrm{MF}(\CL \C^2, W_0) \qquad W_0 = \oint \frac{\diff \theta}{2\pi} Z^- \pd_\theta Z^+\,.
\ee

The same analysis goes through once we consider the $\Z_k$ orbifold, from which we find the category $\wt{\CC}_k$ of line operators in this orbifold is given by
\be
	\wt{\CC}_k \simeq \textrm{MF}(\CL (\C^2/\Z_k), W_0) 
\ee
In addition to restricting to $\Z_k$-equivariant morphisms of matrix factorizations, there are additional line operators that source a non-trivial flat background for the $\C^\times$ flavor symmetry of the underlying hypermultiplet that couples trivially to $\Z_k$-invariants. Such a flat connection takes the form $A_e = a_e \diff \theta = i e \diff \theta$ for $e \in \frac{1}{k}\Z$; the holonomy of $A_e$ is then $g = \exp(2\pi i e)$. The category of line operators with this background is then
\be
	\wt{\CC}_{k, [e]} \simeq \textrm{MF}(\CL \C^2, W_e)^{\Z_k}
\ee
where $W_e$ is obtained by replacing $\pd_\theta$ in $W_0$ by the covariant derivative $\pd_\theta + a_e$. It is useful to expand the fields into Fourier modes $Z^\pm = \sum_n Z^\pm_n e^{i n \theta}$ so this category of matrix factorizations is re-expressed as (removing an overall factor of $i$ from the superpotential)
\be
	\wt{\CC}_{k, [e]} \simeq \textrm{MF}\bigg(\prod_{n \in \Z} \C^2_n, W_e = \sum_n \big(n + e\big)Z^-_{-n} Z^+_n\bigg)^{\Z_k}
\ee
cf. Equations (2.41)--(2.43) of loc. cit. From this presentation of $\wt{\CC}_{k, [e]}$, we see that it only depends on $e$ up to shifts by $\Z$ via redefining the modes, in accordance with our notation. 

Another route to $\wt{\CC}_{k,[e]}$ is by viewing these as twisted sectors as where the fundamental fields $Z^\pm$ have non-integral mode expansions $Z^\pm = \sum_n Z^\pm_n e^{i (n \pm e) \theta}$. This perspective is related to turning on the above flat background connection via a factional large flavor transformation $Z^\pm(\theta) \to e^{\pm i e \theta}Z^\pm(\theta)$; under such a transformation, the superpotential transforms as $W_0 \to W_e$. 

As with a free hypermultiplet, the categories $\wt{\CC}_{k, [e]}$ admit a simple finite-dimensional model by integrating out (most of) the massive modes, i.e. via (an equivariant version of) Kn\"{o}rrer periodicity. Without loss, assume that $0 \leq e < 1$; we then see that the modes $Z^+_n, Z^-_{-n}$ for $n \neq 0$ all have a quadratic potential of the form $(n+e)Z^-_{-n} Z^+_n$. Since $n \neq 0$ implies $n + e \neq 0$, these modes are massive after compactifying and hence many be safely integrated out. The resulting finite-dimensional model is
\be
	\wt{\CC}_{k, [e]} \simeq \textrm{MF}\big(\C^2, W = e Z^- Z^+\big)^{\Z_k}
\ee
where we have replaced $Z^\pm_0 \to Z^\pm$ to simplify the notation.

Finally, we note that the symmetric algebra $\C[Z^\pm]$ with curving $W = e Z^- Z^+$ is Koszul dual to the Clifford/exterior algebra $\textrm{Cl}_e$, but this is exactly the algebra describing the categories $\CC_{k,[e]}$ and $\CC_{(k,1,0),[e]}$! This quadratic-constant Koszul duality then implies that there is a (derived) equivalence
\be
	\CC_{k, [e]} \simeq \CC_{(k,1,0), [e]} \simeq \textrm{Cl}_e\textrm{-mod}^{\Z_k} \simeq \textrm{MF}\big(\C^2, W = eZ^- Z^+\big)^{\Z_k} \simeq \wt{\CC}_{k, [e]}.
\ee
Thus, the categories $\CC_k, \CC_{(k,1,0)}$ of line operators in $\CT_k, \CT_{(k,1,0)}$ are indeed equivalent to the category $\wt{\CC}_k$ of line operators in a 3d $B$-model with target $\C^2/\Z_k$, at least as abelian categories.

\subsubsection{Comparing boundary VOAs}
\label{subsubsec:orbifoldcomparison}
Rozansky-Witten theory with target $\C^2/\Z_k$ has a natural boundary VOA: the $\Z_k$-orbifold of the symplectic fermion VOA $V_{\chi_,\chi_-}^{\mathbb{Z}_k}$. Using the relation between $\CV_{(k,1,0)}$ and $V_{\chi_,\chi_-}^{\mathbb{Z}_k}$ derived in Section \ref{sec:VOAorbifold}, we can compare $\CC_{(k,1,0)}$ and category of modules of this orbifold. The category of modules for this orbifold is expected to be equivalent to the above category of matrix factorizations $\wt{\CC}_k$, where the braiding on the latter uses the $E_2$ structure on the category (Koszul dual to the one) studied in \cite{ben2010integral, gammage2022betti}. Correspondingly, we also denote the category of modules for the orbifold by $\wt{\CC}_k$.

Recall that we can relate the boundary VOA $\CV_{(k,1,0)}$ to $V_{\chi_+,\chi_-}^{\mathbb{Z}_k}$ as
\be
	\CV_{(k,1,0)} \cong (V_{\chi_+,\chi_-}\otimes W)^{\mathbb{Z}_k}
\ee
for the lattice VOA $W = V_{\phi_2} \otimes V_{\phi_3}$ associated to the self-dual lattice $\Z^{1,1}$. To see how $\CC_{(k,1,0)}$ can be related to $\wt{\CC}_k$, consider the following product of orbifolds:
\be
	V=V_{\chi_+,\chi_-}^{\mathbb{Z}_k} \otimes W^{\mathbb{Z}_k}.
\ee
It follows that $\CV_{(k,1,0)}$ is a simple current extension of $V$ and that objects in $\CC_{(k,1,0)}$ can be realized as lifts of objects from $V\textrm{-mod}$, via the usual induction functor $\CV_{(k,1,0)} \times -$, that have trivial monodromy with $\CV_{(k,1,0)}$. The fact that $V$ is a product of VOAs implies the category $V\textrm{-mod}$ can be expressed as
\be
	V\textrm{-mod}\simeq \wt{\CC}_k \boxtimes W^{\mathbb{Z}_k}\textrm{-mod}.
\ee   

The action of $\Z_k$ on $W = V_{\phi_2} \otimes V_{\phi_3}$ only acts on vertex operators built from $\phi_3$, and the operators surviving are those of weight $k$ with respect to $\phi_{3,0}$. It follows that
\be
	W^{\Z_k} \simeq V_{\phi_2} \otimes V_{k\phi_3},
\ee
where $V_{k\phi_3}$ is the extension by the lattice generated by $e^{k\phi_3}$. As abelian categories, there is an equivalence $W^{\mathbb{Z}_k}\textrm{-mod}\cong \text{Vect}_{\mathbb{Z}_{k^2}}$ between modules for $W^{\Z_k}$ and the category of vector spaces graded by $\frac{1}{k} \Z /k \Z \simeq \Z_{k^2}$. As a monoidal category, there is an interesting braided tensor structure coming from the bilinear pairing on $\mathbb{Z}_{k^2}$ induced from the original lattice. The extension to $\CV_{(k,1,0)}$ can be identified as a diagonal embedding
\be
	\text{Vect}_{\mathbb{Z}_k}\longrightarrow \wt{\CC}_k \boxtimes \text{Vect}_{\mathbb{Z}_{k^2}}, 
\ee
i.e. we consider an extension of $V$ by the diagonal $\Z_k$ simple currents. Here, the embedding $\text{Vect}_{\mathbb{Z}_k}\to \text{Vect}_{\mathbb{Z}_{k^2}}$ corresponds to the group homomorphism
\be
	\mathbb{Z}_k \simeq \Z/k\Z \to \mathbb{Z}_{k^2} \simeq \tfrac{1}{k}\Z/k\Z: 1 + k \Z \mapsto 1 + k \Z. 
\ee
The embedding into $\wt{\CC}_k$ corresponds to the $\Z_k$ simple currents extending $V_{\chi_+, \chi_-}^{\Z_k}$ to $V_{\chi_+, \chi_-}$. Putting this together, we can describe the category $\CC_{(k,1,0)}$ of line operators in $\CT_{(k,1,0)}$ in terms of the category $\wt{\CC}_k$ of modules for the orbifold $V_{\chi_+, \chi_-}^{\Z_k}$ as the de-equivariantization by this $\Z_k$:
\be
	\CC_{(k,1,0)} \simeq\left(\wt{\CC}_k\boxtimes \text{Vect}_{\mathbb{Z}_{k^2}}\right)/\mathbb{Z}_k.
\ee
This is in fact an equivalence of braided tensor categories. 

The category of modules for $V$ admits a decomposition
\be
	\wt{\CC}_k \boxtimes \text{Vect}_{\mathbb{Z}_{k^2}}=\bigoplus\limits_{[e],[e']}\wt{\CC}_{k, [e]}\boxtimes \text{Vect}_{\mathbb{Z}_{k^2},[e']}
\ee 
where the subscript $[e] \in \frac{1}{k}\Z/\Z$ denotes the subcategory of $\wt{\CC}_k$ where monodromy with $\CV_{(k,1,0)}$ is $e^{2\pi i e}$, and similarly for the subscript $[e']$. Computation in the free field algebra $W$ shows that
\be
	\text{Vect}_{\mathbb{Z}_{k^2},[e']}=\text{Vect}_{k e' + k\Z},
\ee
where $\text{Vect}_{k e' + k\Z}$ is the subcategory of $\text{Vect}_{\mathbb{Z}_{k^2}}$ whose objects are supported on the coset $k e' + k\Z \subset \frac{1}{k}\Z/k\Z$. For each $[e]$ and $[e']$, the subcategory
\be
	\wt{\CC}_{k, [e]}\boxtimes \text{Vect}_{\mathbb{Z}_{k^2},[e']}
\ee
has monodromy $\exp\big(2\pi i(e+e')\big)$ with $\CV_{(k,1,0)}$. To ensure trivial monodromy with $\CV_{(k,1,0)}$, we therefore need that $[e]+[e']=0$. This leads to the following:
\be
	\CC_{(k,1,0)} \simeq \left(\bigoplus\limits_{[e]}\wt{\CC}_{k, [e]}\boxtimes \text{Vect}_{\mathbb{Z}_{k^2},[-e]}\right)/{\mathbb{Z}_k}.
\ee
The category $\text{Vect}_{\mathbb{Z}_{k^2},[-e]}$ is equivalent, as an abelian category, to $\text{Vect}_{\mathbb Z_k}$, and therefore we find the following equivalence of abelian categories:
\be
	\CC_{(k,1,0)} \simeq \bigoplus\limits_{[e]}  \left(\wt{\CC}_{k, [e]}\boxtimes \text{Vect}_{\mathbb{Z}_{k}}\right)/{\mathbb{Z}_k}\simeq  \bigoplus\limits_{[e]} \wt{\CC}_{k, [e]} \simeq \wt{\CC}_k.
\ee
The second equivalence is because a vector space supported at any other point on $\mathbb{Z}_k$ can be transferred to one supported at the identity via the action of $\mathbb{Z}_k$. This equivalence preserves the decomposition into sectors with given 1-form charge $[e]$. In conclusion: the categories of line operators $\CC_k$, $\CC_{(k,1,0)}$, and $\wt{\CC}_k$ are all equivalent as abelian categories, even as tensor categories fibered over the Pontryagin dual group $\Hom(\Z^{(1)}_k, \C^\times)$ of 1-form charges, but will have different braided tensor structures. 

\subsubsection{The quantum group $\CA_{(k,1,0)}$}\label{subsubsec:Ak10}
We can obtain the quantum group analog of $\CV_{(k,1,0)}$ as in Section \ref{sec:qgroup}. The algebra object corresponding to this extension of $V(\fgl(1|1))$ is given by
\be
	\CM_{(k,1,0)}=\bigoplus\limits_{\mathfrak{m},\mathfrak{n}}  A_{\mathfrak{m}k -\frac{1}{2} \mathfrak{n}+\epsilon(\mathfrak{n}), \mathfrak{n}}
\ee
The condition of trivial monodromy imposes the constraints
\be
	\exp\bigg(2\pi i \big(k E\big)\bigg)=1 \qquad \exp\bigg(2\pi i \big(\wt{N}-\tfrac{E}{2}\big)\bigg)=1.
\ee
Moreover, the action of the exponential generators $\wt{K}_1=q^{2\wt{N}-E}$ and $\wt{K}_2=q^{2kE}$ are well defined. We obtain the superalgebra $\CA_{(k,1,0)}$ generated by $\wt{K}_1, \wt{K}_2$ and $\Psi_\pm$. It turns out that the commutation relations exactly coincide with the commutation relation of Eq. \eqref{eq:A_kcommutation}. Thus, as algebra, $\CA_{(k,1,0)}$ is isomorphic to $\CA_k$. The difference here, as indicated in the Introduction, is a difference of the quasi-Hopf algebra structure. 

To see this, we can simply study the extension at the level of $\C[H, A, B]$ modules. In this case, the lattice $\Lambda_{(k,1,0)}$ is spanned by vectors
\be
\Lambda_{(k,1,0)}=\mathrm{Span}_\Z \left\{ (1,1,0), (0,k,0), (0, \frac{1}{2}, 1) \right\},
\ee
and its dual lattice is given by
\be
\Lambda_{(k,1,0)}'=\mathrm{Span}_\Z \left\{ (1,1,0), (0,1,0), (-\frac{1}{k}, -\frac{1}{2k}, \frac{1}{k}) \right\}.
\ee
The quotient $\Lambda_{(k,1,0)}'/\Lambda_{(k,1,0)}=\Z_k\times \Z_k$ and the blinear form one obtains on the quotient is given by
\be
Q(a,b)=\exp (\frac{2\pi i}{k} ab), \qquad (a, b)\in \Z_k\times \Z_k.
\ee
We can therefore choose the braiding $\sigma ((a,b), (a', b'))=\exp (\frac{2\pi i}{k} ab')$, corresponding to the non-symmetric $R$-matrix $\exp (2\pi i (\wt{N}\otimes E))$, which illiminates the associator. In conclusion, $\CA_{(k, 1,0)}$ can be given a genuine quasi-triangular Hopf algebra structure. It is not difficult to see that it is the double of the Hopf subalgebra $\C[\wt{K}_1, \Psi_+]$.

From the perspective of the algebra objects, we can decompose $\CM_k$ as a direct sum:
\be
	\CM_k=\bigoplus\limits_{\mathfrak{n}} \CM_{k, \mathfrak{n}}
\ee 
where $\CM_{k,\mathfrak{n}}:=\bigoplus\limits_{\mathfrak{m}} A_{\mathfrak{m}k+\mathfrak{n}\big(\frac{k-1}{2}\big)+\epsilon (\mathfrak{n}), \mathfrak{n}}$. Similarly, there is a decomposition of $\CM_{(k,1,0)}$, conveniently expressed as
\be
	\CM_{(k,1,0)}=\bigoplus\limits_{\mathfrak{n}} \CM_{k,\mathfrak{n}}\otimes A_{-\mathfrak{n}\frac{k}{2} ,0},
\ee
which uses $A_{\mathfrak{m}k+\mathfrak{n}\big(\frac{k-1}{2}\big)+\epsilon (\mathfrak{n}), \mathfrak{n}}\otimes  A_{-\mathfrak{n}\frac{k}{2} ,0} = A_{\mathfrak{m}k - \frac{\mathfrak{n}}{2}+ \epsilon (\mathfrak{n}), \mathfrak{n}}$. This shows how to transform one lattice into another with the modules $A_{-\mathfrak{n}\frac{k}{2} ,0}$. Unfortunately, this relation between extensions cannot be applied to local modules thereof in a canonical way, and it remains mysterious to us to see how the local modules of these two superalgebra objects are related. 

As a final comment about quantum groups, we mention that the quantum group $\CA_k$ can be seen to be immediately Koszul dual to the 3d $B$-model. For this, we will treat $K_1$ as a generator of $\mathbb{Z}_k$ and view its role as imposing $\mathbb{Z}_k$ equivariance. Let us Koszul dualize the variable $\Psi_+$, and call the dualized variable $Z^+$. The anticommutation relation is dualized into a differential $Q$:
\be
	\{\Psi_+,\Psi_-\}=K_2-1 \rightsquigarrow Q\Psi_-=(K_2-1)Z^+.
\ee
If we view $Z^+$ as a linear function on $\mathbb{C}$, then this is imposing (in a derived way) the constraint $\zeta v=v$ for $(\zeta,v)\in \mathbb{Z}_k\times \mathbb{C}$. This can be expressed by the following Cartesian diagram:
\be
\begin{tikzcd}
	\CZ \rar\dar & \mathbb{C}\dar{\Delta}\\
	\mathbb{Z}_k\times \mathbb{C}\rar{\rho} & \mathbb{C}\times \mathbb{C}
\end{tikzcd}
\ee
where $\Delta$ is the diagonal embedding and $\rho$ is given by:
\be
	\rho(\zeta,v)=(v, \zeta v). 
\ee
The Koszul duality implies an equivalence of derived categories:
\be
	D^b\CC_k\simeq \text{Coh}(\widehat{\CZ/\mathbb{Z}_k}),
\ee
where the right hand side $\widehat{\CZ/\mathbb{Z}_k}$ is a formal completion of $\CZ/\mathbb{Z}_k$ along $\mathbb{Z}_k\times \{0\}$, $0$ being the origin of $\mathbb C$. The space $\CZ/\mathbb{Z}_k$ is nothing but $\CL(\mathbb{C}/\Z_k)$, the derived (topological) loop space of the stack $\mathbb{C}/\Z_k$, in other words the derived mapping space $\mathrm{Hom}(S^1_{\textrm{dR}}, \C/\Z_k)$ \cite{ben2012loop}. Indeed, the topological loop space is defined by the following derived intersection:
\be
	\CL(\mathbb{C}/\Z_k)\cong \C/\mathbb{Z}_k\times_{\C/\mathbb{Z}_k\times \C/\mathbb{Z}_k} \C/\mathbb{Z}_k.
\ee
Note that if there is a map $X/G\to Y/H$, then the derived intersection $(X/G)\times_{Y/H}(X/G)$ can be alternatively defined by the pull-back diagram:
\be
\begin{tikzcd}
	(X/G)\times_{Y/H}(X/G)\rar\dar & G\setminus X\dar\\
	G\setminus  H\times_G X \rar & G\setminus Y
\end{tikzcd}
\ee
Applying this to $X=\C, Y=\C\times \C, G=\Z_k$ and $H=\Z_k^2$, we find the space $\CZ/\Z_k$ as above. 

The derived loop space as a $B$ model has been investigated by \cite{ben2010integral}  and recently by \cite{gammage2022betti}. If we additionally Koszul dualize the fermion $\Psi_-$, we arrive at precisely the category of matrix factorizations considered above, if we replace $eZ^+Z^-$ by $(e^{2\pi i e}-1)Z^+Z^-$, which does not change the category. In conclusion, we have the following diagram of categories:
\be
\begin{tikzcd}
	\CV_k\textrm{-mod} \arrow[rr, leftrightarrow]\arrow[dr, leftrightarrow]&  & \CA_k\textrm{-mod}\arrow[dl, leftrightarrow]\\
	& \text{Coh}\left(\CL(\mathbb{C}/\Z_k)\right)\arrow[d, leftrightarrow]&\\
	& \text{MF}\big(\mathbb{C}^2\times \mathbb{Z}_k, W=(\zeta-1)Z^+ Z^-\big)^{\mathbb{Z}_k} & 
\end{tikzcd}
\ee
The categories in the first row are braided tensor categories and, as we have shown, might have more than one braided tensor structure. The category on the second row is known to have an $E_2$ structure, and is the Drinfeld center of the category $\mathrm{Coh}(\C/\Z_k)$. This $E_2$ structure is most directly comparable to the one from $\CA_{(k,1,0)}$, since this is the double of $\C\cdot \Z_k\ltimes \C[\Psi_+]$. However, due to the abelian equivalence $\CA_{k}\textrm{-mod}\simeq \CA_{(k,1,0)}\textrm{-mod}$, we came to the puzzle that the category on the second row has more than one braided monoidal structure. 

From a physical perspective, the identification between boundary VOAs
\be
	\CV_{(k,1,0)} \simeq (V_{\chi_+, \chi_-}^{\Z_k} \times V_{k \phi_3})/\Z_k \times V_{\phi_2}
\ee
corresponds to the following identification of theories:%
\footnote{There is a slightly less trivial relation between $\CT_k$ and $\CH^B/\Z_k^{(0)}$, e.g. by first relating $\CT_k$ and $\CT_{(k,1,0)}$ as in Section \ref{sec:introqgroup}, but we will not describe it here.} %
\be
	\CT_{(k,1,0)} \simeq (\CH^B/\Z_k^{(0)} \times U(1)_{-k^2})/\Z_k^{(1)} \times U(1)_1
\ee
where $\CH^B$ denotes the $B$-twist of a free hypermultiplet, with $\CH^B/\Z_k^{(0)}$ its $\Z_k$ orbifold. The factor of $U(1)_1$ is essentially trivial: it has only no non-trivial local operators or line operators. The theory $U(1)_{-k^2}$ is equivalent to $U(1)_{k^2}$ with inverted braiding: the Wilson lines $\CW_{\ell_1}, \CW_{\ell_2}$, with $\ell_1, \ell_2 = 0,..., k^2-1$, braid with a factor $e^{-\frac{2\pi i}{k^2}\ell_1 \ell_2}$. These Wilson lines generate the $\Z_{k^2}^{(1)}$ 1-form symmetry of $U(1)_{-k^2}$. Importantly, the subgroup with $\ell$ divisible by $k$ braids trivially with itself (the symmetry corresponding to that subgroup is non-anomalous) and we gauge the (non-anomalous) subgroup of the $\Z_k^{(1)} \times \Z_{k^2}^{(1)}$ 1-form symmetry of $\CH^B/\Z_k^{(0)} \times U(1)_{-k^2}$ generated by $(1,k)$; this $\Z_k^{(1)}$ symmetry is precisely the diagonally embedded $\text{Vect}_{\mathbb{Z}_k}$ above. The effect of this gauging this 1-form is to dress the line operators in $\CH^B/\Z_k^{(0)}$ by the Wilson line with complementary 1-form charge. For example, a line in $\wt{\CC}_{k,[e]}$ would get dressed by the Wilson line $\CW_{ke}$ and, correspondingly, will alter the braiding morphisms by a factor $\exp(-2\pi i e e')$.

It is an interesting question where these different $E_2$ structures come from mathematically, e.g. from the geometry of the symplectic orbifold $\C^2/\Z_k$, as the usual braiding of lines in Rozansky-Witten theory is determined by the Poisson bivector on the holomorphic-symplectic target \cite{RW}.

\section{Coupling to Flat Connections}
\label{sec:flatconns}

As described in detail in \cite[Section 2.2]{CDGG}, when a theory can be deformed by background flat, complexified connections for a flavor symmetry $G$, the category of line operators $\CC$ can be extended to include line operators compatible with flat connections possessing a given holonomy $g \in G$ around the line operator
\be
	\CC^{\textrm{ext}} = \bigoplus_{g \in G} \CC^{\textrm{ext}}_{g}\,.
\ee
The category $\CC^{\textrm{ext}}_1$ is identified with the category of line operators in the usual sense, i.e. it is the category analyzed in Section \ref{sec:lines}.

Relatedly, we saw in the analysis of Section \ref{sec:Wilson} that there are many modules in $KL$ that do not induce local modules of the boundary VOA $\CV_k$. Indeed, we saw that no module in $KL_e$ induced a local module for $\CV_k$ unless $e \in \frac{1}{k}\Z \subset \C$. Modules with $k e \notin \Z$ should not be ignored because of this; they simply induce non-local modules for $\CV_k$, i.e. they are monodromy defects (a.k.a. twisted modules) for $\CV_k$. 

In the same fashion, the quasi-quantum group $\CA_k$ is a central quotient of a larger algebra $\CA^{\textrm{ext}}_k$. The main difference between the two is that we do not impose the second relation in Eq. \eqref{eq:A_kcommutation}. Instead, $\CA^{\textrm{ext}}_k$ has a large central subalgebra generated by $K_2^{\pm1}$. Via an analog of Schur's Lemma, the category of finite-dimensional representations will decompose into subcategories where $K_2 = \zeta \id$ for $\zeta \in \C^\times$:
\be
	\CA^{\textrm{ext}}_k\textrm{-mod} = \bigoplus_{\zeta \in \C^\times} \CA^{\textrm{ext}}_k\textrm{-mod}^\zeta
\ee 
The monodromy of a module in $\CA^{\textrm{ext}}_k\textrm{-mod}^\zeta$ with $\CM_k$ is precisely $\zeta^k$. We note that the category of modules for $\CA^{\textrm{ext}}_k$ isn't naturally braided, but we expect that it has a natural $\C^\times$-crossed braiding.

These extra modules are avatars of the above phenomenon: our Chern-Simons theory $\CT_k$ can be coupled to a background complexified, flat connection for a $U(1)$ flavor symmetry, implying an extended category of line operators of the form
\be
	\CC^{\textrm{ext}}_{k} = \bigoplus_{g \in \C^\times} \CC^{\textrm{ext}}_{k,g}
\ee
There are natural deformations of the constructions in Section \ref{sec:lines} to include this extended category of line operators:
\be
\label{eq:fiberequiv}
	\CC^{\textrm{ext}}_{k, g} \simeq \bigoplus_{[e]\in \frac{1}{k}\Z/\Z} KL^{N + \frac{k}{2}E, E}_{[e-\lambda]}/\Lambda_k \simeq \bigoplus_{\zeta^k = g} \CA_k^{\textrm{ext}}\textrm{-mod}^{\zeta}
\ee
where $g = e^{2\pi i k\lambda}$. In this identification, the category $KL^{N + \frac{k}{2}E, E}_{[e-\lambda]}$ is the subcategory of $KL$ on which $N_0+\frac{k}{2}E_0$ acts semisimply with integer eigenvalues and $E_0$ acts semisimply with eigenvalues in $[e-\lambda] = (e - \lambda) + \Z$; $KL^{N + \frac{k}{2}E, E}_{[e - \lambda]}/\Lambda_k$ is the de-equivariantization with respect to fusion with the simple currents $\widehat{A}_{k,0}, \Pi^{k+1}\widehat{A}_{\frac{k}{2},1}$ and their duals.

The equivalence 
\be
	KL^{N + \frac{k}{2}E, E}_{[e-\lambda]}/\Lambda_k \simeq \CA_k^{\textrm{ext}}\textrm{-mod}^{\zeta}
\ee
where $\zeta = \exp(2\pi i (e-\lambda))$ can be established in the same way as in Section \ref{sec:qgroup}. In Sections \ref{sec:flatdeform} and \ref{sec:cext} we pursue the VOA portion of the equivalence in Eq. \eqref{eq:fiberequiv}. Section \ref{sec:flatdeform} describes a procedure for coupling a general VOA $\CV$ to a background connection for a $\C^\times$ flavor symmetry, and shows how this procedure works for the simple example of the symplectic fermion VOA $V_{\chi_+, \chi_-}$ as well as our boundary VOA $\CV_k$. We also describe a second perspective on this coupling as a certain large-level limit in analogy with the examples in Section 6.2 of \cite{CDGG}. The recent paper \cite{FLcenter} describes general aspects of these large level limits, focusing on the so-called Feigin-Tipunin algebra $\CF\CT_p(\fg)$ associated to a reductive Lie algebra $\fg$ and $p>0$ an integer; the authors of this paper denote the Feigin-Tipunin algebra $\CW_p(\fg)$ due to the fact that it realizes to the well-known triplet algebras $\CW_p$ for $\fg = \fsl(2)$. In Section \ref{sec:cext}, we turn to the category $\CC^{\textrm{ext}}_{k}$ and provide three complementary approaches that each highlight different aspects of $\CC^{\textrm{ext}}_{k}$.

Before moving on, we note that if we identify the flavor symmetry group $\C^\times$ as $\C/\frac{1}{k}\Z$, with coordinate $\lambda$ modulo shifts by $\frac{1}{k}\Z$, the category $\CC^{\textrm{ext}}_k$ naturally fibers over the $k$-fold cover $\C/\Z$:
\be
	\CC^{\textrm{ext}}_k = \bigoplus_{[\nu] \in \C/\Z} \CC^{\textrm{ext}}_{k, [\nu]}\,, \qquad \CC^{\textrm{ext}}_{k, [\nu]} = KL^{N+\frac{k}{2}E, E}_{[\nu]}/\Lambda_k
\ee
The category $\CC^{\textrm{ext}}_{k,g}$ combines $k$ of these subcategories: those $[\nu]$ that project to $g$ under $\C/\Z \twoheadrightarrow \C/\frac{1}{k}\Z$. Collision of line operators is compatible with this refined fibration:
\be
	\CC^{\textrm{ext}}_{k, [\nu_1]} \times \CC^{\textrm{ext}}_{k, [\nu_2]} \longrightarrow \CC^{\textrm{ext}}_{k, [\nu_1+\nu_2]}
\ee
From the quantum group perspective, this fibration over $\C/\Z$ corresponds to the fibration over the eigenvalue $\zeta$ of the central element $K_2$. Restricting to line operators sourcing connections with trivial holonomy, we see that $[\nu]$ precisely corresponds to the charge for the $\frac{1}{k}\Z/\Z$ 1-form global symmetry. If we denote the kernel of the projection $\C/\Z \to \C/\frac{1}{k}\Z$ by $\CZ$, we thus have a homomorphism $\Hom(\frac{1}{k}\Z/\Z, \C^\times) \to \CZ$; this is dual to the homomorphism $\Hom(\CZ, \C^\times) \to \frac{1}{k}\Z/\Z$ characterizing the mixed 0-form--1-form anomaly of $\CT_k$.%
\footnote{We can determine this mixed anomaly as by following, e.g., Section 5.2 \cite{BBFSanomalies}. To match notation therein, we identify $\CF = \C/\Z$, $F = \C/\frac{1}{k}\Z$, $\CZ = \frac{1}{k}\Z/\Z$; the group $\CF$ is identified with the group scaling the fermions $\psi_\pm$, with $F$ the quotient by the subgroup $\CZ$ that act trivially on local operators $\Hom(\id, \id)$. The mixed 0-form--1-form anomaly is characterized by the homomorphism $\gamma: \Z_k^{(1)} \to \widehat{\CZ} = \Hom(\CZ, \C^\times)$ that measures the weights of local operators at the end of the 1-form symmetry generators under $\CZ$. We start by considering local operators at the end of the 1-form symmetry generators; for the generating Wilson line $\CW$, we find $\Hom(\CW, \id)$ is generated over $\Hom(\id, \id) = \C[Z^\pm]^{\Z_k}$ by $Z^+$ and $(Z^-)^{k-1}$, which have weight $1$, whence $\gamma$ is the identity map.} %

\subsection{Coupling $\CV_k$ to flat $\C^\times$ connections}
\label{sec:flatdeform}

We will start our analysis of the extended category of line operators by describing what it means to deform a VOA $\CV$ by a flat flavor symmetry background. We will focus on flat abelian connections for simplicity; see e.g. \cite[Section 3.6]{GaiottoTwisted} or \cite[Section 2.4]{CDGG} for related examples.

Given a vertex operator algebra $\CV$ with an action of $\mathbb{C}^\times$, coupling to a background connection can be done by coupling the VOA to the lattice vertex algebra $V_\varphi$ of a degenerate boson $\varphi$, where the background connection is identified as $A = \pd \varphi$, and taking an orbifold:
\be
	\CV^{\mathrm{ext}}:=\left(\CV\otimes V_\varphi\right)^{\C^\times},
\ee
where the $\C^\times$ action on $V_\varphi$ gives the vertex operator $\norm{e^{n \varphi}}$ weight $-n$. This orbifold simply dresses operators in $\CV$ of $\C^\times$ weight $n$ by $\norm{e^{n \varphi}}$. Local modules of $\CV^{\mathrm{ext}}$ where $\varphi_0$ acts with eigenvalue $\lambda$ are understood as modules of the VOA sourcing the background connection $A = \frac{\lambda}{z}$. Assuming $\CV$ is a simple current extension of $\CV^{\mathbb{C}^\times}$
\be
	\CV = \bigoplus_{n\in \Z} \CV^n\,,\qquad  \CV^0 = \CV^{\C^\times}\,, \qquad \CV^n \times \CV^m \simeq \CV^{n+m}\,,
\ee
where $\CV^n$ is the subspace of weight $n$, the category $\CC^{\textrm{ext}}_{[\lambda]}$ can then be defined as a de-equivariantization of the category $\CD$ of $\CV^{\C^\times}$ modules whose monodromy with $\CV^1$ is $e^{2\pi i \lambda}$.

We can formulate this procedure in terms of the category of modules in the following way. Coupling to a flat connection is equivalent to de-equivariantization along the diagonally embedded copy of $\Z$
\be
	\Z\hookrightarrow \CD\boxtimes \mathrm{Vect}_{\C}
\ee
where $\mathrm{Vect}_{\C}$ is the category of $\C$ graded vector spaces, and the embedding above is given by:
\be
	n \mapsto \CV^{n}\boxtimes \C_n,
\ee
where $\C_n$ denotes the 1-dimensional vector space with nonzero support only at the grading $n \in \C$. Lifts of objects from $\mathrm{Vect}_{\lambda}$ are understood as modules sourcing the background connection $\frac{\lambda}{z}$.

This extended category of modules will not have a natural braided tensor category structure, again due to nontrivial monodromy. Nonetheless, one can obtain a braided tensor category structure by changing the braiding of $\mathrm{Vect}_{\C}$ to cancel the monodromy coming from $\CD$. For example, we could take the boson $\varphi$ to have a non-zero pairing $(\varphi,\varphi)=1$. It is not difficult to show that the category of $\CH_\varphi$ modules where $\varphi_0$ acts semi-simply is equivalent to the category of $\mathbb{C}$-graded vector spaces; $\mathbb{C}_\lambda$ can be identified with the module of $\CH_\varphi$ generated by $\ket{\lambda \varphi}$. The category $\mathrm{Vect}_{\C}$ then acquires an interesting braiding given by the quadratic form $(\lambda, \mu) = e^{2\pi i \lambda\mu}$, allowing us to stay in the realm of vertex operator algebras and braided tensor categories. The procedure of turning on a non-trivial braiding is essentially the method used in the third description of the extended category, and turning off braiding can be thought of as taking the limit under which the non-degenerate boson becomes degenerate.

\subsubsection{Example: symplectic fermions}
Let us apply the above construction to the simple example of the symplectic fermion VOA $V_{\chi_+,\chi_-}$. This VOA has an outer action of $\C^\times$ that scales $\chi_\pm$ with weight $\pm1$. Following the above, we introduce a degenerate boson $\varphi$ and consider the extension by the vertex operators $\norm{e^{\pm \varphi}}$ to obtain the lattice VOA $V_\varphi$. We then take a $\C^\times$ orbifold of the product $V_{\chi_+, \chi_-} \otimes V_\varphi$; in addition to the boson $A = \pd \varphi$, the generators of this orbifold are the symplectic fermions dressed by vertex operators
\be
	\widehat{\chi}_\pm = \norm{e^{\pm \varphi}} \chi_\pm.
\ee
The only singular OPE of these generators is
\be
\label{eq:flavorsymplectic}
	\widehat{\chi}_+(z) \widehat{\chi}_-(w) = \frac{1}{(z-w)^2}\bigg(1 + (z-w)\pd \varphi(w) + \ldots\bigg) \sim \frac{1}{(z-w)^2} + \frac{A(w)}{z-w}
\ee
exactly reproducing the vertex algebra $V^{\textrm{ext}}_{\chi_+, \chi_-}$ corresponding to symplectic fermions deformed by a background $\C^\times$ connection $A$, cf. \cite[Section 3.6.1]{GaiottoTwisted}.

Inspired by the construction in Sections 6.2.3 and  6.2.4 of \cite{CDGG} and by \cite{FLcenter}, we now rederive this deformation of $V_{\chi_+, \chi_-}$ from a suitable large-level limit. We start with a VOA with two Heisenberg fields $\CH_\varphi\otimes \CH_{\wt{\varphi}}$ and pairings $(\varphi,\varphi)=-(\wt{\varphi}, \wt{\varphi})=\delta^{-1}$. We then couple the symplectic fermion $\chi_\pm$ to the diagonal vertex operator of $\CH_\varphi\otimes \CH_{\wt{\varphi}}$:
\be
	\chi_+^\delta:= \chi_+\norm{e^{\varphi-\wt{\varphi}}},\qquad \chi_-^\delta:=\chi_-\norm{e^{-\varphi+\wt{\varphi}}}
\ee 
This gives the desired OPE
\be
	\chi_+^\delta\chi_-^\delta\sim \frac{1}{(z-w)^2}+\frac{\partial\varphi-\pd\wt{\varphi}}{z-w}
\ee
together with
\be
	\partial\varphi\chi_\pm^\delta\sim \partial\wt{\varphi}\chi_\pm^\delta\sim \frac{\pm \delta^{-1}}{z-w} \chi_\pm^\delta.
\ee
We denote $A=\partial\varphi-\pd\wt{\varphi}$ and $\wt{A} = \frac{1}{2}(\partial\varphi+\pd\wt{\varphi})$. The OPEs for finite $\delta$ can be identified with those of the affine Lie algebra of $V(\fgl(1|1))$. In terms of 4d $\CN=4$ gauge theory, this VOA is realized at the intersection of a Dirichlet boundary condition and an interface between $U(1)_\delta$ SYM theory and $U(1)_{-\delta}$ SYM supporting hypermultiplets valued in $T^*\mathbb{C}$.%
\footnote{We note that it is not possible to couple the hypermultiplet to a single $U(1)$ gauge theory in a way compatible with all $\delta$. As shown by Gaiotto-Witten in \cite{GWjanus}, to realize a consistent coupling between 4d $\CN=4$ $G$ gauge theory and a boundary 3d $\CN=4$ theory with $G$ flavor symmetry requires the flavor symmetry to satisfy a ``fundamental identity,''  cf. Eq. (3.23) of loc. cit. Although the $SU(2)$ flavor symmetry of a free hypermultiplet satisfies this identity, leading to $\mathfrak{osp}(1|2)$ living on a corner of $SU(2)$ gauge theory, the $U(1)$ subgroup does not.} %

In order to consistently take the $\delta \to \infty$ limit, we need to choose the set of generators we keep finite in the limit. We choose to keep the bosons $A,\wt{A}$ and the fermions $\chi^\delta_\pm$. The above OPEs correspond to
\be
	A \wt{A} \sim \frac{\delta^{-1}}{(z-w)^2}\qquad \wt{A} \chi^\delta_\pm  \sim \frac{\pm \delta^{-1}}{z-w} \chi^\delta_\pm \qquad \chi_+^\delta\chi_-^\delta\sim \frac{1}{(z-w)^2}+\frac{A}{z-w}
\ee
Importantly, the OPE coefficients are all polynomial in $\delta^{-1}$, and not $\delta$, so the $\delta \to \infty$ limit makes sense.
As $\delta\to \infty$, we see that $A$ and $\wt{A}$ become degenerate bosons and the only non-trivial OPE is precisely Eq. \eqref{eq:flavorsymplectic}. 

One nice feature of this construction is to realize that the subalgebra generated by the degenerate bosons $A,\wt{A}$ inherits a Poisson structure from the large-level limit, cf. Section 6.2.4 of \cite{CDGG}. Explicitly, we see that the modes of these fields have commutators
\be
	[\wt{A}_n, A_m] = n \delta^{-1} \delta_{n+m}
\ee
which vanish in the ``classical limit'' $\delta \to \infty$. The Poisson bracket inherited by this classical limit then takes the form
\be
	\{\wt{A}_n, A_m\} := \lim_{\delta \to \infty} \delta[\wt{A}_n, A_m] = n \delta_{n+m}
\ee
More generally, the full vertex algebra inherits the structure of a Poisson module: in addition to the (non-singular) OPE of the fields $A, \wt{A}$, the finite-level OPEs above induce an additional action via Poisson bracket. For example, we find
\be
	\{\wt{A}_n,\widehat{\chi}_{\pm,m}\}:=\lim\limits_{\delta \to \infty} \delta [\wt{A}_n,\chi^\delta_{\pm,m}] = \pm \widehat{\chi}_{\pm, n+m},
\ee
In particular, the Poisson bracket with $\wt{A}$ encodes the action of the $\mathbb{C}^\times$ symmetry action on both $\chi_\pm$ and the connection $A$. Note that there is no well-defined bracket between two general elements of the large-level limit VOA -- the above Poisson bracket is finite only when one of the elements belongs to the commutative subalgebra.

\subsubsection{$\CV^{\textrm{ext}}_k$ from a large-level limit}
We can apply the above symplectic fermion computation to the cases we are interested in: the VOAs $V(\fgl(1|1))$ and $\CV_k$. We consider coupling the fermions $\psi_\pm$ of $V(\fgl(1|1))$ with the diagonal vertex operator of $\CH_\varphi\otimes \CH_{\wt{\varphi}}$:
\be\label{eq:gl11flavor}
	\psi_+^\delta:=\psi_+\norm{e^{-\varphi+\wt{\varphi}}},\qquad \psi_-^\delta:=\psi_-\norm{e^{\varphi-\wt{\varphi}}}.
\ee
We will denote the resulting VOA by $V(\fgl(1|1))^\delta$; aside from the usual OPEs in $V(\fgl(1|1))$, the non-trivial OPEs are
\be\label{eq:gl11flavorfinite}
	\psi_+^\delta\psi_-^\delta\sim \frac{1}{(z-w)^2}+\frac{E-A}{z-w}, \qquad \wt{A}\psi_\pm^\delta\sim \frac{\pm \delta^{-1}}{z-w}\psi_\pm^\delta, \qquad \wt{A} A\sim \frac{\delta^{-1}}{(z-w)^2}.
\ee

The VOA $V(\fgl(1|1))^\delta$ inherits simple modules corresponding to spectral flow of $V(\fgl(1|1))$, and we will denote by $\CV^{\delta}_k$ the extension corresponding to $\widehat{A}_{k,0}$ and $\Pi^{k+1}\widehat{A}_{\frac{k}{2},1}$, which we describe in more detail below. Similar to the above example, the $\delta\to \infty$ limit of $\CV^{\delta}_k$ is the VOA $\CV^{\textrm{ext}}_k$ tensored with a degenerate Heisenberg $\wt{A}$, as can be seen from the OPEs in Eq. \eqref{eq:gl11flavorfinite}. The commutative subalgebra generated by $A, \wt{A}$ inherits a vertex Poisson structure and $\CV^{\textrm{ext}}_k$ becomes a module thereof, where Poisson bracket with $\wt{A}$ generates the action of the $\mathbb{C}^\times$ symmetry on $\CV^{\textrm{ext}}_k$.

\subsection{Three ways to the category $\CC^{\textrm{ext}}_{k}$}
\label{sec:cext}
With the above deformation to $\CV_k$ by a background flat connection $A$, we give three alternative ways of describing the extended category of line operators $\CC^{\textrm{ext}}_{k}$ (in the case of regular singular connection) using local modules of some other vertex algebras. These will lead to the fusion structure on $\CC^{\textrm{ext}}_{k}$ and an identification with a de-equivariantization of $KL$.

\subsubsection{Via non-local modules}
\label{sec:nonlocal}
Our first avenue to the category $\CC^{\textrm{ext}}_{k}$ is to follow the general procedure described in Section \ref{sec:flatdeform}. Although this approach is generally applicable, it does not provide much by the way of a concrete description of the category.

Recall the Heisenberg VOA $\CH$ used in Section \ref{sec:freefield}. To obtain $\CV_k$, we extended this Heisenberg by the lattice generated by the states
\be
	\ket{X+\tfrac{k}{2}Y+Z},~\ket{kY},~\ket{Z}.
\ee
We want to consider non-local actions the fermionic operators $\psi_\pm$, which were essentially identified with $\norm{e^{\pm Z}}$ in the free-field realization. For this, let us first extend $\CH$ by the sublattice spanned by $\ket{X+\tfrac{k}{2}Y+Z}$ and $\ket{kY}$, and denote by $\CW_k$ the kernel of the screening operator in this extended VOA. Clearly this is a subVOA of $\CV_k$, and $\CV_k$ is a simple current extension of $\CW_k$. In fact, $\CW_k=\CV_k^{\mathbb{C}^\times}$ for the action of $\mathbb{C}^\times$ induced by $Z_0-Y_0$. We denote the category of modules of $\CW_k$ by $\CD_k$. For each $\lambda \in \mathbb{C}$, denote by $\CD_{k, [\lambda]}$ the subcategory of modules where the monodromy with $\CV^1_k$ is equal to $e^{2\pi i \lambda}$. The lifting functor
\be
	M	\mapsto \CV_k \times M 
\ee
lifts objects in $\CD_{k, [\lambda]}$ to non-local modules of $\CV_k$ where $\psi_\pm$ is $\pm \lambda + \Z$ moded. We thus have the following identification: 
\be
\label{eqCextD}
	\CC^{\textrm{ext}}_{k, [\lambda]} = \CD_{k, [\lambda]}/{\mathbb{Z}}
\ee
Technically speaking, because $\mathbb{Z}$ does not have trivial monodromy with elements in $\CD_{k, [\lambda]}$, this is not a de-equivariantization in the sense of \cite[Section 8.23]{etingof2016tensor}. However, we can still understand it as identifying objects in $\CD_{k, [\lambda]}$ via left multiplication with direct summands of $\CV_k$. The fact that this construction only depends on the image of $\lambda$ in $\C/\Z$ can be understood as identifying connections that differ by large gauge transformations.

The equivalence in Eq. \eqref{eqCextD} immediately implies the following fusion structure on $\CC^{\textrm{ext}}_{k}$:
\be
	\CC^{\textrm{ext}}_{k, [\lambda]}\times \CC^{\textrm{ext}}_{k, [\mu]}\longrightarrow \CC^{\textrm{ext}}_{k, [\lambda+\mu]},
\ee
coming from the fusion rules in $\CD_k/\mathbb{Z}$. This endows the category $\CC^{\textrm{ext}}_{k}$ with the structure of a monoidal category such that $\CC^{\textrm{ext}}_{k, [0]}$ is the center (and thus a braided tensor category). $\CC^{\textrm{ext}}_{k}$ is itself not a braided tensor category since the composition:
\be
\begin{tikzcd}
	\CV_k \times \CC^{\textrm{ext}}_{k, [\lambda]} \rar{R} & \CC^{\textrm{ext}}_{k, [\lambda]}\times \CV_k \rar{R} & \CV_k \times \CC^{\textrm{ext}}_{k, [\lambda]}
\end{tikzcd}
\ee
is given by the action of $e^{2\pi i \lambda}$. Instead, we expect that it is a $\mathbb{C}/\Z$-crossed tensor category in the sense of \cite[Section 8.24]{etingof2016tensor}.

\subsubsection{Via large-level limits}
\label{sec:largelevellimit}
In this second approach, we analyze the category of $\CC^{\textrm{ext}}_{k}$ from the perspective of $\CV^{\textrm{ext}}_k$ as the large-level limit of $\CV^{\textrm{ext}, \delta}_k$ described in the previous section. This route is particular useful because it affords a very concrete description of the extended category and makes manifest some subtleties in the tensor structure.

Recall that $\CV^{\delta}_k$ is an extension of the VOA $V(\fgl(1|1))^\delta$ generated by the bosons $N,E, A, \wt{A}$ and the fermions $\psi^\delta_\pm$ with OPE given by the usual $V(\fgl(1|1))$ OPE together with the deformation given in Eq. \eqref{eq:gl11flavorfinite}. Consider the following redefinition inside $V(\fgl(1|1))^\delta$:
\be
	N' = N+\tfrac{k}{2}A, \quad ~E' = E-A, \quad~\wt{A}' = \wt{A} - 2\delta^{-1}N + (2-k)\delta^{-1}E
\ee
It is not hard to verify that the OPEs of $N', E'$ and $\psi_\pm^\delta$ realize a copy of the VOA $V(\fgl(1|1))$, and it is decoupled from the fields $A$ and $\wt{A}'$. In other words:
\be
	V(\fgl(1|1))^\delta \cong V(\fgl(1|1))\otimes \CH_{A,\wt{A}'}. 
\ee
Under this identification, the VOA  $\CV^{\delta}_k$ can be viewed as an extension in $KL\boxtimes \mathrm{Vect}_{\mathbb{C}^2}$, where $\mathrm{Vect}_{\mathbb{C}^2}$ is the category of vector spaces graded by the weight of $A_0$ and $\wt{A}'_0$. We are only interested in modules of $\CV^{\delta}_k$  where both $A_0$ and $\wt{A}'_0$ act semisimply, and so this category suffices for our needs. As was the case in the absence of a background flat connection, we extend by a $\Z^2$ lattice of simple currents. The first copy of $\Z$ is purely in $KL$, and corresponds to the combination $N'+\frac{k}{2}E' = N + \frac{k}{2}E$, and so the trivial-monodromy condition requires that the combination $N'_0+\frac{k}{2}E'_0 = N_0 + \frac{k}{2}E_0$ acts semisimply with integer eigenvalues. The second copy of $\Z$ corresponds to the element $kE'+kA = k E$, and the trivial-monodromy condition requires that $kE'_0+kA_0 = k E_0$ acts semisimply with integer eigenvalues. In conclusion, if we denote by $KL_{e'}^{N'+\frac{k}{2}E', E'}$ the subcategory of $KL$ where $N'_0+\frac{k}{2}E'_0$ acts semisimply with integer eigenvalues and $E'_0$ acts semisimply with eigenvalue $e'$, then the category
\be
	KL_{e - \lambda}^{N'+\frac{k}{2}E', E'}\boxtimes \mathrm{Vect}_{(\lambda)\times \mathbb{C}}
\ee
has trivial monodromy with $\CV^{\delta}_k$ for any $e \in \frac{1}{k}\Z$ and $\lambda \in \C$. That said, tensoring with the generator of the first copy of $\Z$ changes the weight of $E'_0$ by $1$, so this category is not closed under fusing with the simple currents extending $V(\fgl(1|1))^\delta$; fusion with the generator of the second copy of $\Z$ shifts the weight of $N'$ by $k$ and $\wt{A}'$ by $k \delta^{-1}$. Denote the category of modules of $\CV^{\delta}_k$ by $\CC^{\delta}_{k}$; this category can then be described as the following de-equivariantization:
\be
	\CC^{\delta}_{k} = \bigoplus_{\lambda \in \C} \CC^{\delta}_{k,\lambda}, \qquad \CC^{\delta}_{k, \lambda} \simeq  \bigg(\bigoplus\limits_{e \in \frac{1}{k}\Z}KL_{e-\lambda}^{N'+\frac{k}{2}E', E'}\boxtimes \mathrm{Vect}_{(\lambda)\times \mathbb{C}}\bigg)/\Z^2\,.
\ee

We are now faced with the task of taking the $\delta \to \infty$ limit of the category $\CC^{\delta}_{k, \lambda}$. De-equivariantization with respect to the first copy of $\Z$ is essentially the same as in Section \ref{sec:fibercat}, so we will not describe it in detail. Fusion with the generator of the second copy will identify an object in
\be
	KL_{e-\lambda}^{N'+\frac{k}{2}E', E'}\boxtimes  \mathrm{Vect}_{(\lambda, \mu)}
\ee
with an object in
\be
	KL_{e-\lambda}^{N'+\frac{k}{2}E', E'}\boxtimes  \mathrm{Vect}_{(\lambda, \mu+k\delta^{-1})}.
\ee
Thus, fusion with the generators of $\Z^2$ do not change the eigenvalues of either $A_0$ or $\wt{A}_0$ in the limit $\delta\to \infty$:
\be
	\CC^{\infty}_{k, \lambda} \simeq  \bigoplus_{\mu \in \C}\bigg(\bigoplus\limits_{e\in \frac{1}{k}\Z} KL_{e-\lambda}^{N'+\frac{k}{2}E', E'}\boxtimes \mathrm{Vect}_{(\lambda, \mu)}\bigg)/\Z^2
\ee
As noted above, the large-level limit $\CV^{\infty}_k$ decomposes as $\CV_k^{\textrm{ext}} \otimes \CH_{\wt{A}}$. In order to isolate $\CV_k^{\textrm{ext}}$, we need to set to zero the boson $\wt{A}$ and hence we restrict to those modules such that $\wt{A}'$ acts as zero in the limit. We can rescale $\mu\mapsto \mu/\delta$ to achieve this. Taking the limit, we are led to an equivalence of categories
\be
	\CC^{\textrm{ext}}_{k, \lambda} \simeq  \bigg(\bigoplus\limits_{e \in \frac{1}{k}\Z} KL_{e-\lambda}^{N'+\frac{k}{2}E', E'}\boxtimes \mathrm{Vect}_{(\lambda,0)}\bigg)/\Z^2
\ee
This is precisely the category of modules sourcing the flat connection $A = \frac{\lambda}{z}$.

If we choose to remember only the holonomy $g$ of the connection, rather than the connection itself, we simply remove the factor of Vect$_{(\lambda, 0)}$; this gives us the desired equivalence of abelian categories:
\be
	\CC^{\textrm{ext}}_{k, g} \simeq  \bigg(\bigoplus\limits_{e \in \frac{1}{k}\Z} KL_{e-\lambda}^{N'+\frac{k}{2}E', E'}\bigg)/\Z^2 = \bigoplus_{[e] \in \frac{1}{k}\Z/\Z} KL_{[e-\lambda]}^{N'+\frac{k}{2}E', E'}/\Z^2
\ee
where $g = e^{2\pi i k\lambda}$. Note that when $\lambda = 0$ we arrive at the category of line operators $\CC_k$ derived in Section \ref{sec:Wilson}, as desired.

We argued above that we don't care about $\wt{A}$ in the $\delta \to \infty$ limit. In practice, the advantage of having the degenerate boson $\wt{A}$ is that the limit vertex algebra 
\be
	\CV^{\infty}_k\cong \CV^{\textrm{ext}}_k \boxtimes \CH_{\wt{A}}
\ee 
inherits a Poisson structure on $\CH_{\wt{A}}$. Moreover, modules in $\CC^{\textrm{ext}}_{k,\lambda}$ have the structure of a Poisson module where the bracket 
\be
	\{\wt{A}_n,-\}:=\lim\limits_{\delta\to \infty} \delta [\wt{A}_n, -],
\ee
can be identified with the action of meromorphic gauge transformations.

\subsubsection{Via shifting the current}
\label{sec:shifting}
Our third description of $\CC^{\textrm{ext}}_{k}$ is conceptually the most simple, but doesn't have the general applicability of the previous approaches. Here, we realize the deformation by a background connection $A$ by shifting the current $E$ associated to a central element of the (finite) Lie algebra. Abelian flavor symmetries of this type arise naturally from topological flavor symmetries of the bulk TQFT.

In the definition of $\CV^{\textrm{ext}}_k$, we coupled the VOA to a degenerate boson $A$, and obtained the OPE
\be
	\psi_+\psi_-\sim \frac{1}{(z-w)^2}+\frac{E-A}{z-w}.
\ee
Alternatively, if we view $E - A$ and $N + \frac{k}{2}A$ as new currents $E'$ and $N'$, while keeping the other variables fixed, we arrive back at $V(\fgl(1|1))$. In terms of the free field realization of $V(\fgl(1|1))$ from Section \ref{sec:freefield}, this amounts to the field redefinition:
\be
	Z = Z' + \varphi, \quad Y = Y' + \varphi, \qquad X = X' - \tfrac{k+2}{2}\varphi
\ee
where $A = \pd \varphi$.

If we perform this change of variables, we can view the VOA $\CV^{\textrm{ext}}_k$ as an extension of $V(\fgl(1|1)) \otimes \CH_A$ by modules corresponding to: 
\be
	X'+\tfrac{k}{2}Y'+Z',\quad ~ kY'+ k\varphi.
\ee 
Assume that $\varphi_0$ acts with eigenvalue $\lambda$. The trivial-monodromy condition then requires that both $N_0 ' + \frac{k}{2} E'_0 = N_0+\frac{k}{2}E_0$ and $k E_0' = k(E_0 - A) = k(E_0 - \varphi_0)$ act semisimply with integer eigenvalues, whence $E_0$ acts semisimply with eigenvalues in $-\lambda+ \frac{1}{k}\Z$, cf. the trivial-monodromy constraint encountered above. The category $\CC^{\textrm{ext}}_{k}$ is then obtained by de-equivariantization with respect to fusion with the lattice of simple currents generated by $\widehat{A}_{k,0},\Pi^{k+1}\widehat{A}_{\frac{k}{2},1}$:
\be
	\CC^{\textrm{ext}}_{k} = \bigoplus_{g \in \C^\times} \CC^{\textrm{ext}}_{k, g}, \qquad \CC^{\textrm{ext}}_{k, g} = \bigg(\bigoplus_{e \in \frac{1}{k}\Z} KL^{N'+\frac{k}{2}E', E'}_{e - \lambda}\bigg)/\Z^2
\ee
where $g = e^{2\pi i k \lambda}$, which is precisely the result obtained in the previous subsection. This observation can be rather easily tied back to non-local modules by observing that the above shift of the current can be viewed as shifting the momenta of the vertex operators appearing in modules, which are naturally non-local modules for general choices of the shift.

\subsubsection{Large-level limit vs. non-local modules}
\label{sec:largelevelvsnonlocal}

In Section \ref{sec:nonlocal}, we used the VOA $\CW_k = \CV_k^{\mathbb{C}^\times}$ and its modules that are not local for $\CV_k$ to understand the extended category $\CC^{\textrm{ext}}_{k}$. In this final subsection, we examine carefully the relation between this approach and the large-level limit approach. We note that a systematic study of twisted modules and large centers, albeit in a slightly different setup, recently appeared in work of Feigin-Lentner \cite{Feigin:2025gfj}.

The idea is as follows. We defined the VOA $\CV_k^{\delta}$ by coupling the fields $\psi_\pm$ with $e^{\pm (\varphi - \wt{\varphi})}$ (where $A = \pd\varphi - \pd \wt{\varphi}$). Let $V_A$ be the extension of $\CH_{A,\wt{A}}$ by the fields $e^{\pm (\varphi - \wt{\varphi})}$, then $\CV_k^{\delta}$ is the orbifold:
\be
	\CV_k^{\delta}\cong (\CV_k\otimes V_A)^{\mathbb{C}^\times},
\ee
where the action of $\mathbb{C}^\times$ on $V_A$ is generated by $\delta \wt{A}_0$. As such, $\CV_k^{\delta}$ is a simple current extension of the VOA
\be
	\CV_k^{\mathbb{C}^\times}\otimes V_A^{\mathbb{C}^\times}\cong \CW_k \otimes \CH_{A,\wt{A}}.
\ee
Therefore, we see that modules of $\CV_k^{\delta}$ can be obtained from de-equivariantization of the category
\be
	\CD_k \boxtimes \mathrm{Vect}_{\mathbb{C}^2}.
\ee
A similar argument as in Section \ref{subsubsec:orbifoldcomparison} tells us that, in the limit $\delta \to \infty$ where $A$ and $\wt{A}$ degenerate, we have an equivalence of abelian categories:
\be
	\CC^{\textrm{ext}}_{k, [\lambda]}\simeq \CD_{k, [\lambda]}/\Z. 
\ee

\pagebreak

\appendix

\section{Aspects of $U(1|1)$ Chern-Simons theories}
\label{sec:u11CS}

In this Appendix, we review aspects of Chern-Simons theories based on $U(1|1)$. %In terms of the discrete data presented in Section \ref{sec:introglobal}, we are considering the theory based on the cocharacter lattice $\Lambda = \Lambda_{(k,1,\frac{k}{2})} =\langle k E, N + \frac{k}{2}E\rangle$.
In Section \ref{sec:global} we review the analysis of \cite{Mikhaylov} that determines the data that encodes such a theory and then in Section \ref{sec:fieldtheoryVOA} we extract the boundary VOAs $\CV_{(\kappa, \nu, \xi)}$ from a field theory analysis.
Finally, in Section \ref{sec:locopsGQ} we apply the analysis of \cite{Zeng} to derive the vector space of local operators from geometric quantization.

%We use a slightly different convention to the one presented in the main body of the text: we rescale $E \to \frac{1}{k} E$ so that the pairing used in the action functional has an overall factor of $k$ and the cocharacter lattice is generated $E$ and $N + \frac{1}{2}E$.

Recall that the Lie superalgebra $\mathfrak{gl}(1|1)$ is generated by two bosonic elements $N,E$ and two fermionic elements $\psi_\pm$ with non-vanishing brackets
\be
	[N, \psi_\pm] = \pm \psi_\pm \,, \qquad \{\psi_+, \psi_-\} = E\,.
\ee
We can represent these generators as the following supermatrices:
\be
	N \leftrightarrow \begin{pmatrix}
		\tfrac{1}{2} & 0\\ 0 & -\tfrac{1}{2}\\
	\end{pmatrix}\quad E \leftrightarrow \begin{pmatrix}
		1 & 0\\ 0 & 1\\
	\end{pmatrix}\quad \psi_+ \leftrightarrow \begin{pmatrix}
		0 & 1\\ 0 & 0\\
	\end{pmatrix}\quad \psi_- \leftrightarrow \begin{pmatrix}
		0 & 0\\ 1 & 0\\
	\end{pmatrix}
\ee
There is a natural (graded) symmetric bilinear form on this Lie superalgebra given by the supertrace in the above representation, which we denote by $(-,-)$
\be
	(N, E) = (E,N) = 1\,, \qquad (\psi_\pm, \psi_\mp) = \pm1\,.
\ee

\subsection{Formulation as a BV field theory}
We now formulate $\mathfrak{gl}(1|1)$ Chern-Simons theory in the BV formalism. Let $\CM$ denote spacetime, a smooth (framed) 3-manifold. Working for the moment around the trivial gauge bundle, the fields of the theory can be organized into a superconnection 1-form $A = A_\mu \diff x^\mu$. We decompose this superconnection in terms of the above basis as follows:
\be
	A = A^N N + A^E E + \Psi^+ \psi_+ + \Psi^- \psi_-\,.
\ee
If we restrict to bosonic gauge transformations, we can identify the bosonic fields $A^\pm_\mu$ with separate abelian connections and the fermionic fields $\Psi^\pm_\mu$ with 1-forms valued in an associated $\Pi \C_{(\pm1,0)}$ bundle, where $\C_{(q_N,q_E)}$ denotes the 1-dimensional representation of charge $(q_N, q_E)$ and $\Pi$ denotes a reversal of fermionic parity. We will use the convention that the coordinate 1-forms $\diff x^\mu$ are fermionic, whence $A^\pm$ are naturally fermionic while $\Psi^\pm$ are naturally bosonic. We denote the de Rham differential by $\diff$ and the covariant exterior derivative by $\diff_A$, so that
\be
	\diff_A \Psi^\pm = \diff \Psi^\pm \pm A^N \Psi^\pm\,.
\ee

The action for this supergroup Chern-Simons theory is given by
\be
\begin{aligned}
	S & = \tfrac{1}{4\pi} \int \Tr \bigg(A, \diff A + \tfrac{1}{3}[A,A]\bigg)\\
	& = \tfrac{1}{2\pi}\int A^E \diff A^N + \Psi^- \diff_A \Psi^+
\end{aligned}
\ee
from which we find that the equations of motion say that the fermionic fields $\Psi^\pm$ are covariantly constant (with respect to $\diff_A$), $A^N$ is flat, and that the curvature $\diff A^E$ is constrained by the fermionic fields
\be
\diff A^N = 0\,, \qquad \diff A^E + \Psi^- \Psi^+ = 0\,, \qquad \diff_A \Psi^\pm = 0\,.
\ee
Note the equations of motion do not localize to a flat bosonic connection. Rather, they localize to flat superconnections: $\diff A + \tfrac{1}{2}[A,A] = 0$.

In the BV formalism, we are instructed to extend the above space of fields so that we may implement the above equations of motion and their redundancies in a cohomological fashion. The above equations of motion are subject to the usual gauge redundancy
\be
\delta A^N = \diff a^N\,, \qquad \delta A^E = \diff a^E + \Psi^- \alpha^+ + \Psi^+ \alpha^-\,, \qquad \delta \Psi^\pm = \diff_A \alpha^\pm\,,
\ee
where $a^\pm$ are gauge parameters for the bosonic part of the gauge algebra and $\alpha^\pm$ are gauge parameters for the fermionic part. We introduce (fermionic) ghosts $c^\pm$ associated to bosonic gauge transformations and (bosonic) ghosts $Z^\pm$ associated to fermionic gauge transformations. The BV/BRST variation of these ghost fields and the physical fields encodes the brackets discussed above as well as the above gauge transformation:
\be
\begin{aligned}
	Q c^N & = 0 \qquad & Q c^E & = Z^- Z^+ \qquad & Q Z^\pm &= \pm c^N Z^\pm\\
	Q A^N & = \diff c^N \qquad & Q A^E & = \diff c^E + \Psi^- Z^+ + \Psi^+ Z^- \qquad & Q \Psi^\pm & = \pm c^N \Psi^\pm + \diff_A Z^\pm\\
\end{aligned}
\ee

We now introduce anti-fields for each of the above fields and ghosts; the latter are often called anti-ghosts. Roughly, the BV/BRST variation of an anti-field encodes the equations of motion of its corresponding field as well as its gauge variation (with parameters $c^\pm, Z^\pm)$. For example, the anti-fields to the fermionic field $\Psi^+_\mu$ is a bosonic 2-form field $\Psi^{+,*}_{\mu \nu}$ whose BV/BRST variation is given by
\be
Q \Psi^{+,*} = c^N \Psi^{+,*} + \diff_A \Psi^+ + A^{N,*}Z^+\,,
\ee
where the first term encodes the fact that $\Psi^{+,*}$ has charge $(q_N, q_E) = (1,0)$ and the second term is the equation of motion for $\Psi^+$. We can concisely encode these fields, ghosts, anti-fields, and anti-ghosts by combining them into fields with inhomogeneous form degree:
\be
\begin{array}{c}
	\CA^N = c^N + A^N + A^{N,*} + c^{N,*} \qquad \CA^E = c^E + A^E + A^{E,*} + c^{E,*}\\
	\CZ^\pm = Z^\pm + \Psi^\pm + \Psi^{\pm,*} + Z^{\pm,*}
\end{array}
\ee
In this notation, the BV/BRST variation then takes the following simple form:
\be
Q \CA^N = \diff \CA^N\,, \qquad Q \CA^E = \diff \CA^E + \CZ^- \CZ^+\,, \qquad Q \CZ^\pm = \diff_{\CA} \CZ^\pm\,,
\ee
where $\diff_{\CA} \CZ^\pm = \diff \CZ^\pm \pm \CA^N \CZ^\pm$. In terms of the components of homogeneous form degree, the covariant derivative $\diff_{\CA} \CZ^\pm$ reads
\be
\begin{aligned}
	\diff_{\CA} \CZ^\pm & = \bigg(\overset{0\textrm{-form}}{\pm c^N Z^\pm}\bigg) + \bigg(\overset{1\textrm{-form}}{\pm c^N \Psi^\pm + \diff_A Z^\pm}\bigg) + \bigg(\overset{2\textrm{-form}}{\pm c^N \Psi^{\pm,*} + \diff_A \Psi^\pm \pm A^{N,*} Z^\pm} \bigg)\\
	& \quad + \bigg(\overset{3\textrm{-form}}{\pm c^N Z^{\pm,*} + \diff_A \Psi^{\pm,*} \pm A^{N,*} \Psi^\pm \pm c^{N,*}Z^\pm}\bigg)\,.
\end{aligned}
\ee
This notation also makes the action functional of our theory particularly simple
\be
\begin{aligned}
	S = \tfrac{1}{4\pi}\int \Tr(\CA, \diff \CA + \tfrac{1}{3}[\CA, \CA]) = \tfrac{1}{2\pi}\int \CA^E \diff \CA^N + \CZ^- \diff_\CA \CZ^+\,,
\end{aligned}
\ee
where, by definition, the integral can only be non-vanishing on the 3-form component of its integrand.

The extended space of fields used in the above BV formulation has several $\Z$ gradings in addition to differential form degree. First, there is the BV ghost number: physical fields $A^\pm_\mu, \Psi^\pm_\mu$ have vanishing ghost number $\textrm{gh} = 0$; the ghosts $c^\pm, Z^\pm$ sit in ghost number $\textrm{gh} = 1$; the anti-fields $A^{\pm,*}_\mu, \Psi^{\pm,*}_\mu$ in $\textrm{gh}=-1$; and the anti-ghosts in $\textrm{gh}=-2$. It follows that $S$ has vanishing ghost number and $Q$ increases ghost number by $1$.

We also note that the extended space of fields is naturally $(-1)$-shifted symplectic. The symplectic form is given by
\be
\Omega = \tfrac{1}{2\pi}\int \delta \CA^E \delta \CA^N + \delta \CZ^- \delta \CZ^+\,,
\ee
and has ghost number $\textrm{gh} = -1$. A straightforward computation implies that the BV/BRST supercharge $Q$ is identified with the Hamiltonian vector field associated to the action functional
\be
\iota_Q \Omega = \delta S\,.
\ee
Together with the fact that $Q$ is then nilpotent, this implies $S$ solves the (classical) master equation
\be
\{S,S\} = 0\,,
\ee
where $\{-,-\}$ denotes the (odd) Poisson bracket induced by the symplectic form $\Omega$.

\subsection{Choices of global form}
\label{sec:global}

As emphasized in \cite{Mikhaylov}, there are infinitely many choices of global form of the gauge group for a Chern-Simons theory based on the Lie superalgebra $\fgl(1|1)$, even if we require the bosonic gauge group is the compact torus $U(1)^2$. Although all of these theories are perturabtively equivalent, their differences become apparent once non-perturbative considerations (e.g. monopoles) are taken into account.

The Lie superalgebra $\fgl(1|1)$ has several natural bases that we will frequently go between. In addition to the one mentioned above, we will also make use of a different basis that replaces the bosonic generators $N,E$ with the combinations $T_\pm = \tfrac{1}{2}E \pm N$. With a fixed gauge-invariant pairing on $\fgl(1|1)$, which we take to be
\be
	(N,E) = (E,N) = 1 \qquad (\psi_+, \psi_-) = -(\psi_-, \psi_+) = 1,
\ee
the gauge theory is determined uniquely by a choice of cocharacter lattice $\Lambda$ for the bosonic subgroup. Not any lattice will do: as described in Section 4.2 of \cite{Mikhaylov}, we need to ensure that this is an integer lattice in $\R^{1,1}$ whose dual lattice contains the weight of the fermionic generators $\psi_\pm$. Any such lattice is uniquely specified by positive integers $\kappa, \nu \in \Z_{>0}$ and a half-integer $\xi$ defined mod $\kappa$, $\xi \in \frac{1}{2}\Z/\kappa \Z$; the lattice is generated by the vectors
\be
\label{eq:cocharlattice}
	v_1  = \tfrac{\kappa}{\nu} E \qquad v_2 = \tfrac{\xi}{\nu} E + \nu N
\ee
cf. Section 4.3 of \cite{Mikhaylov}. We will denote this cocharacter lattice $\Lambda_{(\kappa, \nu, \xi)}$. In other words, we declare that \be
    A^1 = \tfrac{\nu}{\kappa} A^E - \tfrac{\xi}{\nu \kappa} A^N \qquad A^2 = \tfrac{1}{\nu} A^N
\ee
are connections on $U(1)$ principal bundles $P_1, P_2 \to \CM$ and $\Psi^\pm$ are sections of suitable associated bundles. The action takes the following form when expressed in terms of these properly-quantized fields:
\be
    S = \frac{1}{2\pi} \int \kappa \CA^1 \diff \CA^2 + \xi \CA^2 \diff \CA^2 + \CZ^- \bigg(\diff \CZ^+ + \nu \CA^2 \CZ^+\bigg)~.
\ee

There are two distinguished choices forms mentioned in \cite{Mikhaylov}. The most natural choice corresponds to taking $\nu = 1$ together with $\kappa = 2 \xi =: k$. We find (after rescaling the $\CZ^\pm$) a theory of $U(1)_k \times U(1)_{-k}$ Chern-Simons gauge fields $\CA^+ = -\CA^1$ and $\CA^- = -(\CA^1 + \CA^2)$ coupled to a ($B$-twisted) hypermultiplets $\CZ^+$ and$\CZ^-$ of weights $(1,-1)$ and $(-1,1)$:
\be
	\CT_{(k,1,\frac{k}{2})} \rightsquigarrow S = \int \tfrac{k}{4\pi}\big(\CA^+ \diff \CA^+ - \CA^- \diff \CA^-\big) + \CZ^- \diff_\CA \CZ^+
\ee
where $\diff_\CA \CZ^{\pm} = \diff \CZ^\pm \pm (\CA^+ - \CA^-) \CZ^\pm$. Another natural choice comes from taking $\nu = 1$, $\xi = 0$ (we again set $\kappa = k$). Correspondingly, there is a pair of $U(1)$ Chern-Simons gauge fields coupled to a ($B$-twisted) hypermultiplet $\CZ^+, \CZ^-$ of weight $(1,0) \oplus (-1,0)$:
\be
	\CT_{(k,1,0)} \rightsquigarrow S = \int \tfrac{k}{2\pi} \CA^1 \diff \CA^2 + \CZ^- \diff_\CA \CZ^+
\ee
where $\diff_\CA \CZ^{\pm} = \diff \CZ^\pm \pm \CA^2 \CZ^\pm$.

Although there are an infinite number of choices, each of the global forms can be connected to one another via gauging 0-form and 1-form global symmetries. The main point is that we can extend the cocharacter lattice by gauging a 1-form electric symmetry and we can restrict to a sublattice by gauging a 0-form topological symmetry. See Fig. \ref{fig:globalforms} for an illustration of how to relate the theory $\CT_{(\kappa, \nu, \xi)}$ to the reference theory $\CT_{(1,1,0)}$; note that each of the steps is invertible despite the fact we use directed arrows -- extending the cocharacter lattice (gauging a 1-form electric symmetry) and restricting to a sublattice (gauging a 0-form topological symmetry) are inverse operations. At the level of cocharacter lattices, the upper path (where $2 \xi$ is an even integer) is as follows. First extend $\Lambda_{(\kappa, \nu, \xi)}$ by $\frac{1}{\nu}E$. The resulting lattice is generated by $\frac{1}{\nu}E$ and $\frac{\xi}{\nu}E + \nu N$ or equivalently $\frac{1}{\nu}E$ and $\nu N$, yielding the lattice $\Lambda_{(1,\nu,0)}$.%
\footnote{When $2\xi$ is odd, the result of extending $\Lambda_{(\kappa, \nu, \xi)}$ by $\frac{1}{\nu}E$ is instead equivalent to the lattice generated by $\frac{1}{\nu}E$ and $\frac{1}{2\nu}E + \nu N$, i.e. $\Lambda_{(1,\nu,\scriptstyle{\frac{1}{2}})}$. This is the reason why we need to consider two different cases.} %
We then restrict to the sublattice generated by $E$ and $\nu N$, i.e., we restrict to $\Lambda_{(\nu,\nu,0)}$, and then, finally, extend by the vector $N$ to arrive at the cocharacter lattice $\Lambda_{(1,1,0)}$.

\begin{figure}[H]
	\centering
	\begin{tikzpicture}
		\draw (-2.5,0) node {$\CT_{(\kappa, \nu, \xi)}$};
		
		\draw (0,1) node {$\CT_{(1, \nu, 0)}$};
		\draw (0,-1) node {$\CT_{(1, \nu, \frac{1}{2})}$};
		
		\draw (2.5,1) node {$\CT_{(\nu, \nu, 0)}$};
		\draw (2.5,-1) node {$\CT_{(2, 2\nu, 0)}$};
		
		\draw (5,1) node {$\CT_{(1, 1, 0)}$};
		\draw (5.15,-1) node {$\CT_{(2\nu, 2\nu, 0)}$};
		
		\draw (7.7,-1) node {$\CT_{(1, 1, 0)}$};
		
		\draw[->] (-1.75,0.25) -- (-0.75, 1);
		\draw (-1.75, 0.8) node {$\Z_{\kappa}^{(1)}$};
		
		\draw[->] (-1.75,-0.25) -- (-0.75, -1);
		\draw (-1.75, -0.8) node {$\Z_{\kappa}^{(1)}$};
		
		\draw[dashed, ->] (0.75,1) -- (1.75,1);
		\draw (1.25,1.5) node {$\Z_\nu^{(0)}$};
		
		\draw[dashed, ->] (0.75,-1) -- (1.75,-1);
		\draw (1.25,-1.5) node {$\Z_2^{(0)}$};
		
		\draw[->] (3.25,1) -- (4.25,1);
		\draw (3.75, 1.5) node {$\Z_\nu^{(1)}$};
		
		\draw[dashed, ->] (3.25,-1) -- (4.25,-1);
		\draw (3.75, -1.5) node {$\Z_\nu^{(0)}$};
		
		\draw[->] (6,-1) -- (7, -1);
		\draw (6.5, -1.5) node {$\Z_{2\nu}^{(1)}$};
		
	\end{tikzpicture}
	\caption{Schematic relation between the global forms $\CT_{(\kappa, \nu, \xi)}$ and $\CT_{(1,1,0)}.$ The upper (resp. lower) path corresponds to $2\xi$ is even (resp. odd). The solid arrows correspond to gauging a 1-form symmetry and the dashed arrows correspond to gauging a 0-form topological symmetry.}
	\label{fig:globalforms}
\end{figure}

As will be shown below, we will use a uniform method to study this infinite family of theories, that of simple current extensions. Therefore, although we will state our results below in terms of the infinite family, we do not feel the need of presenting the calculation for a generic theory. On the contrary, we will focus on the theory $\CT_k := \CT_{(k,1,\frac{k}{2})}$ since it captures the subtleties of a general theory in the family, with the exception that we will also look at the theory $\CT_{(k,1,0)}$ with the purpose of comparing to results of \cite{Mikhaylov} on the 3d $B$-model with orbifold target $\C^2/\Z_k$.

\subsection{Deriving the boundary VOA from field theory}
\label{sec:fieldtheoryVOA}
In order to introduce a holomorphic boundary condition, we split the gauge fields $\CA, \CZ$ as
\be
	\CA^N = \BA^N + 2\pi \BB_E \qquad \CA^E = \BA^E + 2\pi \BB_N \qquad \CZ^\pm = \BZ^\pm \mp 2\pi \BPsi_\mp
\ee
as in \cite{AganagicCostelloMcNamaraVafa}, where $\BB, \BPsi$ contain the $\diff z$ parts of the connection 1-forms. We similarly split the exterior differential $\diff = \diff' + \pd$, where $\diff' = \pd_{\ol{z}} \diff \ol{z} + \pd_t \diff t$. Such a splitting is compatible with spacetime 3-manifold possessing a transverse holomorphic foliation. With this splitting, the above action takes the form
\be
\begin{aligned}
	S & = \int \BB_N \diff' \BA^N + \BB_E \diff' \BA^E + \BPsi_+ \diff'_\BA \BZ^+ + \BPsi_- \diff'_\BA \BZ^-\\
	& \qquad + \int \tfrac{1}{4\pi} \big(\BA^N \pd \BA^E + \BA^E \pd \BA^N + \BZ^- \pd \BZ^+ + \BZ^+ \pd \BZ^-) + \BB_E \BZ^+ \BZ^-
\end{aligned}
\ee

With this splitting, there is a natural holomorphic Dirichlet boundary condition where we require the boundary values of the gauge fields $\BA^\pm$ and $\BZ^\pm$ to vanish. We denote the boundary values of the remaining bosonic fields by $N, E$, respectively, and the fermionic fields $\psi_\pm$; the equations of motion of these fields imply that correlation functions involving these operators only depend holomophically on their insertion points. The quantum-corrected identification between the affine currents $J_a$ and the gauge fields $\BB^a_z$ is $B_a = K^{\textrm{eff}}_{ab} A^b_z$. Here we have 
\be
    K^{\textrm{eff}}_{ab} = \begin{pmatrix}
    		1 & 1\\ 1 & 0
    \end{pmatrix} \qquad \Rightarrow \qquad N := \BB_N = A^N_z + A^E_z,\, E := \BB_E = A^N_z.
\ee
The perturbative analyses of, e.g., Section 7.1 of \cite{CostelloDimofteGaiotto-boundary} or Section 3.2.2 of \cite{topCSM}, can then be applied to determine the OPEs of these perturbative local operators. Perhaps unsurprisingly, the result is a $\fgl(1|1)$ current algebra $V(\fgl(1|1))$:
\begin{gather}
	\label{eq:opegl11standard}
	N(z) N(w) \sim  \frac{1}{(z-w)^2} \qquad  N(z) E(w) \sim \frac{1}{(z-w)^2}\nonumber \\
	N(z) \psi_\pm(w) \sim \frac{\pm \psi_\pm(w)}{z-w}\\
	\psi_+(z) \psi_-(w) \sim \frac{1}{(z-w)^2} + \frac{E(w)}{z-w} \nonumber
\end{gather}

We now move to the full non-perturbative algebra of boundary local operators. As sketched in \cite[Section 7.2]{CostelloDimofteGaiotto-boundary}, this VOA should arise as Dolbeault homology of the affine Grassmannian $\textrm{Gr}_G$ with values in a certain associated VOA bundle. Thankfully, when $G$ is abelian the (closed points of the) affine Grassmannian is (are) simply a collection of isolated points, themselves identified with cocharacters of $G$, and one does not need to contend with the difficult geometry of these VOA bundles over $\textrm{Gr}_G$. Indeed, having fixed the bilinear form appearing in the above action, and hence the above perturbative algebra, the remaining data needed to determine the $U(1|1)$ Chern-Simons gauge theory is a choice of cocharacter lattice $\Lambda$. The pairing on the bosonic Lie algebra can be used to identify $\Lambda$ with a lattice in $\R^{1,1}$; ensuring the action $S$ is invariant under large gauge transformations requires $\Lambda$ to be integral. Moreover, the dual lattice to $\Lambda$, i.e. the character lattice, must contain the weights of the fermionic generators $\psi_\pm$. As shown in Section 4.3 of \cite{Mikhaylov}, any such lattice can be described by a triple of numbers $(\kappa, \nu, \xi)$, where $\kappa$ and $\nu$ are positive integers and $\xi$ is a half-integer defined mod $\kappa$. We denote the corresponding lattice $\Lambda_{(\kappa, \nu, \xi)}$, it is generated by the two vectors%
%
%\footnote{In the basis $v_1, v_2$ of the bosonic subalgebra adapted to $\CT_{(\kappa, \nu, \xi)}$, we find $U(1) \times U(1)$ Chern-Simons gauge fields $\CA_1, \CA_2$ with pairing $$(v_1, v_1) = 2 \xi \qquad (v_1, v_2) = (v_2, v_1) = \kappa$$ coupled to the hypermultiplets $\CZ^\pm$ of weights $(\nu,0)\oplus(-\nu,0)$.} %
%
\be
	v_1 = \tfrac{\kappa}{\nu} E \qquad v_2 = \tfrac{\xi}{\nu} E + \nu N~.
\ee

Under the identification of spectral flow and large gauge transformations, it's clear that the included spectral flow modules must be compatible with the global structure of the gauge group. For the case at hand, we are thus interested in the spectral flow modules for corresponding to
\be
	A^N_z \to A^N_z - \frac{\mathfrak{n} \nu}{z} \qquad A^E_z \to A^E_z - \frac{\mathfrak{m} \tfrac{\kappa}{\nu} + \mathfrak{n}\tfrac{\xi}{\nu}}{z}
\ee
for $\mathfrak{m},\mathfrak{n} \in \Z$. Translating this to the currents $N,E$, we find
\be
	N \to N - \frac{(\mathfrak{m}\tfrac{\kappa}{\nu} + \mathfrak{n}(\tfrac{\xi}{\nu}+\nu))}{z} \qquad E \to E - \frac{\mathfrak{n}\nu}{z}~.
\ee
In the traditional generators $\wt{N} = N - \tfrac{1}{2}E, E, \psi_\pm$ the basic spectral flows with $(\mathfrak{m}, \mathfrak{n}) = (1,0)$ and $(0,1)$ are given by
\be
\label{eq:spectral}
\begin{aligned}
	\sigma_1(\wt{N}) & = \wt{N} - \frac{\tfrac{\kappa}{\nu}}{z} \qquad & \sigma_1(E) & = E \qquad & \sigma_1(\psi_\pm) & = \psi_\pm\\
	\sigma_2(\wt{N}) & = \wt{N} - \frac{\tfrac{\xi}{\nu} + \tfrac{\nu}{2}}{z} \qquad & \sigma_2(E) & = E - \frac{\nu}{z} \qquad & \sigma_2(\psi_\pm) & = z^{\mp \nu} \psi_\pm
\end{aligned}
\ee
The full, non-perturbative algebra of local operators is then given by a sum over the lattice of spectral flows of the vacuum module.

It is worth verifying that the above space of triples $(\kappa, \nu, \xi)$ indeed captures the space of all extensions of $V(\fgl(1|1))$ by a rank-2 lattice of spectral flows of the vacuum. We first note that that the fusion is additive in $e$, so it suffices to consider the case where we look at the lattice generated by, say, $\widehat{A}_{n,e}$ and $\widehat{A}_{n',0}$ with $e$ nonzero, cf. Section 4.3 of \cite{Mikhaylov}. Fusing $\widehat{A}_{n,e}$ with $\widehat{A}_{n',0}$ shifts $n \to n+n'$ and so we will get the same lattice of spectral flow modules for $n$ and $n+n'$.

The constraint that the lattice of modules generated by these (and their contragredients) has the structure of a VOA is that this forms a commutative (super)algebra object and in particular they have trivial monodromy with themselves and one another.The monodromy of $\widehat{A}_{n,e}$ and $\widehat{A}_{n',e'}$ is given by (see Section \ref{sec:compmonodromy})
\be
    \exp\left(2\pi\left(e(n'+\tfrac{e'+1}{2}) + e'(n+\tfrac{e+1}{2})\right)\right)\,.
\ee
This constraint implies that both $e(2n + e - 1)$ and $n'e$ are integers. As $e$ is itself integral, we see that so too is $2ne$; when $2ne$ is odd we find that $\widehat{A}_{n,e}$ is mutually fermionic with itself and, accordingly, should be replaced by $\Pi \widehat{A}_{n,e}$. The space of rank-two extensions of $V(\fgl(1|1))$ is then parametrized by the integers $e$ and $n' e$ as well as the half-integer $n e$, defined modulo $n' e$. This is precisely the data needed to define the parent $U(1|1)$ Chern-Simons theory \cite{Mikhaylov}; tracing through the identification of parameters given above, we find
\be
    \kappa = n' e \qquad \nu = e \qquad \xi = e\left(n+\tfrac{e-1}{2}\right)~.
\ee

\subsection{Bulk local operators in $\CT_k$ from geometric quantization}
\label{sec:locopsGQ}
Unlike Chern-Simons theories based on a purely bosonic gauge group, supergroup Chern-Simons theories can admit non-trivial local operators. We will describe the vector space of local operators by way of a state-operator correspondence relating the vector space of state on a sphere $S^2$. There are several ways to derive this state space. In this section we determine it by geometric quantization of the phase space on a spacetime of the form $\CM = \R^3 - \{0\}$ in radial quantization. In Section \ref{sec:bdyVOA} we approach this same problem by way of a boundary vertex operator algebra (VOA).

If we identify $\R^3 \simeq \C \times \R$ then we can use the above description of $\CT_k$ and analysis of \cite{Zeng} to explicitly write down the vector space of local operators. The space of solutions to the equations of motion thus admits a decomposition into components labeled a given magnetic charge $(\mathfrak{m},\mathfrak{n}) \in \Z^2$, identified with the cocharacter $\mathfrak{m} E + \mathfrak{n} (N+\frac{1}{2}E)$.

In the perturbative sector $\mathfrak{m} = \mathfrak{n} = 0$, we find that local operators can be described by functions of the lowest components $c^N, b_N,c^E, b_E, Z^\pm, \psi_\pm$ subject to the following differential
\be
\begin{aligned}
	Q c^N & = 0 \qquad & Q b_N & = -\mu + \frac{k}{2\pi}\pd c^E\\
	Q c^E & = Z^- Z^+ \qquad & Q b_E & = \tfrac{k}{2\pi}\pd c^N\\
	Q Z^\pm & = \pm c^N Z^\pm \qquad & Q \psi_\pm & = \mp c^N \psi_\pm \mp \tfrac{k}{2\pi}\pd Z^\mp + b_E Z^\mp\\
\end{aligned}
\ee
It will be convenient to introduce the bosonic covariant derivative $\CD = \pd + \tfrac{2\pi}{k}b_E$, from which it follows that
\be
\begin{aligned}
	Q \CD^n c^N & = 0 \qquad & Q \CD^n b_N & = - \CD^n \mu_\pm + \tfrac{k}{2\pi}\CD^{n+1} c^E\\
	Q \CD^n c^E & = \CD^n(Z^- Z^+) \qquad & Q \CD^n b_E & = \tfrac{k}{2\pi}\CD^{n+1} c^N\\
	Q \CD^n Z^\pm & = \pm c^N \CD^n Z^\pm \qquad & Q \CD^n\psi_\pm & = \mp c^N \CD^n\psi_\pm \mp \tfrac{k}{2\pi}\CD^{n+1} Z^\mp\\
\end{aligned}
\ee

The perturbative part of the algebra of local operators is then generated by $\C^\times \times \C^\times$-invariant functions of the algebra generated by the bosonic fields $\CD^n Z^\pm$, $\CD^n b_N$ and $\CD^n b_E$ as well as the fermionic fields $\CD^n \psi_\pm$, $\CD^{n+1} c^N$ and $\CD^{n+1} c^E$ subject to the above differential. As described in \cite[Section 6.2]{CostelloDimofteGaiotto-boundary}, one should not include the zero-mode of the fermionic ghosts $c^N, c^E$ due to the fact that the group $G = \C^\times \times \C^\times$ of global gauge transformations is a reductive group and therefore taking $G$ invariants is an exact functor and should not be done with ghosts. More physically, the derivative $\pd c$ of the ghost is cohomologous to a physical gaugino; only the modes of this gaugino contribute to the algebra of local operators.

A straightforward computation shows that all $\C^\times \times \C^\times$-invariant and $Q$-closed functions involving $\CD$ are $Q$-exact, and the only remaining local operator is the moment map operator $\nu = Z^- Z^+$:
\be
\Ops^{(0,0)} = H^\bullet\bigg(\C[\CD^n Z^\pm, \CD^n b_N, \CD^n b_E, \CD^n \psi_\pm, \CD^{n+1} c^N, \CD^{n+1} c^E]^{\C^\times \times \C^\times}, Q\bigg) \simeq \C[\nu]\,.
\ee
Note that it is crucial that we ignore the zeromodes $c^N, c^E$ to obtain this result. A similar phenomenon happens in the $A$-twist of a standard $\CN=4$ gauge theory, where the complex scalar $\phi$ of the $\CN=4$ vector multiplet appears in the BV/BRST variation of the $c$-ghost. The perturbative algebra in that case is exactly $G$-invariant polynomials in the complex scalar. We will see that this analogy can be made quite sharp. Indeed, we will find that the moment map operator $\nu$ serves as the (a complex scalar component of the) current (supermultiplet) for the topological flavor symmetry measuring monopole charge $\mathfrak{m}$.

Now consider monopole sectors with general magnetic charge $(\mathfrak{m},\mathfrak{n})$. We can again apply the analysis of \cite{Zeng} to write down the vector space of such local operators. If $\mathfrak{n} \geq 0$, the vector space of local operators is then given by the $Q$ cohomology of the $\C^\times \times \C^\times$ subspace of
\be
	\C[\CD^{n+\mathfrak{n}} Z^+, \CD^{n} Z^-, \CD^n b_N, \CD^n b_E, \CD^{n} \psi_+, \CD^{n+\mathfrak{n}} \psi_-, \CD^{n+1} c^N, \CD^{n+1} c^E]v_{(\mathfrak{m}, \mathfrak{n})}
\ee
When $\mathfrak{n}<0$, we simply exchange $Z^+ \leftrightarrow Z^-$ and $\psi_+ \leftrightarrow \psi_-$ in this expression. The Chern-Simons terms induce an electric charge $(k (\mathfrak{m}+\frac{1}{2}\mathfrak{n}), k \mathfrak{n})$ on bare monopole operators. It immediately follows that there are no $\C^\times \times \C^\times$ invariant elements unless $\mathfrak{n}=0$, and therefore we only need to consider these cohomology groups, cf. \cite[Section 3.2]{KapustinSaulina-CSRW}. Up to multiplication by perturbative local operators, the $\C^\times \times \C^\times$ invariant operators of magnetic charge $(\mathfrak{m}, 0)$ are generated by the bare monopole $v_{(\mathfrak{m},0)}$ dressed by a degree $k \mathfrak{m}$ polynomial in $\CD^n Z^-$ and $\CD^n \psi_+$ (if $k \mathfrak{m} > 0$) or $\CD^n Z^+$ and $\CD^n \psi_-$ (if $k \mathfrak{m} < 0$). Just as in the perturbative sector, none of the operators involving the fields $\CD^n b_N$, $\CD^n b_E$, $\CD^{n+1}$, $c^N$, $\CD^{n+1}c^N$, $\CD^{n+1} Z^\pm$, or $\CD^n \psi_\pm$ survive cohomology and we find
\be
\label{eq:GQops}
\Ops^{(\mathfrak{m},\mathfrak{n})} = \begin{cases}
	\C[\nu] (Z^+)^{-\mathfrak{m} k} v_{(\mathfrak{m}, 0)} & \mathfrak{n} = 0, k \mathfrak{m} \leq 0\\
	\C[\nu] (Z^-)^{\mathfrak{m} k} v_{(\mathfrak{m}, 0)} & \mathfrak{n} = 0, k \mathfrak{m} \geq 0\\
	0 & \textrm{else}
\end{cases}
\ee

It is important to note that the above description does not account for the algebraic relations satisfied by these operators. Indeed, an explicit calculation of the Euler character of this state space, e.g. as an appropriate twisted index, yields
\be
\chi(\Ops) = \frac{k}{(1-(-1)^k y^2)(1-(-1)^k/y^2)}\,,
\ee
where $y$ is the fugacity for the topological flavor symmetry. Note that this exactly matches the $t \to -1$ limit of the Hilbert series for the ring of functions on the orbifold $\C^2/\Z_k$:
\be
HS(\C[\C^2/\Z_k]) = \frac{1-t^{2k}}{(1-t^k y^2)(1-t^2)(1-t^k y^{-2})}\,, \quad \chi(\Ops) = HS(\C[\C^2/\Z_k])\big|_{t \to -1}\,.
\ee
In Section \ref{sec:locops} we will show that this ring of functions arises as the algebra of local operators in two ways: exactly matches the algebra of self-extensions of the vacuum module of the boundary VOA introduced in Section \ref{sec:bdyVOA}.

\section{Representation theory of $\fgl(1|1)$ and $V(\fgl(1|1))$}
\label{sec:reptheory}

In this Appendix we review some general aspects of the representation theory of $\fgl(1|1)$ and $V(\fgl(1|1))$. In the case of $V(\fgl(1|1))$, we pay particular attention to how to realize these modules from the free-field realization presented in Section \ref{sec:freefield}. In Section \ref{sec:compmonodromy} we use these free-field modules to compute their monodromies with the simple currents $\widehat{A}_{n,e}$ of $V(\fgl(1|1))$.

\subsection{Representations and Verma modules}

Let us recall some facts about representations of $\fgl(1|1)$ and $V(\fgl(1|1))$, following \cite[Section 2.2]{CMY20}. Let us first consider $\fgl(1|1)$. Denote by $\CG$ the category of finite-dimensional modules of $\fgl(1|1)$. Since $E$ is central and any object $V$ in $\CG$ is finite dimensional, $V$ decomposes into generalized eigenspaces of $E$. We thus have a decomposition:
\be
\CG=\bigoplus_{e\in \mathbb{C}}\CG_e
\ee
where $\CG_e$ denotes the subcategory consisting of objects on which $E-e$ is nilpotent. For each $e$, we will denote by $V_{n,e}$ the two-dimensional Verma module of $\fgl(1|1)$ generated by a bosonic vector $v$ such that $\psi_+v=0$, $Nv=(n+\frac{1}{2})v$ and $Ev=ev$. The number $n$ denotes the average of the $N$ eigenvalues on $V_{n,e}$. We note that there are also Verma modules $\Pi V_{n,e}$ where the highest-weight vector $v$ is fermionic.

When $e\ne 0$, the two-dimensional module $V_{n+1/2, e}$ is irreducible; when $e=0$, they fit into short exact sequences:
\be
\begin{tikzcd}
	0 \rar  & \Pi A_{n-\frac{1}{2},0} \rar & V_{n, 0}\rar & A_{n+\frac{1}{2},0}\rar & 0
\end{tikzcd}
\ee
where $\Pi$ denotes a parity shift.
The modules $A_{n,0}$ are one-dimensional (hence simple) and characterized by an eigenvector of $N$ of weight $n$. The modules $V_{n, e}$ ($e \neq 0$) and $A_{n,0}$ are the only simples and generate the category $\CG$. These simple modules have interesting self extensions and extensions with each other, and it is not easy to characterize all the indecomposable objects in $\CG$. 

We now introduce generalized Verma modules that have ``as few relations as possible," such that any element in $\CG$ is a sub-quotient of a finite direct sum of such modules. For each $n,e\in \mathbb{C}$ and each $p,q\in \mathbb{N}$, we will define $V_{n,e}^{p,q}$ essentially by requiring that $N$ acts on the generator with generalized weight $n+1/2$ and Jordan blocks of size $p$, $E$ acts on the generator with generalized weight $e$ and Jordan block of size $q$, as well as that the generator is a highest-weight vector.

Let $v$ be a vector. The vector space $V_{n,e}^{p,q}$ is spanned by the following vectors:
\be
	(N-n-\frac{1}{2})^s(E-e)^t v,~ \psi_-(N-n-\frac{1}{2})^s(E-e)^tv, 
\ee
where $0\leq s\leq p-1$ and $0\leq t\leq q-1$. The action of $\psi_+$ is defined by $\psi_+v=0$ and commutation relations. Clearly there is an embedding $V_{n,e}\to V_{n,e}^{p,q}$ whose image is generated by the vector $(N-n-\frac{1}{2})^{p-1}(E-e)^{q-1}v$. 

We claim that any object $M\in \CG$ is a sub-quotient (namely a quotient of a submodule) of a finite direct sum of $V_{n,e}^{p,q}$ for various $n,e,p$ and $q$. This will allow us to compute monodromy in the next section. To prove this statement, we may assume WLOG that $M$ is generated by a single element $m\in M$ that is a generalized eigenvector of $N$ with weight $n+1/2$ and $E$ with weight $e$. Assume that $(N-n-1/2)^pm=(E-e)^qm=0$. Using the commutation relation, it is clear that $M$ is spanned by vectors of the following form:
\be\label{eq:generatorM}
\begin{aligned}
&(N-n-\frac{1}{2})^s(E-e)^t m,~ \psi_-(N-n-\frac{1}{2})^s(E-e)^tm, ~\\ & \psi_+(N-n-\frac{1}{2})^s(E-e)^tm,~
 \psi_-\psi_+(N-n-\frac{1}{2})^s(E-e)^tm,\qquad (s<p, t<q)
\end{aligned}
\ee
Let us consider the direct sum $V_{n,e}^{p,q}\oplus V_{n+1, e}^{p,q+1}$. Denote by $v$ and $w$ the generators of the two copies of the Verma modules. Note that the element $v$ and $w$ both satisfies $\psi^+v=\psi^+w=0$. Consider the element $u=v+\psi^-w$. This satisfies that $(N-n-1/2)^{p}u=(E-e)^{q+1}u=0$, and moreover, $\psi^-u=\psi^-v$ and $\psi^+u=Ew=(E-e)w+ew$. Let $U$ be the submodule of $V_{n,e}^{p,q}\oplus V_{n+1, e}^{p,q+1}$ generated by $u$, then clearly it is also spanned by a similar set of vectors:
\be\label{eq:generatorU}
\begin{aligned}
&(N-n-\frac{1}{2})^s(E-e)^t u,~ \psi_-(N-n-\frac{1}{2})^s(E-e)^tu, ~\\ & \psi_+(N-n-\frac{1}{2})^s(E-e)^tu,~
 \psi_-\psi_+(N-n-\frac{1}{2})^s(E-e)^tu,\qquad (s<p, t<q)\\
 & (N-n-\frac{1}{2})^s (E-e)^q u,\qquad (s<p)
\end{aligned}
\ee
But the difference here is that these vectors are linearly independent, since they are inside $V_{n,e}^{p,q}\oplus V_{n+1, e}^{p,q+1}$. Therefore, one can produce a map $U\to M$ by sending $u\to m$, sending the corresponding vectors in $U$ in equation \eqref{eq:generatorU} to those in $M$ in equation \eqref{eq:generatorM} for $s<p$ and $t<q$, and sending those vectors in the last line of equation \eqref{eq:generatorU} to zero. It is well-defined thanks to linear-independence, and is clearly a $\fgl(1|1)$ homomorphism because the action of $\fgl(1|1)$ on both $U$ and $M$ are given by commutation relations on these two sets of vectors. We thus produced a surjection from a submodule of $V_{n,e}^{p,q}\oplus V_{n+1, e}^{p,q+1}$ to $M$, making $M$ a sub-quotient of two copies of the generalized Verma modules.

Let us now turn to $V(\fgl(1|1))$ and its Kazhdan-Lusztig category $KL$. Since $E_0$ is still central, we have a decomposition:
\be
KL=\bigoplus_{e\in \mathbb{C}} KL_e.
\ee
For each $M \in \CG$, one can build a highest-weight module $\widehat{M}$ of $V(\fgl(1|1))$, which is called the induction of $M$. Induction is clearly a functor:
\be\label{eq:GKLequiv}
\mathrm{Ind}: \CG_e\longrightarrow KL_e.
\ee
It was proved in \cite{BN22} that when $e=0$ or $e \notin \mathbb{Z}\backslash\{0\}$, the functor $\mathrm{Ind}$ in the above equation is an equivalence of categories.  This means that generically, one can study the category $KL$ simply by studying the category $\CG$. For $e \in \Z$, one can study $KL_e$ via fusion with a simple current $\widehat{A}_{0,-e}$, to be described below; $\widehat{A}_{0,-e} \times -$ gives an equivalence between $KL_e$ and $KL_0$. 

As a consequence of these equivalences, the induction of a Verma module $\widehat{V}_{n, e}$ is still simple for $e\notin \mathbb{Z}$. When $e=0$, we still have a short exact sequence:
\be
\begin{tikzcd}
	0 \rar  & \Pi \widehat{A}_{n-\frac{1}{2},0} \rar &\widehat{V}_{n, 0}\rar & \widehat{A}_{n+\frac{1}{2},0}\rar & 0
\end{tikzcd}
\ee
When $e\in \mathbb{Z}$, the module $\widehat{V}_{n, e}$ is no longer irreducible, and fits into one of the following short exact sequences:
\be
\begin{tikzcd}
	0\rar & \Pi \widehat{A}_{n+1, e} \rar & \widehat{V}_{n,e} \rar & \widehat{A}_{n,e}\rar & 0~ \text{ if } e>0\\
	0\rar & \Pi \widehat{A}_{n-1, e} \rar & \widehat{V}_{n,e} \rar & \widehat{A}_{n,e}\rar & 0~ \text{ if } e<0
\end{tikzcd}
\ee
The modules $\widehat{A}_{n,e}$ can be obtained as spectral flows of the vacuum module $\widehat{A}_{0,0}$ and they realize simple currents for $\mathbb{Z}\times \mathbb{C}$, where $n\in \mathbb{C}$ can be an arbitrary complex number. The lattice of simple currents we used to extend from $V(\fgl(1|1))$ to $\CV_{k}$ is generated by $\Pi^{k+1}\widehat{A}_{\scriptstyle{\frac{k}{2}}, 1}$ and $\widehat{A}_{k, 0}$, as well as their contragredients, and can be embedded into the Verma modules $\Pi^{k} \widehat{V}_{\scriptstyle{\frac{k}{2}}-1, 1}$ and $\widehat{V}_{k-\scriptstyle{\frac{1}{2}}, 0}$, respectively. 

It turns out that $\widehat{V}_{n,e}^{p,q}$ can be seen from free field realizations.%
\footnote{Strictly speaking, we can realize a fixed parity of this module with free fields. From a physical perspective, it is not clear whether one should restrict to the parities that arise in the free field realization, so we will consider both parities in our analysis.} %
Consider the free field realization of $V(\fgl(1|1))$ in terms of chiral bosons $X, Y$, and $Z$ as in Section \ref{sec:freefield}. For each $\mu=aX+bY$, consider the module of the VOA $\CV_{Z}$ generated by:
\be
v=Y^{p-1}X^{q-1}\vert aX+b Y\rangle. 
\ee
We denote this by $\CV_{Z,\mu}^{p,q}$. Define a grading $\Delta$ on $\CV_{Z,\mu}^{p,q}$ by:
\be
\Delta (v)=\Delta (c_{-1}v)=0,\qquad \Delta (b_{-1}v)=\Delta (X_{-1}v)=\Delta (Y_{-1}v)=1.
\ee
Here $b(z)=\norm{e^Z}$ and $c(z)=\norm{e^{-Z}}$. Moreover, $\Delta (A_{n}B)=\Delta (A)+\Delta (B)-n-1$. The module $\CV_{Z,\mu}^{p,q}$ under this grading is positively graded, and the minimal degree part of $\CV_{Z,\mu}^{p,q}$ under this is spanned by vectors of the form:
\be
Y^{t}X^{s}\vert aX+b Y\rangle, \qquad c_{-1}Y^{t}X^{s}\vert aX+b Y\rangle, \qquad s\leq p-1, t\leq q-1.
\ee
Consider the vector $c_{-1}v$. We claim that $\psi_{0}^-c_{-1}v=0$. Indeed, by definition, since $\Delta (\psi_{0}^-c_{-1}v)=0$ the vector $\psi_{0}^-c_{-1}v$ is a linear combination of the  vectors in the above equation. However, vectors in the above equation have generalized eigenvalue $b-a/2-1$ or $b-a/2$ under the action of $\wt{N}_0$, while $\psi_{0}^-c_{-1}v$ is of generalized eigenvalue $b-a/2-2$. Therefore it has to be zero. Consequently, the $\fgl(1|1)$ module generated by $c_{-1}v$ is a copy of $V_{b-\scriptstyle{\frac{a+1}{2}}, a}^{p,q}$. From the universal property of induction functor, we obtain a morphism:
\be\label{eqfreefieldverma}
	\widehat{V}_{b-\scriptstyle{\frac{a+1}{2}}, a}^{p,q}\longrightarrow \CV_{Z,\mu}^{p,q}.
\ee
When $a\notin \Z$ or $a=0$, one can show, following a similar method as in \cite[Proposition 2.4]{creutzig2021duality}, that this is an isomorphism of modules. 

Using this, we claim that any module of $V(\fgl(1|1))$ is a quotient of a finite direct sum of $\CV_{Z,\mu}^{p,q}$. Let $M$ be a module such that $E_0$ acts with generalized weight $e$. If $e=0$ or $e\notin \Z$ then the statement is true since we have the equivalence of Eq. \eqref{eq:GKLequiv}. When $e$ is a non-zero integer, then we can consider the module $\sigma_{0, -e}(M)$, on which $E_0$ has generalized eigenvalue $0$. Using the above arguments, we can show that $\sigma_{0, -e}(M)$ is a quotient of a finite direct sum of  $\CV_{Z,\mu}^{p,q}$, for $\mu=bY$. We simply need to show that $\sigma_{0, e}(\CV_{Z,\mu}^{p,q})$ is also of this form. This is clear since the spectral flow $\sigma_{0, e}(\CV_{Z,\mu}^{p,q})$ can be alternatively defined as the module of $\CV_Z$ generated by $X^{p-1}Y^{q-1}\vert e(X-\frac{1}{2}Y+Z)+bY\rangle$, which is isomorphic to the module $\CV_{Z,\mu'}^{p,q}$ where $\mu':=\vert e(X-\frac{1}{2}Y)+bY\rangle$.

\subsection{Computing monodromy}
\label{sec:compmonodromy}

We now use the isomorphism of Eq. \eqref{eqfreefieldverma} to compute monodromy with the simple currents $\widehat{A}_{n,e}$. We will focus on the example of $\Pi^{k+1} \widehat{A}_{\scriptstyle{\frac{k}{2}}, 1}$. We have claimed that having trivial monodromy with $\Pi^{k+1} \widehat{A}_{\scriptstyle{\frac{k}{2}}, 1}$ is equivalent to that $N_0+k E_0/2$ acts semi-simply with integer eigenvalues. Let $V$ be any object in $KL$. We will in fact show that the morphism:
\be
\begin{tikzcd}
	\Pi^{k+1}\widehat{A}_{\scriptstyle{\frac{k}{2}}, 1}\times V\rar{R} & V\times \Pi^{k+1}\widehat{A}_{\scriptstyle{\frac{k}{2}}, 1}\rar{R} & \Pi^{k+1}\widehat{A}_{\scriptstyle{\frac{k}{2}}, 1}\times V
\end{tikzcd}
\ee
is equal to $\mathrm{Id}\times \exp(2\pi i (N_0+\tfrac{k}{2}E_0))$. 

Let us comment that to show this, it is enough to show this for $V = \CV_{Z,\mu}^{p,q}$ since any other object is a subquotient, and monodromy is functorial with respect to morphisms. The module $\Pi^{k+1}\widehat{A}_{\scriptstyle{\frac{k}{2}}, 1}$ is the unique submodule of $\Pi^k \CV_{Z,X+\scriptstyle{\frac{k}{2}}Y}$, and for any $\CV_{Z,\mu}^{p,q}$, fusion over free field algebra gives an equivalence:
\be
	\Pi^{k+1}\widehat{A}_{\scriptstyle{\frac{k}{2}}, 1}\times  \CV_{Z,\mu}^{p,q}\cong \Pi^k\CV_{Z,X+\scriptstyle{\frac{k}{2}}Y}\times \CV_{Z,\mu}^{p,q}. 
\ee
We can then compute monodromy using the explicit free field intertwining operator. Let us denote by $\ket{\lambda_+}$ the vector corresponding to $X+\tfrac{k}{2}Y+Z$ and $\ket{\mu}$ the vector corresponding to $aX+bY$. Recall that the module $ \CV_{Z,\mu}^{p,q}$ is generated by the vector $Y^{p-1}X^{q-1}\ket{\mu}$. Applying the free field intertwining operator on this vector gives:
\be
\CY(\ket{\lambda_+}, z)Y^{p-1}X^{q-1}\ket{\mu}=e^{X+\tfrac{k}{2}Y+Z} z^{X_0+\tfrac{k}{2}Y_0}  \exp (\cdots) \cdot Y^{p-1}X^{q-1}\ket{\mu}. 
\ee
Here $\exp(\cdots)$ are the normal-ordered products of other modes. They do not contribute to logarithmic terms and so we omit them from the expression. The logarithmic part of this comes from:
\be
\exp\left({ \log (z) }(X_0+\tfrac{k}{2}Y_0+Z_0)\right) Y^{p-1}X^{q-1}\ket{\mu}. 
\ee
Since $[X_0, Y]=[Y_0,X]=1$, we find, after commuting the exponential with $Y^{p-1}X^{q-1}$, the following expression:
\be
(Y+ \log(z))^{p-1}(X+\tfrac{k}{2}\log (z))^{q-1} \exp\left({ \log (z) }(X_0+\tfrac{k}{2}Y_0+Z_0)\right)\ket{\mu}.
\ee
Now since $\mu=aX+bY$, the above is simply:
\be
(Y+ \log(z))^{p-1}(X+\tfrac{k}{2}\log (z))^{q-1} \exp\left({ \log (z) }(b+\tfrac{k}{2}a)\right)\ket{\mu}.
\ee
Double braiding is expressed in logarithmic intertwining operators as mapping $z\mapsto e^{2\pi i}z$. It has the effect of turning $\log(z)\mapsto \log (z)+2\pi i$, which turns the above into:
\be
(Y+ \log(z)+2\pi i)^{p-1}(X+\tfrac{k}{2}\log (z)+k\pi i)^{q-1} \exp\left({ \log (z) }(b+\tfrac{k}{2}a)\right)\exp (2\pi i (b+\tfrac{k}{2}a))\ket{\mu}.
\ee 
The effect of this can be accounted for by the application of $\exp (2\pi i (N_0+\tfrac{k}{2}E_0))$:
\be
\exp (2\pi i (N_0+\tfrac{k}{2}E_0)) Y^{p-1}X^{q-1} \ket{\mu}=(Y+2\pi i)^{p-1}(X+k\pi i)^{q-1}\exp (2\pi i (b+\tfrac{k}{2}a))\ket{\mu}.
\ee
This implies the following identity:
\be
\CY(\ket{\lambda}, e^{2\pi i}z) Y^{p-1}X^{q-1} \ket{\mu}=\CY(\ket{\lambda}, z)  \exp (2\pi i (N_0+\tfrac{k}{2}E_0)) Y^{p-1}X^{q-1} \ket{\mu},
\ee
which is the statement that monodromy is given by $\mathrm{Id}\times \exp (2\pi i (N_0+\tfrac{k}{2}E_0))$. 

\section{Declarations} 

\begin{itemize}
    \item The authors have no relevant financial or non-financial interests to disclose.

    \item The first author acknowledges support from the University of Washington.

    \item The authors have no financial or proprietary interests in any material discussed in this article.

    \item The authors declare that the data supporting the findings of this study are available within the paper. 
    
\end{itemize}

\bibliography{sn-bibliography}% common bib file

@article{Garner:2024yin,
    author = "Garner, Niklas and Geer, Nathan and Young, Matthew B.",
    title = "{B-Twisted Gaiotto{\textendash}Witten Theory and Topological Quantum Field Theory}",
    eprint = "2401.16192",
    archivePrefix = "arXiv",
    primaryClass = "math.RT",
    doi = "10.1007/s00220-024-05211-3",
    journal = "Commun. Math. Phys.",
    volume = "406",
    number = "2",
    pages = "29",
    year = "2025"
}

@article{Lentner:2025hae,
    author = "Lentner, Simon D.",
    title = "{A conditional algebraic proof of the logarithmic Kazhdan-Lusztig correspondence}",
    eprint = "2501.10735",
    archivePrefix = "arXiv",
    primaryClass = "math.QA",
    month = "1",
    year = "2025"
}

@article{Feigin:2025gfj,
    author = "Feigin, Boris L. and Lentner, Simon D.",
    title = "{Coupling a vertex algebra to a large center}",
    eprint = "2504.12808",
    archivePrefix = "arXiv",
    primaryClass = "math.QA",
    month = "4",
    year = "2025"
}

@article{HLZ1,
	author = "Huang, Yi-Zhi and Lepowsky, James and Zhang, Lin",
	title = "{Logarithmic Tensor Category Theory for Generalized Modules for a Conformal Vertex Algebra, I: Introduction and Strongly Graded Algebras and their Generalized Modules}",
	eprint = "1012.4193",
	archivePrefix = "arXiv",
	primaryClass = "math.QA",
	month = "12",
	year = "2010"
}

@article{HLZ2,
	author = "Huang, Yi-Zhi and Lepowsky, James and Zhang, Lin",
	title = "{Logarithmic Tensor Category Theory, II: Logarithmic Formal Calculus and Properties of Logarithmic Intertwining Operators}",
	eprint = "1012.4196",
	archivePrefix = "arXiv",
	primaryClass = "math.QA",
	month = "12",
	year = "2010"
}

@article{creutzig2024KL,
    author = "Creutzig, Thomas and Niu, Wenjun",
    title = "{Kazhdan-Lusztig correspondence for vertex operator superalgebras from abelian gauge theories}",
    eprint = "2403.02403",
    archivePrefix = "arXiv",
    primaryClass = "hep-th",
    doi = "10.1112/jlms.70328",
    journal = "J. Lond. Math. Soc.",
    volume = "112",
    number = "4",
    pages = "e70328",
    year = "2025"
}

@article{HLZ3,
	author = "Huang, Yi-Zhi and Lepowsky, James and Zhang, Lin",
	title = "{Logarithmic Tensor Category Theory, III: Intertwining Maps and Tensor Product Bifunctors}",
	eprint = "1012.4197",
	archivePrefix = "arXiv",
	primaryClass = "math.QA",
	month = "12",
	year = "2010"
}

@article{HLZ4,
	author = "Huang, Yi-Zhi and Lepowsky, James and Zhang, Lin",
	title = "{Logarithmic Tensor Category Theory, IV: Constructions of Tensor Product Bifunctors and the Compatibility Conditions}",
	eprint = "1012.4198",
	archivePrefix = "arXiv",
	primaryClass = "math.QA",
	month = "12",
	year = "2010"
}

@article{HLZ5,
	author = "Huang, Yi-Zhi and Lepowsky, James and Zhang, Lin",
	title = "{Logarithmic Tensor Category Theory, V: Convergence Condition for Intertwining Maps and the Corresponding Compatibility Condition}",
	eprint = "1012.4199",
	archivePrefix = "arXiv",
	primaryClass = "math.QA",
	month = "12",
	year = "2010"
}

@article{HLZ6,
	author = "Huang, Yi-Zhi and Lepowsky, James and Zhang, Lin",
	title = "{Logarithmic Tensor Category Theory, VI: Expansion Condition, Associativity of Logarithmic Intertwining Operators, and the Associativity Isomorphisms}",
	eprint = "1012.4202",
	archivePrefix = "arXiv",
	primaryClass = "math.QA",
	month = "12",
	year = "2010"
}

@article{HLZ7,
	author = "Huang, Yi-Zhi and Lepowsky, James and Zhang, Lin",
	title = "{Logarithmic Tensor Category Theory, VII: Convergence and Extension Properties and Applications to Expansion for Intertwining Maps}",
	eprint = "1110.1929",
	archivePrefix = "arXiv",
	primaryClass = "math.QA",
	month = "10",
	year = "2011"
}

@article{HLZ8,
	author = "Huang, Yi-Zhi and Lepowsky, James and Zhang, Lin",
	title = "{Logarithmic Tensor Category Theory, VIII: Braided Tensor Category Structure on Categories of Generalized Modules for a Conformal Vertex Algebra}",
	eprint = "1110.1931",
	archivePrefix = "arXiv",
	primaryClass = "math.QA",
	month = "10",
	year = "2011"
}

@article{Nagatomo:2009xp,
	author = "Nagatomo, Kiyokazu and Tsuchiya, Akihiro",
	title = "{The Triplet Vertex Operator Algebra $W_{(p)}$ and the Restricted Quantum Group $\bar U_{q}(sl_2)$ at $q = e ^{\pi i/p}$}",
	eprint = "0902.4607",
	archivePrefix = "arXiv",
	primaryClass = "math.QA",
	month = "2",
	year = "2009"
}

@misc{mcrae2021structure,
	title={Structure of Virasoro tensor categories at central charge $13-6p-6p^{-1}$ for integers $p > 1$}, 
	author={Robert McRae and Jinwei Yang},
	year={2021},
	eprint={2011.02170},
	archivePrefix={arXiv},
	primaryClass={math.QA}
}

@article {KondoSaito,
	AUTHOR = {Kondo, Hiroki and Saito, Yoshihisa},
	TITLE = {Indecomposable decomposition of tensor products of modules
	over the restricted quantum universal enveloping algebra
	associated to {${\mathfrak{sl}}_2$}},
	JOURNAL = {J. Algebra},
	FJOURNAL = {Journal of Algebra},
	VOLUME = {330},
	YEAR = {2011},
	PAGES = {103--129},
	ISSN = {0021-8693},
	MRCLASS = {17B37},
	MRNUMBER = {2774620},
	MRREVIEWER = {Peter W. Tingley},
	DOI = {10.1016/j.jalgebra.2011.01.010},
	URL = {https://doi-org.login.ezproxy.library.ualberta.ca/10.1016/j.jalgebra.2011.01.010},
}

@article{Gainutdinov:2015lja,
	author = "Gainutdinov, A. M. and Runkel, I.",
	title = "{Symplectic fermions and a quasi-Hopf algebra structure on $\bar{U}_i sl(2)$}",
	eprint = "1503.07695",
	archivePrefix = "arXiv",
	primaryClass = "math.QA",
	doi = "10.1016/j.jalgebra.2016.11.026",
	journal = "J. Algebra",
	volume = "476",
	pages = "415--458",
	year = "2017"
}

@article{Creutzig:2020zvv,
	author = "Creutzig, Thomas and Jiang, Cuipo and Orosz Hunziker, Florencia and Ridout, David and Yang, Jinwei",
	title = "{Tensor categories arising from the Virasoro algebra}",
	eprint = "2002.03180",
	archivePrefix = "arXiv",
	primaryClass = "math.RT",
	doi = "10.1016/j.aim.2021.107601",
	journal = "Adv. Math.",
	volume = "380",
	pages = "107601",
	year = "2021"
}

@article {GPT,
	AUTHOR = {Geer, Nathan and Patureau-Mirand, Bertrand and Turaev,
	Vladimir},
	TITLE = {Modified quantum dimensions and re-normalized link invariants},
	JOURNAL = {Compos. Math.},
	FJOURNAL = {Compositio Mathematica},
	VOLUME = {145},
	YEAR = {2009},
	NUMBER = {1},
	PAGES = {196--212},
	ISSN = {0010-437X},
	MRCLASS = {57M27 (17B37 18D10)},
	MRNUMBER = {2480500},
	MRREVIEWER = {Daniel David Moskovich},
	DOI = {10.1112/S0010437X08003795},
	URL = {https://doi.org/10.1112/S0010437X08003795},
}

@article{DGGPR,
    author = "De Renzi, Marco and Gainutdinov, Azat M. and Geer, Nathan and Patureau-Mirand, Bertrand and Runkel, Ingo",
    title = "{3-Dimensional TQFTs from non-semisimple modular categories}",
    eprint = "1912.02063",
    archivePrefix = "arXiv",
    primaryClass = "math.GT",
    doi = "10.1007/s00029-021-00737-z",
    journal = "Selecta Math.",
    volume = "28",
    number = "2",
    pages = "42",
    year = "2022"
}

@misc{derenzi2021extended,
	title={Extended TQFTs From Non-Semisimple Modular Categories}, 
	author={Marco De Renzi},
	year={2021},
	eprint={2103.04724},
	archivePrefix={arXiv},
	primaryClass={math.GT}
}

@article{ben2012loop,
	author = {Ben-Zvi, David and Nadler, David},
	journal = {Journal of Topology},
	number = {2},
	pages = {377--430},
	publisher = {Oxford University Press},
	title = {Loop spaces and connections},
	volume = {5},
	year = {2012}}

@article{creutzig2018logarithmic,
	author = {Creutzig, Thomas and Milas, Antun and Rupert, Matt},
	journal = {Journal of Pure and Applied Algebra},
	number = {10},
	pages = {3224--3247},
	publisher = {Elsevier},
	title = {Logarithmic link invariants of $U_q^H (\mathfrak{sl}(2))$ and asymptotic dimensions of singlet vertex algebras},
	volume = {222},
	year = {2018}}

@article{creutzig2014false,
	author = {Creutzig, Thomas and Milas, Antun},
	journal = {Advances in Mathematics},
	pages = {520--545},
	publisher = {Elsevier},
	title = {False theta functions and the Verlinde formula},
	volume = {262},
	year = {2014}}

@article{costantino2015some,
	author = {Costantino, Francesco and Geer, Nathan and Patureau-Mirand, Bertrand},
	journal = {Journal of Pure and Applied Algebra},
	number = {8},
	pages = {3238--3262},
	publisher = {Elsevier},
	title = {Some remarks on the unrolled quantum group of sl (2)},
	volume = {219},
	year = {2015}}

@article{creutzig2021duality,
	author = {Creutzig, Thomas and Genra, Naoki and Nakatsuka, Shigenori},
	journal = {Advances in Mathematics},
	pages = {107685},
	publisher = {Elsevier},
	title = {Duality of subregular W-algebras and principal W-superalgebras},
	volume = {383},
	year = {2021}}

@article{gammage2022betti,
	author = {Gammage, Benjamin and Hilburn, Justin},
	journal = {arXiv preprint arXiv:2210.06548},
	title = {Betti Tate's thesis and the trace of perverse schobers},
	year = {2022}}

@article{ben2010integral,
	author = {Ben-Zvi, David and Francis, John and Nadler, David},
	journal = {Journal of the American Mathematical Society},
	number = {4},
	pages = {909--966},
	title = {Integral transforms and Drinfeld centers in derived algebraic geometry},
	volume = {23},
	year = {2010}}

@article{descent,
    author = "Beem, Christopher and Ben-Zvi, David and Bullimore, Mathew and Dimofte, Tudor and Neitzke, Andrew",
    title = "{Secondary products in supersymmetric field theory}",
    eprint = "1809.00009",
    archivePrefix = "arXiv",
    primaryClass = "hep-th",
    doi = "10.1007/s00023-020-00888-3",
    journal = "Annales Henri Poincare",
    volume = "21",
    number = "4",
    pages = "1235--1310",
    year = "2020"
}

@incollection{KashaevReshetikhin,
	author = {Kashaev, R. and Reshetikhin, N.},
	booktitle = {Graphs and patterns in mathematics and theoretical physics},
	doi = {10.1090/pspum/073/2131015},
	mrclass = {57M27 (18D10 20G42 57R56)},
	mrnumber = {2131015},
	mrreviewer = {Justin Sawon},
	pages = {151--172},
	publisher = {Amer. Math. Soc., Providence, RI},
	series = {Proc. Sympos. Pure Math.},
	title = {Invariants of tangles with flat connections in their complements},
	url = {https://doi.org/10.1090/pspum/073/2131015},
	volume = {73},
	year = {2005}}

@article{RW,
	archiveprefix = {arXiv},
	author = {Rozansky, L. and Witten, Edward},
	doi = {10.1007/s000290050016},
	eprint = {hep-th/9612216},
	journal = {Selecta Math.},
	pages = {401-458},
	primaryclass = {hep-th},
	reportnumber = {IASSNS-HEP-96-128},
	slaccitation = {%%CITATION = HEP-TH/9612216;%%},
	title = {{HyperKahler geometry and invariants of three manifolds}},
	volume = {3},
	year = {1997}}

@article{FGR,
	author = {Feigin, Boris and Gukov, Sergei and Reshetikhin, Nikolai},
	journal = {https://www.birs.ca/events/2021/5-day-workshops/21w5121/schedule},
	title = {{talks at the BIRS workshop Quantum Field Theories and Quantum Topology Beyond Semisimplicity}},
	year = {2021}}

@article{CostantinoGukovPutrov,
    author = "Costantino, Francesco and Gukov, Sergei and Putrov, Pavel",
    title = "{Non-Semisimple TQFT's and BPS $q$-Series}",
    eprint = "2107.14238",
    archivePrefix = "arXiv",
    primaryClass = "math.GT",
    doi = "10.3842/SIGMA.2023.010",
    journal = "SIGMA",
    volume = "19",
    pages = "010",
    year = "2023"
}

@article{GHNPPS,
    author = "Gukov, Sergei and Hsin, Po-Shen and Nakajima, Hiraku and Park, Sunghyuk and Pei, Du and Sopenko, Nikita",
    title = "{Rozansky-Witten geometry of Coulomb branches and logarithmic knot invariants}",
    eprint = "2005.05347",
    archivePrefix = "arXiv",
    primaryClass = "hep-th",
    doi = "10.1016/j.geomphys.2021.104311",
    journal = "J. Geom. Phys.",
    volume = "168",
    pages = "104311",
    year = "2021"
}

@article{ADO,
	author = {Akutsu, Yasuhiro and Deguchi, Tetsuo and Ohtsuki, Tomotada},
	doi = {10.1142/S0218216592000094},
	fjournal = {Journal of Knot Theory and its Ramifications},
	issn = {0218-2165},
	journal = {J. Knot Theory Ramifications},
	mrclass = {57M25},
	mrnumber = {1164114},
	number = {2},
	pages = {161--184},
	title = {Invariants of colored links},
	url = {https://doi.org/10.1142/S0218216592000094},
	volume = {1},
	year = {1992}}

@article{Hennings,
	author = {Hennings, Mark},
	doi = {10.1112/jlms/54.3.594},
	fjournal = {Journal of the London Mathematical Society. Second Series},
	issn = {0024-6107},
	journal = {J. London Math. Soc. (2)},
	mrclass = {57M25 (57N10)},
	mrnumber = {1413901},
	mrreviewer = {Louis H. Kauffman},
	number = {3},
	pages = {594--624},
	title = {Invariants of links and {$3$}-manifolds obtained from {H}opf algebras},
	url = {https://doi.org/10.1112/jlms/54.3.594},
	volume = {54},
	year = {1996}}

@article{Lyubashenko,
	author = {Lyubashenko, Volodymyr V.},
	fjournal = {Communications in Mathematical Physics},
	issn = {0010-3616},
	journal = {Comm. Math. Phys.},
	mrclass = {57N10 (17B37)},
	mrnumber = {1354257},
	number = {3},
	pages = {467--516},
	title = {Invariants of {$3$}-manifolds and projective representations of mapping class groups via quantum groups at roots of unity},
	url = {http://projecteuclid.org/euclid.cmp/1104274312},
	volume = {172},
	year = {1995}}

@article{CLR,
	archiveprefix = {arXiv},
	author = {Creutzig, Thomas and Lentner, Simon and Rupert, Matthew},
	eprint = {2104.13262},
	month = {4},
	primaryclass = {math.QA},
	title = {{Characterizing braided tensor categories associated to logarithmic vertex operator algebras}},
	year = {2021}}

@article{GN,
	archiveprefix = {arXiv},
	author = {Gannon, Terry and Negron, Cris},
	eprint = {2104.12821},
	month = {4},
	primaryclass = {math.QA},
	title = {{Quantum SL(2) and logarithmic vertex operator algebras at (p,1)-central charge}},
	year = {2021}}

@article{FT,
	archiveprefix = {arXiv},
	author = {Feigin, B. L. and Tipunin, I. Yu.},
	eprint = {1002.5047},
	month = {2},
	primaryclass = {math.QA},
	title = {{Logarithmic CFTs connected with simple Lie algebras}},
	year = {2010}}

@article{Kausch,
	author = {Kausch, H. G.},
	doi = {10.1016/0370-2693(91)91655-F},
	journal = {Phys. Lett. B},
	pages = {448--455},
	reportnumber = {DAMTP-90-27, NSF-ITP-90-200},
	title = {{Extended conformal algebras generated by a multiplet of primary fields}},
	volume = {259},
	year = {1991}}

@article{FGST1,
	archiveprefix = {arXiv},
	author = {Feigin, B. L. and Gainutdinov, A. M. and Semikhatov, A. M. and Tipunin, I. Yu.},
	doi = {10.1007/s11232-006-0113-6},
	eprint = {math/0512621},
	journal = {Theor. Math. Phys.},
	pages = {1210--1235},
	title = {{Kazhdan-Lusztig correspondence for the representation category of the triplet W-algebra in logarithmic CFT}},
	volume = {148},
	year = {2006}}

@article{FGST2,
	archiveprefix = {arXiv},
	author = {Feigin, B. L. and Gainutdinov, A. M. and Semikhatov, A. M. and Tipunin, I. Yu.},
	doi = {10.1016/j.nuclphysb.2006.09.019},
	eprint = {hep-th/0606196},
	journal = {Nucl. Phys. B},
	pages = {303--343},
	title = {{Logarithmic extensions of minimal models: Characters and modular transformations}},
	volume = {757},
	year = {2006}}

@article{RT,
	author = {Reshetikhin, N. and Turaev, V. G.},
	doi = {10.1007/BF01239527},
	fjournal = {Inventiones Mathematicae},
	issn = {0020-9910},
	journal = {Invent. Math.},
	mrclass = {57N10 (17B37 57M25 81R50)},
	mrnumber = {1091619},
	mrreviewer = {Louis H. Kauffman},
	number = {3},
	pages = {547--597},
	title = {Invariants of {$3$}-manifolds via link polynomials and quantum groups},
	url = {https://doi.org/10.1007/BF01239527},
	volume = {103},
	year = {1991}}

@article{KL-I,
	author = {Kazhdan, D. and Lusztig, G.},
	doi = {10.2307/2152745},
	fjournal = {Journal of the American Mathematical Society},
	issn = {0894-0347},
	journal = {J. Amer. Math. Soc.},
	mrclass = {17B67 (17B37)},
	mrnumber = {1186962},
	mrreviewer = {Ya. S. So\u{\i}bel\cprime man},
	number = {4},
	pages = {905--947, 949--1011},
	title = {Tensor structures arising from affine {L}ie algebras. {I}, {II}},
	url = {https://doi.org/10.2307/2152745},
	volume = {6},
	year = {1993}}

@article{Jones,
	author = {Jones, V. F. R.},
	doi = {10.1090/S0273-0979-1985-15304-2},
	journal = {Bull. Am. Math. Soc.},
	pages = {103--111},
	title = {{A polynomial invariant for knots via von Neumann algebras}},
	volume = {12},
	year = {1985}}

@article{GYtqft,
	archiveprefix = {arXiv},
	author = {Geer, Nathan and Young, Matthew B.},
	eprint = {2210.04286},
	month = {10},
	primaryclass = {math.QA},
	title = {{Three dimensional topological quantum field theory from $U_q(\mathfrak{gl}(1 \vert 1))$ and $U(1 \vert 1)$ Chern--Simons theory}},
	year = {2022}}

@article{GWjanus,
	author = {Gaiotto, Davide and Witten, Edward},
	doi = {10.1007/jhep06(2010)097},
	issn = {1029-8479},
	journal = {Journal of High Energy Physics},
	month = {Jun},
	number = {6},
	publisher = {Springer Science and Business Media LLC},
	title = {Janus configurations, Chern-Simons couplings, and The $\theta$-Angle in $ \mathcal{N} = 4 $ super Yang-Mills theory},
	url = {http://dx.doi.org/10.1007/JHEP06(2010)097},
	volume = {2010},
	year = {2010}}

@article{AGPS,
	archiveprefix = {arXiv},
	author = {Aghaei, Nezhla and Gainutdinov, Azat M. and Pawelkiewicz, Michal and Schomerus, Volker},
	eprint = {1811.09123},
	month = {11},
	primaryclass = {hep-th},
	reportnumber = {DESY-18-203, ZMP-HH/18-19, Hamburger Beitrage zur Mathematik 750},
	title = {{Combinatorial Quantisation of $GL(1|1)$ Chern-Simons Theory I: The Torus}},
	year = {2018}}

@article{KWemduality,
	archiveprefix = {arXiv},
	author = {Kapustin, Anton and Witten, Edward},
	doi = {10.4310/CNTP.2007.v1.n1.a1},
	eprint = {hep-th/0604151},
	journal = {Commun. Num. Theor. Phys.},
	pages = {1--236},
	title = {{Electric-Magnetic Duality And The Geometric Langlands Program}},
	volume = {1},
	year = {2007}}

@article{RSwzw,
	archiveprefix = {arXiv},
	author = {Rozansky, L. and Saleur, H.},
	doi = {10.1016/0550-3213(93)90326-K},
	eprint = {hep-th/9203069},
	journal = {Nucl. Phys. B},
	pages = {365--423},
	reportnumber = {YCTP-P10-92, UTTG-07-92},
	title = {{S and T matrices for the super $U(1,1)$ WZW model: Application to surgery and three manifolds invariants based on the Alexander-Conway polynomial}},
	volume = {389},
	year = {1993}}

@article{RSpoly,
	author = {Rozansky, L. and Saleur, H.},
	doi = {10.1016/0550-3213(92)90118-U},
	journal = {Nucl. Phys. B},
	pages = {461--509},
	title = {{Quantum field theory for the multivariable Alexander-Conway polynomial}},
	volume = {376},
	year = {1992}}

@article{RStorsion,
	archiveprefix = {arXiv},
	author = {Rozansky, L. and Saleur, H.},
	doi = {10.1016/0393-0440(94)90022-1},
	eprint = {hep-th/9209073},
	journal = {J. Geom. Phys.},
	pages = {105--123},
	reportnumber = {YCTP-P35-92, UTTG-21-92},
	title = {{Reidemeister torsion, the Alexander polynomial and $U(1,1)$ Chern-Simons Theory}},
	volume = {13},
	year = {1994}}

@article{WittenJones,
	author = {Witten, Edward},
	doi = {10.1007/BF01217730},
	editor = {Mitra, Asoke N.},
	journal = {Commun. Math. Phys.},
	pages = {351--399},
	reportnumber = {IASSNS-HEP-88-33},
	title = {{Quantum Field Theory and the Jones Polynomial}},
	volume = {121},
	year = {1989}}

@article{BCDN,
    author = "Ballin, Andrew and Creutzig, Thomas and Dimofte, Tudor and Niu, Wenjun",
    title = "{3d mirror symmetry of braided tensor categories}",
    eprint = "2304.11001",
    archivePrefix = "arXiv",
    primaryClass = "hep-th",
    month = "4",
    year = "2023"
}

@article{creutzig2022uprolling,
	author = {Creutzig, Thomas and Rupert, Matthew},
	journal = {Communications in Contemporary Mathematics},
	number = {04},
	pages = {2150023},
	publisher = {World Scientific},
	title = {Uprolling unrolled quantum groups},
	volume = {24},
	year = {2022}}

@article{creutzig2020quasi,
	author = {Creutzig, Thomas and Gainutdinov, Azat M and Runkel, Ingo},
	journal = {Communications in Contemporary Mathematics},
	number = {03},
	pages = {1950024},
	publisher = {World Scientific},
	title = {A quasi-Hopf algebra for the triplet vertex operator algebra},
	volume = {22},
	year = {2020}}

@article{BBFSanomalies,
    author = "Bhardwaj, Lakshya and Bullimore, Mathew and Ferrari, Andrea E. V. and Schafer-Nameki, Sakura",
    title = "{Anomalies of Generalized Symmetries from Solitonic Defects}",
    eprint = "2205.15330",
    archivePrefix = "arXiv",
    primaryClass = "hep-th",
    doi = "10.21468/SciPostPhys.16.3.087",
    journal = "SciPost Phys.",
    volume = "16",
    pages = "087",
    year = "2024"
}

@article{FLcenter,
    author = "Feigin, Boris L. and Lentner, Simon D.",
    title = "{Vertex algebras with big centre and a Kazhdan-Lusztig correspondence}",
    eprint = "2210.13337",
    archivePrefix = "arXiv",
    primaryClass = "hep-th",
    doi = "10.1016/j.aim.2024.109904",
    journal = "Adv. Math.",
    volume = "457",
    pages = "109904",
    year = "2024"
}

@incollection{etingof2016tensor,
	author = {Etingof, Pavel and Gelaki, Shlomo and Nikshych, Dmitri and Ostrik, Victor},
	optisbn = {978-1-4704-3441-0},
	optnumber = {205},
	optvolume = {Mathematical surveys and monographs},
	publisher = {American Mathematical Society},
	title = {{Tensor Categories}},
	year = {2015}}

@article{GaiottoTwisted,
	archiveprefix = {arXiv},
	author = {Gaiotto, Davide},
	doi = {10.1007/JHEP02(2019)061},
	eprint = {1611.01528},
	journal = {JHEP},
	pages = {061},
	primaryclass = {hep-th},
	title = {{Twisted compactifications of 3d $ \mathcal{N} $ = 4 theories and conformal blocks}},
	volume = {02},
	year = {2019}}

@article{CGP,
	author = {Costantino, Francesco and Geer, Nathan and Patureau-Mirand, Bertrand},
	doi = {10.1112/jtopol/jtu006},
	fjournal = {Journal of Topology},
	issn = {1753-8416},
	journal = {J. Topol.},
	mrclass = {57M27 (17B37)},
	mrnumber = {3286896},
	mrreviewer = {Leandro Vendramin},
	number = {4},
	pages = {1005--1053},
	title = {Quantum invariants of 3-manifolds via link surgery presentations and non-semi-simple categories},
	url = {https://doi.org/10.1112/jtopol/jtu006},
	volume = {7},
	year = {2014}}

@article{CDGG,
    author = "Creutzig, Thomas and Dimofte, Tudor and Garner, Niklas and Geer, Nathan",
    title = "{A QFT for non-semisimple TQFT}",
    eprint = "2112.01559",
    archivePrefix = "arXiv",
    primaryClass = "hep-th",
    doi = "10.4310/ATMP.2024.v28.n1.a4",
    journal = "Adv. Theor. Math. Phys.",
    volume = "28",
    number = "1",
    pages = "161--405",
    year = "2024"
}

@article{SS06,
	archiveprefix = {arXiv},
	author = {Schomerus, Volker and Saleur, Hubert},
	doi = {10.1016/j.nuclphysb.2005.11.013},
	eprint = {hep-th/0510032},
	journal = {Nucl. Phys. B},
	pages = {221--245},
	reportnumber = {SPHT-T05-152, DESY-05-186},
	title = {{The GL$(1|1)$ WZW model: From supergeometry to logarithmic CFT}},
	volume = {734},
	year = {2006}}

@article{CMY22,
	archiveprefix = {arXiv},
	author = {Creutzig, Thomas and McRae, Robert and Yang, Jinwei},
	doi = {10.1142/S0219199721500334},
	eprint = {2006.09711},
	journal = {Commun. Contemp. Math.},
	number = {02},
	pages = {2150033},
	primaryclass = {math.QA},
	title = {{Direct limit completions of vertex tensor categories}},
	volume = {24},
	year = {2022}}

@article{CKM17,
    author = {Creutzig, Thomas and Kanade, Shashank and Mcrae, Robert},
    year = {2017},
    month = {05},
    pages = {},
    title = {Tensor Categories for Vertex Operator Superalgebra Extensions},
    volume = {295},
    journal = {Memoirs of the American Mathematical Society},
    doi = {10.1090/memo/1472}
}

@article{AW22,
	archiveprefix = {arXiv},
	author = {Allen, Robert and Wood, Simon},
	doi = {10.1007/s00220-021-04305-6},
	eprint = {2001.05986},
	journal = {Commun. Math. Phys.},
	number = {2},
	pages = {959--1015},
	primaryclass = {math.QA},
	title = {{Bosonic Ghostbusting: The Bosonic Ghost Vertex Algebra Admits a Logarithmic Module Category with Rigid Fusion}},
	volume = {390},
	year = {2022}}

@article{CMY20,
	archiveprefix = {arXiv},
	author = {Creutzig, Thomas and McRae, Robert and Yang, Jinwei},
	doi = {10.1093/imrn/rnab080},
	eprint = {2009.00818},
	month = {9},
	primaryclass = {math.QA},
	title = {{Tensor structure on the Kazhdan-Lusztig category for affine $\mathfrak{gl}(1|1)$}},
	year = {2020}}

@article{CR13a,
	archiveprefix = {arXiv},
	author = {Creutzig, Thomas and Ridout, David},
	doi = {10.1016/j.nuclphysb.2013.04.007},
	eprint = {1107.2135},
	journal = {Nucl. Phys. B},
	pages = {348--391},
	primaryclass = {hep-th},
	title = {{Relating the Archetypes of Logarithmic Conformal Field Theory}},
	volume = {872},
	year = {2013}}

@article{CR13b,
	archiveprefix = {arXiv},
	author = {Creutzig, Thomas and Ridout, David},
	doi = {10.1007/978-4-431-54270-4_24},
	editor = {Dobrev, Vladimir},
	eprint = {1111.5049},
	journal = {Springer Proc. Math. Stat.},
	pages = {349--367},
	primaryclass = {hep-th},
	title = {{W-Algebras Extending Affine $\mathfrak{gl}(1|1)$}},
	volume = {36},
	year = {2013}}

@article{BN22,
    author = "Ballin, Andrew and Niu, Wenjun",
    title = "{3D mirror symmetry and the \ensuremath{\beta}\ensuremath{\gamma} VOA}",
    eprint = "2202.01223",
    archivePrefix = "arXiv",
    primaryClass = "hep-th",
    doi = "10.1142/S0219199722500699",
    journal = "Commun. Contemp. Math.",
    volume = "26",
    number = "01",
    pages = "2250069",
    year = "2024"
}

@article{Zeng,
    author = "Zeng, Keyou",
    title = "{Monopole operators and bulk-boundary relation in holomorphic topological theories}",
    eprint = "2111.00955",
    archivePrefix = "arXiv",
    primaryClass = "hep-th",
    doi = "10.21468/SciPostPhys.14.6.153",
    journal = "SciPost Phys.",
    volume = "14",
    number = "6",
    pages = "153",
    year = "2023"
}

@article{CostelloDimofteGaiotto-boundary,
    author = "Costello, Kevin and Dimofte, Tudor and Gaiotto, Davide",
    title = "{Boundary Chiral Algebras and Holomorphic Twists}",
    eprint = "2005.00083",
    archivePrefix = "arXiv",
    primaryClass = "hep-th",
    doi = "10.1007/s00220-022-04599-0",
    journal = "Commun. Math. Phys.",
    volume = "399",
    number = "2",
    pages = "1203--1290",
    year = "2023"
}

@article{KapustinSaulina-CSRW,
	author = {Kapustin, Anton and Saulina, Natalia},
	doi = {10.1016/j.nuclphysb.2009.07.006},
	issn = {0550-3213},
	journal = {Nuclear Physics B},
	month = {Dec},
	number = {3},
	pages = {403--427},
	publisher = {Elsevier BV},
	title = {{Chern-Simons-Rozansky-Witten topological field theory}},
	url = {http://dx.doi.org/10.1016/j.nuclphysb.2009.07.006},
	volume = {823},
	year = {2009}}

@article{AganagicCostelloMcNamaraVafa,
	archiveprefix = {arXiv},
	author = {Mina Aganagic and Kevin Costello and Jacob McNamara and Cumrun Vafa},
	eprint = {1706.09977},
	primaryclass = {hep-th},
	title = {Topological Chern-Simons/Matter Theories},
	year = {2017}}

@article{DimofteGaiottoPaquette,
	author = {Dimofte, Tudor and Gaiotto, Davide and Paquette, Natalie M.},
	doi = {10.1007/jhep05(2018)060},
	issn = {1029-8479},
	journal = {Journal of High Energy Physics},
	month = {May},
	number = {5},
	publisher = {Springer Science and Business Media LLC},
	title = {Dual boundary conditions in 3d SCFT's},
	url = {http://dx.doi.org/10.1007/JHEP05(2018)060},
	volume = {2018},
	year = {2018}}

@article{topCSM,
    author = "Garner, Niklas",
    title = "{Vertex operator algebras and topologically twisted Chern-Simons-matter theories}",
    eprint = "2204.02991",
    archivePrefix = "arXiv",
    primaryClass = "hep-th",
    doi = "10.1007/JHEP08(2023)025",
    journal = "JHEP",
    volume = "08",
    pages = "025",
    year = "2023"
}

@article{twisted,
    author = "Garner, Niklas",
    title = "{Twisted formalism for 3d ${\mathcal {N}}=4$ theories}",
    eprint = "2204.02997",
    archivePrefix = "arXiv",
    primaryClass = "hep-th",
    doi = "10.1007/s11005-023-01758-9",
    journal = "Lett. Math. Phys.",
    volume = "114",
    number = "1",
    pages = "16",
    year = "2024"
}

@article{Mikhaylov,
	archiveprefix = {arXiv},
	author = {Mikhaylov, Victor},
	eprint = {1505.03130},
	month = {5},
	primaryclass = {hep-th},
	title = {{Analytic Torsion, 3d Mirror Symmetry And Supergroup Chern-Simons Theories}},
	year = {2015}}

@article{creutzig2023algebraic,
    archiveprefix = {arXiv},
eprint = {2306.11492},
month = {6},
  title="{An algebraic theory for logarithmic Kazhdan-Lusztig correspondences}",
  author={Creutzig, Thomas and Lentner, Simon and Rupert, Matthew},
primaryclass = {math-QA},
  year={2023}
}

@article{Finkelberg1,
	author = {Finkelberg, Michael},
	date = {1996/03/01},
	doi = {10.1007/BF02247887},
	id = {Finkelberg1996},
	isbn = {1420-8970},
	journal = {Geometric \& Functional Analysis GAFA},
	number = {2},
	pages = {249--267},
	title = {An equivalence of fusion categories},
	url = {https://doi.org/10.1007/BF02247887},
	volume = {6},
	year = {1996}
}

@article{Finkelberg2,
	author = {Finkelberg, Michael},
	date = {2013/04/01},
	doi = {10.1007/s00039-013-0230-y},
	id = {Finkelberg2013},
	isbn = {1420-8970},
	journal = {Geometric and Functional Analysis},
	number = {2},
	pages = {810--811},
	title = {Erratum to: An Equivalence of Fusion Categories},
	url = {https://doi.org/10.1007/s00039-013-0230-y},
	volume = {23},
	year = {2013}
}

\end{document}